\documentclass[twocolumn]{aastex7}

\usepackage{amsmath}

\def\muG{\mu\mathrm{G}}
\def\kms{\mathrm{km\, s^{-1}}}

\received{}
\revised{}
\accepted{}

\submitjournal{The Astrophysical Journal}

\shorttitle{Binary Galaxy Cluster Mergers}

\begin{document}

\title{Simulation Study of Binary Mergers of Galaxy Clusters I: Properties of Merger Shocks and Radio Emission}

\author[0000-0002-4674-5687]{Hyesung Kang}
\affiliation{Department of Earth Sciences, Pusan National University, Busan 46241, Korea}
\email{hskang@pusan.ac.kr}

\author[0000-0002-5455-2957]{Dongsu Ryu}
\affiliation{Department of Physics, College of Natural Sciences, UNIST, Ulsan 44919, Korea}
\email{dsryu@unist.ac.kr}

\author[0000-0002-5550-8667]{Jeongbhin Seo}
\affiliation{Department of Physics, College of Natural Sciences, UNIST, Ulsan 44919, Korea}
\affiliation{Los Alamos National Laboratory, Theoretical Division, Los Alamos, NM 87545, USA}
\email{jseo@lanl.gov}

\correspondingauthor{Dongsu Ryu}\email{dsryu@unist.ac.kr}

\begin{abstract}

We investigate binary mergers of galaxy clusters, the formation of shocks, and the resulting radio relics using three-dimensional simulations. The initial setup consists of two idealized spherical subclusters with a mass ratio below three, each permeated by turbulent magnetic fields, and we follow their merger with a high-order accurate magnetohydrodynamic (MHD) code. In parallel, we track the acceleration of cosmic-ray electrons (CRe) via diffusive shock acceleration (DSA) at merger-driven shocks, together with radiative cooling and Fermi-II (turbulent) acceleration in the postshock region, employing a {newly developed Eulerian} Fokker–Planck solver. Synchrotron emission is computed from the simulated CRe distribution and magnetic fields. In this paper, we detail these numerical approaches and present the first results obtained with them. Two prominent axial shocks emerge along the merger axis; the shock ahead of the heavier subcluster systematically attains a higher Mach number, although it is more compact, than that ahead of the lighter subcluster. Turbulent magnetic fields, which are both inherited from the initial conditions and amplified during the merger, produce patchy, fine-scale structures in the radio surface brightness. Because of the combined effects of turbulent acceleration, spatially nonuniform magnetic fields, and the curved geometry of merger shocks, the volume-integrated radio spectra show deviations from the canonical power-law steepening expected for a planar shock with a uniform field. Reacceleration of preexisting fossil CRe enhances the surface brightness. Our results highlight the coupled roles of merger dynamics, MHD turbulence, and CRe physics in shaping the observed properties of radio relics in cluster outskirts.

\end{abstract}


\section{Introduction}\label{s1}

Galaxy clusters assemble over cosmic time via hierarchical clustering within the large-scale structure (LSS) of the Universe, in which subclusters and groups continuously merge. These merger events drive shock waves and turbulent flows in the intracluster medium (ICM). The resulting dissipation of gravitational energy not only heats the ICM but also amplifies magnetic fields and accelerates cosmic rays (CRs) \citep[e.g.,][]{sarazin2002,ryu2003,ryu2008,brunetti2014}.

When two subclusters of comparable mass collide with a relatively small impact parameter, the encounter is termed a major binary merger. Following pericenter passage of the dark matter (DM) cores, such mergers generate pairs of bow shocks, commonly referred to as merger shocks \citep[e.g.,][]{markevitch2007}. Numerical simulations of idealized mergers, in which the initial subclusters are taken to be spherical and in hydrostatic equilibrium (HSE), have shown that {the strength, morphology, and asymmetry of merger-driven shocks are influenced by merger parameters, such the mass ratio, impact parameter, and orbital geometry, along with projection effects \citep[e.g.,][]{evrard1990,schindler1993,roettiger1993,ricker2001,gabichi2003,springel2007,vanweeren2011a,zuhone2011,zuhone2018,donnert2017,molnar2017}.}

In the broader context of LSS formation, various populations of shocks, including merger shocks, are induced within galaxy clusters during their formation {\citep[e.g.,][]{ryu2003,hoeft2008,paul2011,skillman2011,schmidt2017,ha2018,lokas2023,leewk2024,leewk2025,lee2025}.} Cosmological simulations reveal that merger shocks, which dissipate the largest fraction of gravitational energy among different populations of cosmologically induced shocks, exhibit a wide range of strengths and a variety of shapes, reflecting the diversity of cluster assembly histories. Bow shocks formed along the merger axis (hereafter axial shocks) can extend up to several Mpc into the cluster outskirts, whereas weaker shocks develop in the plane perpendicular to the merger axis (hereafter equatorial shocks) \citep[e.g.,][]{ha2018}. The asymmetry between axial shocks propagating ahead of the heavier and lighter subclusters arises from their different gravitational potentials, resulting in different spatial extents and Mach number distributions \citep[e.g.,][]{lee2025}.

Merger shocks have been proposed as sites for the acceleration of relativistic particles via diffusive shock acceleration (DSA). Although the sonic Mach numbers of these shocks are typically modest \citep[$M_s \lesssim 5$; see, e.g., Table 3 of][]{lee2025}, they should nevertheless be capable of producing CRs, once their Mach number exceeds the critical value, $M_{s,\rm crit}\approx2.3$ \citep{ha2018b,kang2019,ha2021,ha2022}. In particular, the electron population of CRs can emit synchrotron radiation at radio frequencies in the presence of $\mu$G-level magnetic fields. This mechanism is widely invoked to explain the origin of ``radio relics'' observed in cluster outskirts \citep[e.g.,][]{ensslin1998,hoeft2007,kang2012,pinzke2013}.

The morphology, spectrum, and polarization of radio relics can provide valuable diagnostics for probing the properties of underlying shocks and the physical conditions of the ambient medium \citep[e.g.,][]{vanweeren2019}. However, the interplay among shock dynamics, particle acceleration, and magnetic field distribution is complex and hence remains not fully understood. The impacts of additional physical processes such as the reacceleration of preexisting fossil CR electrons (CRe) at shocks and postshock Fermi-II (turbulent) acceleration also need to be investigated \citep[e.g.,][]{kang2017,brunetti2014,kang2024}.

As mentioned above, along with shocks, turbulent flow motions are naturally induced in the ICM during the formation of galaxy clusters \citep[e.g.,][]{miniati2015,vazza2017}. This turbulence is typically subsonic, with a turbulent Mach number $M_{s,\rm turb} \lesssim 0.5$, and characterized by outer-scale eddies of size $L \sim 100-500$~kpc \citep[e.g.,][]{gaspari2014,hofmann2016}. Magnetic fields in the ICM are amplified via the turbulent dynamo, producing intermittent, filamentary structures \citep[e.g.,][]{ryu2008,cho2009,porter2015}. In addition, turbulence is further generated behind shocks through vorticity production, which enhances magnetic fields in the postshock region \citep[see, e.g.,][]{porter2015}. The combined effects of turbulent dynamo and shock compression lead to substantial amplification of magnetic fields downstream of shocks. Since these magnetic fields strongly influence the observed spectra, polarization, and fine-scale structure of synchrotron emission \citep[e.g.,][]{bruggen2012,dominguez-fernandez2021}, reproducing radio relics in merger-shock simulations requires high-quality magnetohydrodynamic (MHD) modeling.

With low Mach numbers, DSA at merger shocks is expected to operate in the test-particle regime \citep[e.g.,][]{caprioli2014}. The non-detection of gamma-ray emission from galaxy clusters places a stringent constraint on the acceleration efficiency of CR protons (CRp), i.e., $\eta_p\equiv \varepsilon_{\rm CRp,2} u_2/(0.5\rho_{g,1} V_s^3)\lesssim 0.03$, where the numerator represents the postshock CRp energy flux and the denominator corresponds to the shock kinetic energy flux \citep[e.g.,][]{ha2020,wittor2020}. Under these conditions, the momentum distributions of CRp and CRe downstream of merger shocks can be approximated by a DSA power-law form, $f(p)\propto p^{-q_{\rm sh}}$, where the slope depends solely on the Mach number as $q_{\rm sh}=4M_s^2/(M_s^2-1)$.

Furthermore, within the thermal-leakage injection model, the DSA power-law spectra for both CRp and CRe are expected to emerge from their respective Maxwellian distributions at the injection momenta, parameterized as $p_{\rm inj,p} = Q_p p_{\rm th,p}$ and $p_{\rm inj,e} = Q_e p_{\rm th,e}$. {Here, $p_{\rm th,p}=(2m_p k_B T_2)^{1/2}$ and $p_{\rm th,e}=(2m_e k_B T_2)^{1/2}$ denote the postshock thermal momenta for protons and electrons, respectively \citep[e.g.,][]{kang2020,kang2024}. The proton injection parameter is expected to lie in the range $Q_p\approx 3.5-3.8$, based on results from plasma hybrid and particle-in-cell (PIC) simulations \citep[e.g.,][]{caprioli2014,ha2018b}. Although electron preacceleration from $p_{\rm th,e}$ up to $p_{\rm inj,p}$ has not been fully explored due to the computational limitations of PIC simulations, it is commonly assumed that various micro-instabilities and wave-particle interactions in the shock transition zone lead to $Q_e\sim Q_p$ \citep[e.g.,][]{kang2019,trotta2019,arbutina2021}.}

Within the thermal-leakage framework, previous studies have proposed that the postshock gas temperature, $T_2 = R_T T_{2,0}$, and the injection parameters, $Q_p$ and $Q_e$, are self-regulated so as to maintain the test-particle condition, as thermal energy is transferred to CRp energy \citep{ryu2019,kang2024}. Here, $T_{2,0}$ denotes the postshock temperature estimated from the Rankine-Hugoniot relation, and $R_T \le 1$ is the temperature reduction factor. In this work, we adopt the analytic fitting forms for $Q_e(M_s)$ and $R_T(M_s)$ proposed in \citet{kang2024} to model the injection and acceleration of CRe at weak shocks.

{The CRe produced at merger shocks may undergo additional energization through Fermi-II acceleration in the downstream region \citep{donnert2014,brunetti2014,kang2024}. This process can occur through scattering off compressible fast-mode waves via transit-time–damping (TTD) resonance \citep{brunetti2007} or off Alfv\'en waves via gyroresonance at kinetic plasma scales \citep[e.g.,][]{brunetti2004,fujita2015}. Such waves may originate from the cascade of cluster-scale MHD turbulence as well as from vorticity production behind curved shocks \citep[e.g.,][]{porter2015,vazza2017}. Preexisting turbulence can also be amplified as shocks propagate through the ICM, for example, through shock-surface rippling and related instabilities \citep[e.g.,][]{guo2015,trotta2023}. Furthermore, kinetic-scale plasma waves may be excited directly within the shock transition zone by microinstabilities driven by reflected ions and temperature anisotropies, which are subsequently advected downstream \citep[e.g.,][]{treumann2009,marcowith2016,kim2021}.}

{For Fermi-II acceleration, we here adopt a simplified prescription, in which the momentum-diffusion coefficient is defined as $D_{\rm pp}=4p^2/\tau_{\rm pp}$, where $\tau_{\rm pp}=0.1-1$~Gyr represents the turbulent acceleration timescale. The rationale for this approach is twofold: (1) the microphysics governing wave generation and wave-particle interactions remains poorly understood, particularly in the high-plasma-beta ($\beta_p \gtrsim 50$) ICM \citep[e.g.,][]{ryu2008,brunetti2014}; and (2) these processes, especially at kinetic scales, cannot be adequately resolved in fluid simulations with limited numerical resolution. Consequently, this work should be regarded as an initial exploratory study of the impact of turbulent acceleration on the CRe population in merging clusters.}

In this study, we extend our previous investigations of merger shocks \citep[e.g.,][]{ha2018,leewk2025} by performing simulations of binary mergers of idealized, spherical subclusters initially in HSE. Each subcluster is initialized with turbulent magnetic fields. We follow the formation of merger shocks and the amplification of turbulent magnetic fields, as well as the DSA, subsequent turbulent acceleration, advection, and cooling of CRe. The resulting synchrotron emission behind the shocks is then computed based on the simulated CRe population and magnetic field distribution. 

{To achieve this, our numerical framework incorporates the following components: (1) a three-dimensional (3D) solver for the ideal MHD equations on a uniform Eulerian grid; (2) gravitational acceleration from fixed potentials representing spherical halos in HSE, containing both dark matter and gas; and (3) an ``Eulerian'' Fokker--Planck solver for the CRe momentum distribution defined on the MHD grid.}

{We point out that many MHD simulations of LSS formation achieve high spatial resolution using schemes such as Lagrangian Smoothed Particle Hydrodynamics (SPH) \citep{springel2005,dolag2009}, Adaptive Mesh Refinement (AMR) \citep{berger1989, bryan2014,mignone2007,fryxell2000,lee2009}, or quasi-Lagrangian moving-mesh method \citep{springel2010,pakmor2011}. In contrast, our numerical framework employs the HOW-MHD code, a code based on the high-order accurate Weighted Essentially Non-Oscillatory (WENO) scheme implemented on a fixed uniform grid \citep{seo2023}. Our simulations of high-order accuracy could be complementary to those that primarily focus on achieving high spatial resolution.}

{A common strategy for modeling the acceleration and subsequent evolution of CRe in LSS formation simulations is to evolve either their momentum distribution function $f(p,\mbox{\boldmath$x$})$ or the number density $n(p,\mbox{\boldmath$x$})=4\pi p^2 f$ by solving the Fokker–Planck equation. This solver is coupled to the background fluid dynamics computed with HD/MHD codes and incorporates adiabatic effects, turbulent acceleration, and radiative cooling of CRe, as well as localized injection or reacceleration at shocks.}

{For the Fokker–Planck solver, some studies have adopted a ``Lagrangian'' tracer or macro-particle approach, in which passive tracers (or SPH particles) are advected with the flow \citep[e.g.,][]{pinzke2013,donnert2014,wittor2017,winner2019,vaidya2018,kundu2021}. Each tracer carries a discretized CRe spectrum that is updated along its trajectory, either during the simulation or in post-processing. In this approach, the tracers capture fluid-dynamical history along their paths, including repeated shock encounters, while the computational cost and memory requirements can be controlled by limiting the number of tracers.}

{In other studies, ``Eulerian'' approaches have been adopted for the Fokker–Planck solver, in which the CRe momentum distribution is assigned on the Eulerian grid as fluid quantities. In one of such approaches, the Fokker–Planck equation is solved assuming a piecewise power-law distribution for $f(p,\mbox{\boldmath$x$})$ within each momentum bin; then, either $n(p,\mbox{\boldmath$x$})$ and the slope of $f(p,\mbox{\boldmath$x$})$ in the momentum bin, $q_i(p,\mbox{\boldmath$x$})$, or $n(p,\mbox{\boldmath$x$})$ and $e_c(p,\mbox{\boldmath$x$})=4\pi p^3 f(p,\mbox{\boldmath$x$})$ (the CRe energy density) are followed \citep[e.g.][]{jones1999,miniati2001,girichidis2020,ogrodnik2021}. With these methods, the CRe spectrum can be reasonably reproduced with a limited number of momentum bins, typically of order 10. This feature makes 3D MHD simulations with the CRe momentum distribution (so effectively four-dimensional in phase space) computationally feasible, although the detailed implementation is somewhat involved.} 

{In contrast, our Eulerian Fokker-Planck solver evolves $e_c(p,\mbox{\boldmath$x$})$, defined in every spatial grid zone, assuming it is piecewise constant within logarithmic momentum bins. This scheme is more straightforward and intuitive than those following $n(p,\mbox{\boldmath$x$})$-$q_i(p,\mbox{\boldmath$x$})$ or $n(p,\mbox{\boldmath$x$})$-$e(p,\mbox{\boldmath$x$})$, although it requires a relatively large number of momentum bins, $N_p$.  We typically employ $N_p=64$, spanning the CRe momentum range $\gamma_e=p/m_ec=10-10^5$. Compared to the approach based on Lagrangian tracers, ours entails high computational cost and memory usage; however, it can track turbulent acceleration, as well as the evolution of preexisting CRe population, throughout the entire computational domain.}

In this paper, we present our numerical framework and, as its first application, investigate {the generic characteristics} of idealized binary mergers. In particular, we examine the strength and shape of merger shocks, the production and evolution of CRe, and the synchrotron radio emission behind the shocks under varying merger parameters. We expect that this work will provide theoretical insights that help interpret the diverse radio relics observed in merging clusters. 

The paper is organized as follows. Section~\ref{s2} describes the numerical details and simulations. Section~\ref{s3} presents the simulation results. Section~\ref{s4} provides a summary. The appendix includes technical details.

\begin{deluxetable*}{cccccccccc}\label{t1}
\tablecolumns{10}
\tablewidth{0pt}
\tablecaption{Parameters of the initial subclusters}
\tablehead{
\colhead{Name} &
\colhead{$M_{200}$$^a$} &
\colhead{$r_{200}$$^b$} &
\colhead{$T_{200}$$^c$} &
\colhead{$T_{\rm cen}$$^c$} &
\colhead{$\langle T \rangle$$^c$} &
\colhead{$c_{\rm s,200}$$^d$} &
\colhead{$V_{\rm circ,200}$$^e$} &
\colhead{$\langle M_{s,\rm turb} \rangle$$^f$} &
\colhead{$\langle \beta_p \rangle$$^g$} \\
\colhead{} & 
\colhead{($M_{\sun}$)} & 
\colhead{(Mpc)} & 
\colhead{(keV)} & 
\colhead{(keV)} & 
\colhead{(keV)} & 
\colhead{(km\,s$^{-1}$)} & 
\colhead{(km\,s$^{-1}$)} & 
\colhead{} & 
\colhead{} 
}
\startdata
\texttt{m2} & $2\times10^{14}$ & 1.2 & 1.56 & 5.76 & 1.82 & $6.37\times 10^2$ & $0.85\times 10^3$ & 0.25 & 108 \\ 
\texttt{m4} & $4\times10^{14}$ & 1.5 & 2.47 & 9.18 & 2.81 & $8.02\times 10^2$ & $1.07\times 10^3$ & 0.26 & 85 \\ 
\texttt{m6} & $6\times10^{14}$ & 1.7 & 3.12 & 12.0 & 3.63 & $9.22\times 10^2$ & $1.22\times 10^3$ & 0.26 & 100
\enddata
\tablenotetext{a}{Total mass (DM plus gas) within $r\le r_{200}$.}
\tablenotetext{b}{Radius within which the mean total (DM plus gas) density of the cluster attains 200 times the critical density of the Universe.}
\tablenotetext{c}{Gas temperatures at $r_{200}$ and at the cluster center, and the volume-averaged mean temperature within $r\le r_{200}$.}
\tablenotetext{d}{Sound speed at $r_{200}$.}
\tablenotetext{e}{Circular velocity at $r_{200}$, $V_{\rm circ,200}=(GM_{200}/r_{200})^{1/2}$.}
\tablenotetext{f}{{Root-mean-square (rms) turbulent Mach number, $\langle M_{s,\rm turb} \rangle \equiv \langle (u_{\rm turb}/c_s)^2 \rangle^{1/2}$, within $r\le r_{200}$, where $u_{\rm turb}$ and $c_s$ are the local turbulent velocity and sound speed, respectively.}}
\tablenotetext{g}{{Volume-averaged  mean plasma beta, $\langle \beta_p \rangle \equiv \langle P_g/P_B\rangle$, within $r\le r_{200}$, where $P_g$ is the gas pressure and $P_B$ is the magnetic pressure with the turbulent component of magnetic fields.}}
\vskip -0.8cm
\end{deluxetable*}

\section{Numerical Details and Simulations}\label{s2}

This section describes the following: (1) the setup of the initial subclusters; (2) the merger simulations, including MHD gas flow dynamics and gravity; (3) the Fokker–Planck solver, together with CRe injection at shocks and Fermi II (turbulent) acceleration; and (4) the calculation of radio synchrotron emission and surface brightness.

\subsection{Physical Constants and Normalizations}\label{s2.1}

We begin by listing the physical constants that appear below: the gravitational constant $G$, the speed of light $c$, the proton and electron masses $m_p$ and $m_e$, the electron charge $q_e$, and the Boltzmann constant $k_B$. The reference units for the physical quantities used are as follows:
\begin{align*}
n_0 &= 10^{-3}~\mathrm{cm}^{-3}, \\
\rho_0 &= 2.34 \times 10^{-27}~\mathrm{g\,cm}^{-3}, \\
r_0 &= 10~\mathrm{Mpc}, \\
t_0 &= (G \rho_0)^{-1/2} = 2.54~\mathrm{Gyr}, \\
u_0 &= r_0/t_0 = 3.86 \times 10^3~\mathrm{km\,s^{-1}}, \\
P_0 &= \rho_0 u_0^2 = 3.49 \times 10^{-10}~\mathrm{erg\,cm}^{-3}, \\
k_B T_0 &= \mu m_p P_0/\rho_0 = 94.8~\mathrm{keV}, \\
B_0 &= (4\pi \rho_0)^{1/2}\, u_0 = 66.2~\mu\mathrm{G},\\
L_0 &= \rho_0^2 T_0^{1/2} r_0,\\
S_0 &= 1 ~\mathrm{mJy~beam^{-1}}.
\end{align*}
{Here, $L_0$ is for the bolometric bremsstrahlung emission integrated along the line of sight (LoS), and $S_0$ is for the synchrotron surface brightness per beam ($1~\mathrm{mJy}=10^{-26}~\mathrm{erg\,s^{-1}\,cm^{-2}\,Hz^{-1}}$). The reference beam size is $\theta_{\rm beam}^2 = 1'' \times 1''$. These units are used to normalize the quantities in figures, unless otherwise specified.} 

\subsection{Model for Initial Subclusters in HSE}\label{s2.2}

Each subcluster is initially spherically symmetric and initialized as follows. The virial radius is defined as
\begin{equation}
r_{200}\equiv\left[\frac{3M_{200}}{4\pi \cdot 200 \rho_{\rm crit}}\right]^{1/3},
\label{r200}
\end{equation}
where $\rho_{\rm crit}=3H_0^2/8\pi G$ is the critical density of the Universe and $H_0=67.4~\mathrm{km\,s^{-1}\,Mpc^{-1}}$ is the present-day Hubble constant. Each merging subcluster is characterized by its total mass (DM plus gas), $M_{200}$, enclosed within $r_{200}$.

The DM density distribution is modeled by the Navarro–Frenk–White (NFW) profile,
\begin{equation}
\rho_{\rm DM}(r) = \frac{\rho_s}{(r/r_s)(1+r/r_s)^2},
\label{rhoDM}
\end{equation}
where $r$ is the radial distance from the cluster center. Here, $r_s=r_{200}/c_{200}$, where $c_{200}$ is the concentration parameter, and we adopt $c_{200}=5$ \citep{navarro1997}. The characteristic density $\rho_s$ is chosen such that the DM mass enclosed within $r_{200}$ equals $f_{\rm DM}M_{200}$, where the DM mass fraction is $f_{\rm DM}=0.86$. 

The gas density follows a beta profile with $\beta=1$,
\begin{equation}
\rho_g(r)=\frac{\rho_c}{[1+(r/r_c)^2]^{3/2}},
\label{rhog}
\end{equation}
where the core radius is $r_c=0.2\ r_{200}$ \citep{cavaliere1976}. The core density $\rho_c$ is set so that the gas mass enclosed within $r_{200}$ equals $f_{\rm g}M_{200}$, with the gas mass fraction $f_{\rm g}=0.14$.

The gas temperature at $r_{200}$ is assumed to be
\begin{equation}
k_B T_{200} = 0.7\ T_{\rm vir}=0.7\left(\frac{ \mu m_p G M_{200}}{2 r_{200}}\right),
\label{T200}
\end{equation}
where the mean molecular weight is $\mu=0.592$ \citep{poole2006,zuhone2011}. The numerical factor 0.7 is chosen to ensure a smooth transition of the initial HSE across $r_{200}$. The temperature profile $T(r)$ is then obtained by integrating the HSE condition inward from $r_{200}$,
\begin{equation}
g_{\rm HSE}(r) = \frac{GM(\le r)}{r^2}
               = \frac{1}{\rho_g}\left|\frac{dP_g}{dr}\right|,
\label{HSE}
\end{equation}
where $M(\le r)=4\pi \int_0^r \big(\rho_{\rm DM}+\rho_g\big) r'^2 \, dr'$ is the mass enclosed within $r$, and the gas pressure is given by $P_g(r)= (k_B/\mu m_p)\rho_g T$. Beyond the virial radius, the ICM is assumed to be isothermal with $T(r_{200})=T_{200}$. The resulting parameters for the model subclusters \texttt{m2}, \texttt{m4}, and \texttt{m6} are listed in Table~\ref{t1}.

\begin{figure}[t]
\vskip 0.1cm
\includegraphics[width=0.97\linewidth]{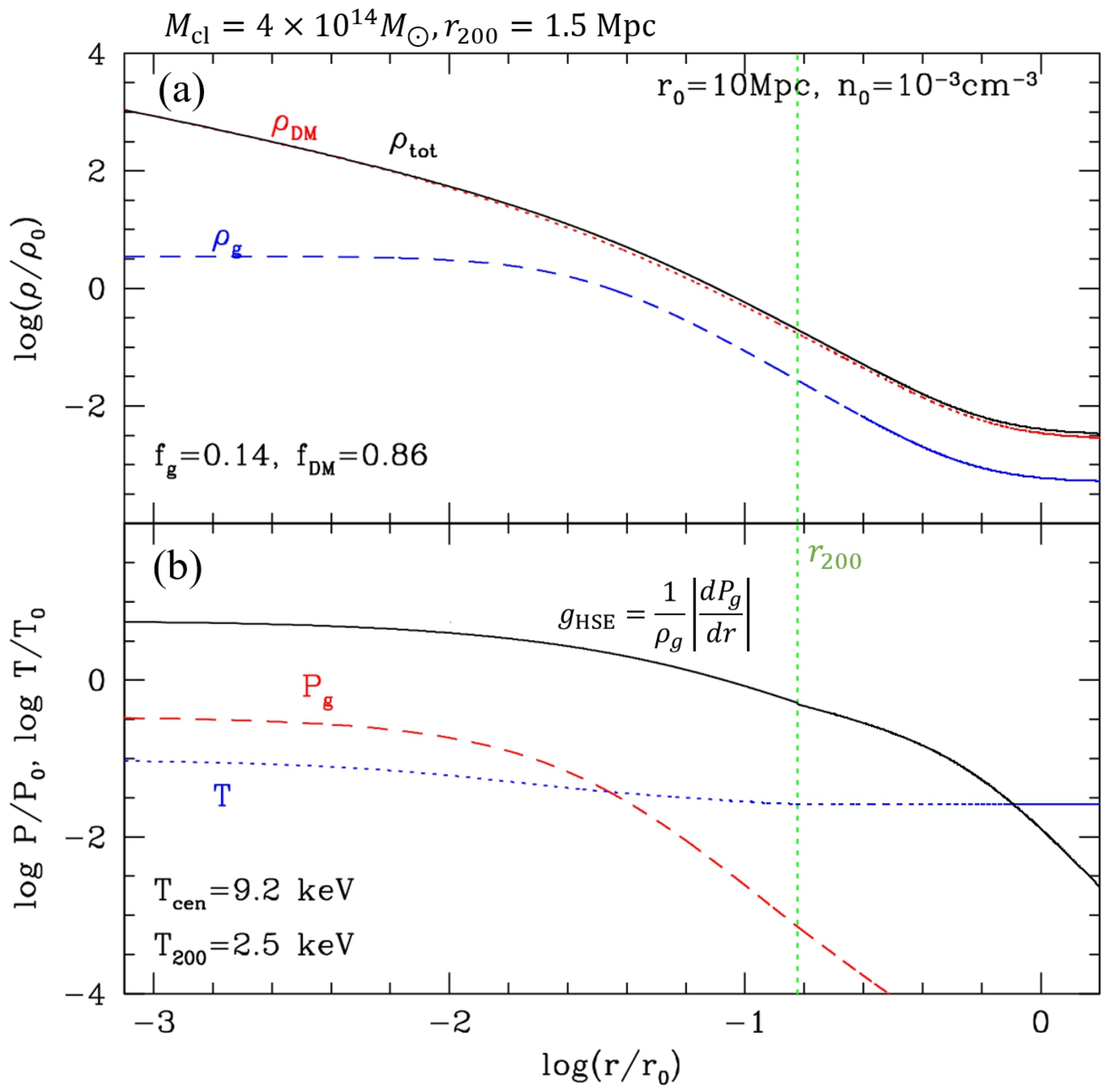}
\vskip -0.1cm
\caption{{(a) Radial profiles of the DM density (red dotted), gas density (blue dashed), and total density (black solid) for the \texttt{m4} subcluster with $M_{200}=4\times 10^{14}\,M_{\sun}$. At small radii, the red dotted and black solid lines nearly overlap. (b) Radial profiles of the gas temperature (blue dotted), gas pressure (red dashed), and the HSE gravitational acceleration (black solid). The green vertical dotted line marks $r_{200}$. The normalization constants, $r_0$, $\rho_0$, $P_0$, and $k_BT_0$, are given in Section \ref{s2.1}.}}\label{f1}
\end{figure}

{Figure~\ref{f1} shows the radial profiles of $\rho_{\rm DM}$, $\rho_g$, $\rho_{\rm tot}$, $T$, $P_g$, and $g_{\rm HSE}$ for the \texttt{m4} model with $M_{200}=4\times10^{14}\,M_{\sun}$. These profiles define the initial equilibrium state of each isolated subcluster. The adopted radial profiles are chosen to be broadly consistent with the observed properties of typical clusters. For example, the \texttt{m4} model has a central electron density of $n_e \approx 3.8 \times 10^{-3}~\mathrm{cm}^{-3}$ and a central temperature of $k_B T \approx 9.2$~keV, corresponding to an initial central entropy of $K_0 = k_B T n_e^{-2/3} \approx 380~\mathrm{keV\,cm}^2$. This relatively high entropy is characteristic of non--cool-core clusters \citep[e.g.,][]{cavagnolo2009}.}

\begin{figure*}[t]
\hskip 0.5 cm
\includegraphics[width=0.92\linewidth]{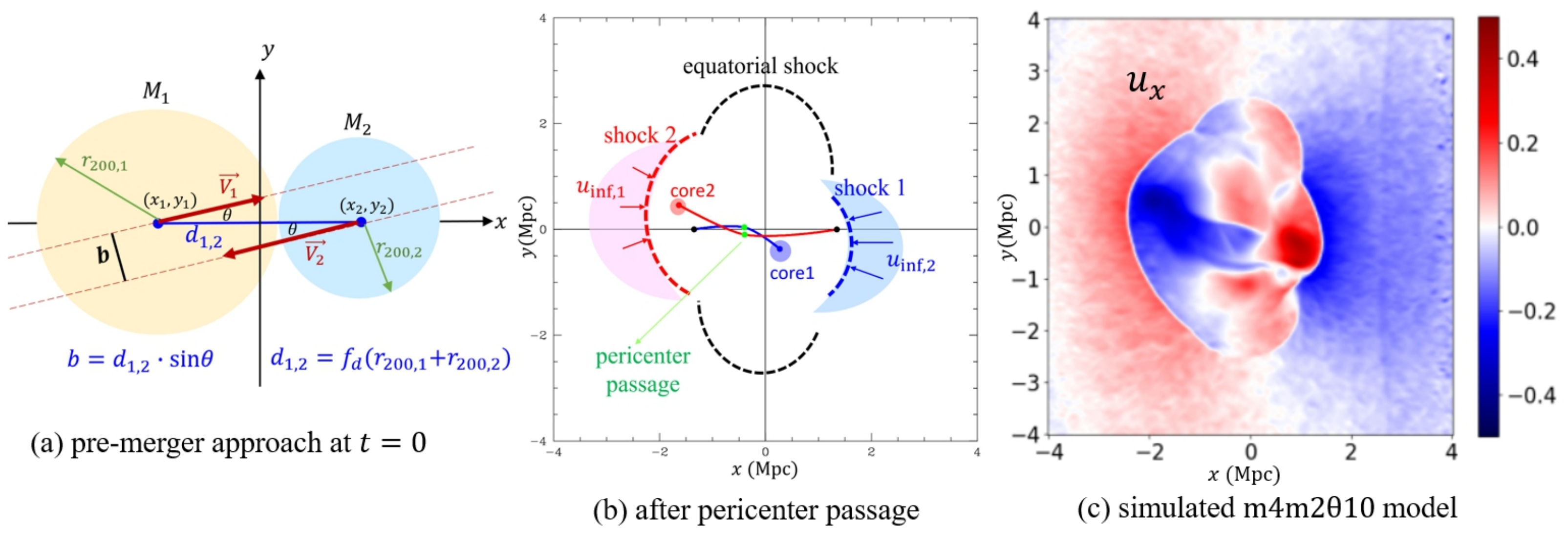}
\vskip -0.1 cm
\caption{(a) Schematic diagram of a binary merger at $t=0$. The initial separation of the two subclusters is $d_{1,2} = f_d \cdot (r_{200,1}+r_{200,2})$ with $f_d = 1$. The impact parameter is $b = d_{1,2} \sin \theta$, where $\theta$ is the velocity inclination angle. (b) Illustration of the shock geometry in the merger plane ($z=0$) containing the merger axis (the $x$-axis) at $\sim1$ Gyr after pericenter passage in the \texttt{m4m2$\theta$10} model simulation with \texttt{m4} and \texttt{m2} subclusters and $\theta=10^\circ$. Axial shock~1 (blue) propagates ahead of the heavier subcluster, axial shock~2 (red) propagates ahead of the lighter subcluster, and equatorial shocks (black) expand in the direction perpendicular to the merger axis. The solid blue (red) curve traces the trajectory of the heavier (lighter) gravity halo. Black dots mark the initial core positions. Pinkish (bluish) regions denote inflows associated with the heavier (lighter) subcluster; the lighter subcluster inflow is faster ($u_{\rm inf,2} > u_{\rm inf,1}$). (c) 2D distribution of $u_x/u_0$ at $\sim1$ Gyr after pericenter passage in the \texttt{m4m2$\theta$10} model simulation.}\label{f2}
\end{figure*}

Focusing on the gas dynamics relevant to merger-shock formation, we approximate the gravitational force acting on the ICM by introducing two ``gravity halos’’ whose fields are prescribed by the initial HSE profiles in equation~(\ref{HSE}), namely $g_{\rm HSE,1}(r_1)$ and $g_{\rm HSE,2}(r_2)$ for Subclusters~1 and 2. Here, $r_1=\big[(x-x_1)^2+(y-y_1)^2+(z-z_1)^2\big]^{1/2}$ and $r_2=\big[(x-x_2)^2+(y-y_2)^2+(z-z_2)^2\big]^{1/2}$ denote the distances from the respective halo centers. The motion of each gravity halo is followed by evolving its center position, $(x_1,y_1,z_1)$ or $(x_2,y_2,z_2)$, under the influence of these prescribed gravitational fields.

{We adopt this simplified treatment rather than solving for self-consistent gravity from the evolving density field. This choice is motivated by our primary objective of investigating the generic properties of merger shocks and their associated particle acceleration and radio emission, rather than reproducing the detailed evolution of individual systems. We acknowledge that this approach does not fully capture the complex gravitational feedback associated with the deformation of DM and gas halos, particularly during and after pericenter passage. Studies aimed at modeling specific observed clusters would therefore require a fully self-consistent treatment of gravity.}

{On the other hand, we expect that this simplification has a rather minor impact on the properties of merger shocks and their radio signatures in cluster outskirts. As demonstrated by \citet{vanweeren2011a}, although gravitational interactions such as dynamical friction and tidal stripping can modify the trajectories of subclusters, the primary characteristics of merger-driven shocks are largely determined by the initial merger parameters, such as the mass ratio and impact parameter. In the low-density peripheral regions of the ICM, where the gravitational potential varies smoothly, shock dynamics are governed mainly by the conversion of kinetic energy into thermal and nonthermal components. Therefore, our approach should provide a reasonable and physically motivated description of the merger-shock properties examined in this study.}

\subsection{MHD Simulations}\label{s2.3}

We calculate the flow dynamics of the gas component using the HOW-MHD code by solving the set of {ideal MHD equations with an adiabatic equation of state.} The code is based on a High-Order WENO (HOW) scheme combined with a strong stability-preserving Runge–Kutta (SSPRK) time integrator, and it implements a high-order version of the constrained transport (CT) scheme ensuring the divergence-free condition of the magnetic field. Further details of the HOW-MHD code can be found in \citet{seo2023}. CRe are treated as test particles, since their energy remains only a small fraction of the gas energy and thus their presence would not influence the dynamics of the gas flow. (The CRe solver is described in Section \ref{s2.5}.)

Each simulation consists of two stages: (1) a pre-merger MHD run to generate turbulent flow motions and magnetic fields for each subcluster, in which turbulence is driven for $\sim 2$~Gyr, and (2) a binary-merger MHD run including the CRe component, in which the collision of the two subclusters is followed for $\sim 3.8$~Gyr.

\subsubsection{Simulations for Pre-merger Turbulence Generation}\label{s2.3.1}

The ICM in merging subclusters is expected to be permeated by subsonic turbulence with $M_{s,\rm turb} \lesssim 0.5$ \citep[e.g.,][]{gaspari2014,hofmann2016}. To emulate such conditions, we first evolve each subcluster in isolation under the gravitational field $g_{\rm HSE}(r)$, while driving turbulence for $\sim2$ Gyr prior to the merger. {The computational domain is a cubic box with periodic boundaries and a side length of $r_0 = 10~\mathrm{Mpc}$, discretized with $512^3$ uniform grid zones. The subcluster initially at rest is placed at the center of the box. During this pre-merger stage, only MHD simulations are performed, without including the CRe component.} 

Velocity perturbations, $\delta\mbox{\boldmath$u$}$, are drawn from a Gaussian random field with a power spectrum
\begin{equation}
|\delta u_k|^2 \propto k^6 \exp(-8k/k_{\rm inj}),
\end{equation}
where $k_{\rm inj}=40k_0$ and $k_0=2\pi/r_0$ \citep[e.g.,][]{stone1998,maclow1999,roh2019}. {The forcing is hence mixed, including both solenoidal (divergence-free) and compressive (curl-free) modes with comparable power \citep[e.g.,][]{federrath2008,cho2022}. Random phases are assigned so that the forcing is temporally uncorrelated. The peak scale of the forcing corresponds to $\sim r_0/40=250~\mathrm{kpc}$.}

{The perturbations are further scaled as $\delta\mbox{\boldmath$u$} \propto \rho_g^{w}$ with $w\geq0$, motivated by the expectation that the amplitude of velocity fluctuations is higher in the central regions of clusters \citep[e.g.,][]{vazza2017}. However, direct observational constraints on the radial scaling of turbulent velocity fluctuations in the ICM remain limited, although X-ray measurements consistently indicate subsonic turbulence in cluster cores \citep[e.g.,][]{hitomi2016,zhuravleva2014,hofmann2016}. Hence, the value of $w$ may be regarded as a free parameter, and we here present simulations using $w=0.25$ (see Appendix \ref{sc1} for a comparison of MHD turbulence with different values of $w$). The amplitude of the perturbations is adjusted such that the volume-averaged turbulence Mach number within $r_{200}$ attains $\langle M_{s,\rm turb} \rangle = \langle (u_{\rm turb}/c_s)^2 \rangle^{1/2} \approx 0.25$, where $u_{\rm turb}$ and $c_s$ are the local turbulent velocity and sound speed, respectively. Table \ref{t1} lists $\langle M_{s,\rm turb} \rangle$ in the three model subclusters \texttt{m2}, \texttt{m4}, and \texttt{m6}.}

In generating turbulent magnetic fields in each subcluster, a uniform background field along the $y$-direction, $B_{\rm bg}$, is imposed in the pre-merger simulations. Its strength is controlled such that the resulting turbulent component yields a volume-averaged plasma beta within $r_{200}$, close to $\langle \beta_p \rangle \approx 100$; {in practice, $B_{\rm bg}=0.5 \muG$ is adopted.} The corresponding values of $\langle \beta_p \rangle$ for the three model subclusters are listed in Table \ref{t1}. To avoid introducing large-scale biases, for each binary merger simulation, the two subclusters are initialized having uniform background fields of equal magnitude but opposite direction: $\mbox{\boldmath$B$}_{\rm bg}=+B_{\rm bg}\hat{y}$ for Subcluster 1 and $\mbox{\boldmath$B$}_{\rm bg}=-B_{\rm bg}\hat{y}$ for Subcluster 2. With this configuration, while the $\mbox{\boldmath$\nabla$}\cdot \mbox{\boldmath$B$}=0$ constraint is preserved, the net background field cancels out.

\begin{figure*}[t]
\vskip 0.1 cm
\hskip -0.5 cm
\centering
\includegraphics[width=0.9\linewidth]{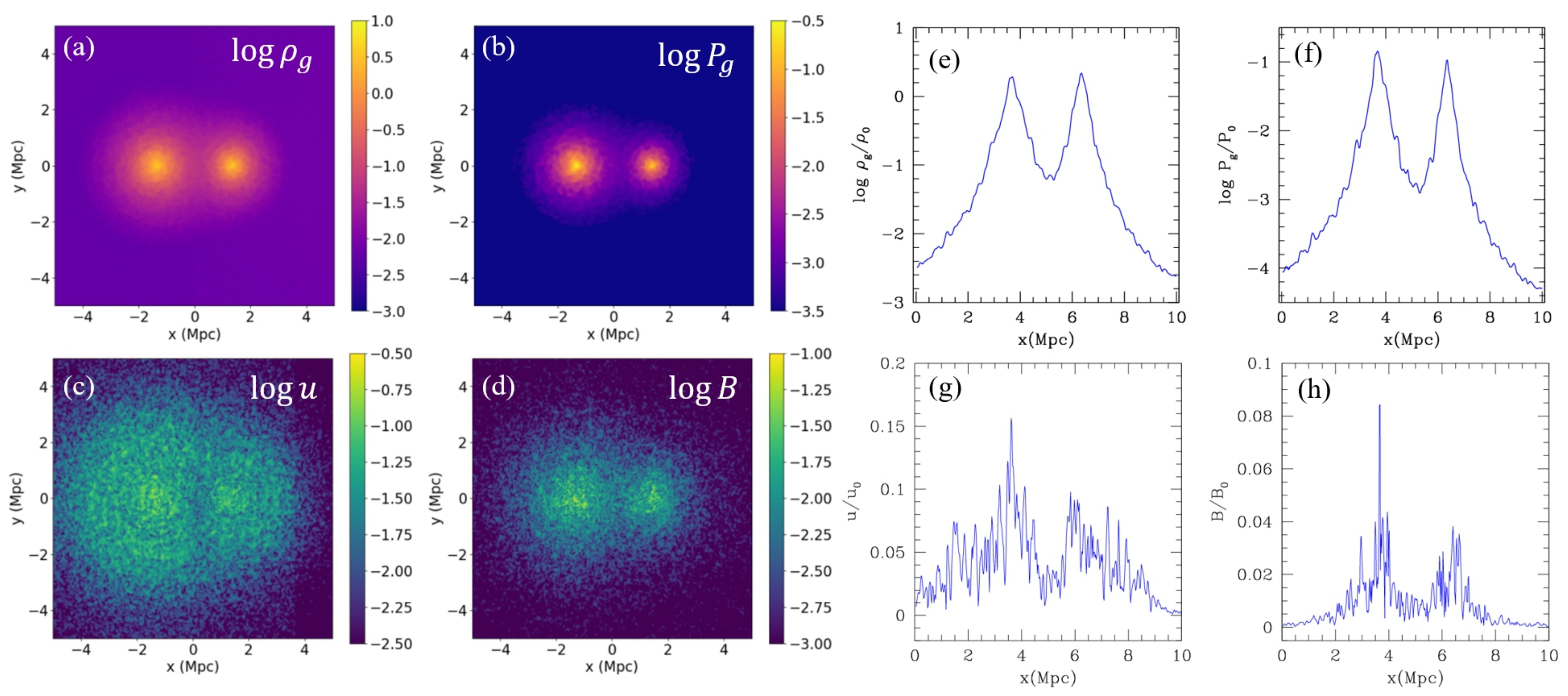}
\vskip -0.1 cm
\caption{Initial state at $t=0$ for the \texttt{m4m2$\theta$0} model simulation. (a)--(d) 2D distributions of $\rho_g$, $P_g$, velocity magnitude, $u=(u_x^2+u_y^2+u_z^2)^{1/2}$, and magnetic field strength, $B=(B_x^2+B_y^2+B_z^2)^{1/2}$, in the $x-y$ merger plane ($z=0$). (e)--(h) Corresponding 1D profiles along the merger axis (the $x$-axis). All quantities are normalized by their respective reference units in Section \ref{s2.1}.}\label{f3}
\end{figure*}

{\subsubsection{Binary Merger Simulations}\label{s2.3.2}}

{The binary merger system consists of two subclusters with either equal or unequal masses. Figure~\ref{f2}(a) illustrates the initial configuration for an unequal-mass merger case. We label the left subcluster as Subcluster~1 with mass $M_1 = M_{200,1}$, and the right subcluster as Subcluster~2 with mass $M_2 = M_{200,2}$.}

{The initial separation between the centers of the two subclusters is set to $d_{1,2} = f_d\,(r_{200,1}+r_{200,2})$, where $f_d$ is a distance factor.} The heavier subcluster is placed at $x_1=-0.5\,d_{1,2}$ (left) and the lighter one at $x_2=+0.5\,d_{1,2}$ (right). The gravity halos approach each other with velocities $\mbox{\boldmath$V_1$}$ and $\mbox{\boldmath$V_2$}$ at an inclination angle $\theta$ {with respect to the merger axis.}\footnote{Throughout this paper, $\mbox{\boldmath$u$}$ denotes the gas (flow) velocity, $\mbox{\boldmath$V$}_1$ and $\mbox{\boldmath$V$}_2$ indicate the initial approach velocities of the two halos, and $V_s$ denotes the shock speed.} The initial approach speeds are set to a fraction of the corresponding circular velocities, $V_1= f_V(GM_2/d_{1,2})^{1/2}$ and $V_2=f_V(GM_1/d_{1,2})^{1/2}$, where $f_V$ is a velocity factor. We adopt $f_d=1$ and $f_V=0.7$. This provides a simplified but reasonable approximation to the early stage of subcluster infall; however, realistic cosmological environments allow a broader range of encounter velocities, and we illustrate the impact of varying $f_V$ in Appendix \ref{sc2}. The inclination angle $\theta$ may be treated as a free parameter, but here we present models only with $\theta=0^\circ$ (head-on merger) and $10^\circ$ (off-axis merger). Merger model names encode both the subcluster masses and the velocity inclination; for example, the \texttt{m4m2$\theta$10} model combines the \texttt{m4} and \texttt{m2} subclusters with a velocity inclination of $\theta=10^\circ$.

{Figure~\ref{f3} shows the initial state of the \texttt{m4m2$\theta$0} merger model, constructed by combining the pre-merger simulations of two subclusters. The turbulent flow speed remains subsonic throughout the subclusters, as shown in Figure \ref{f3}(g), since the sound speed in the ICM is $c_s/u_0 \approx 0.2$. The turbulent magnetic field strength reaches $B\approx 2 - 3 \muG$ in the cores of the subclusters and decreases to to $B\approx0.1 - 0.2 \muG$ beyond $r_{200}$, as shown in Figure \ref{f3}(h).}

{For the merger simulations, we adopt the same computational domain as in the pre-merger simulations: a cubic box with periodic boundaries and a side length of $r_0 = 10~\mathrm{Mpc}$. The main suite of simulations employs $512^3$ uniform grid zones, corresponding to a spatial resolution of $\Delta x = 19.5~\mathrm{kpc}$. Additional runs with $256^3$ zones ($\Delta x = 39~\mathrm{kpc}$) are performed to assess parameter dependence (Figure~\ref{f17}) and numerical resolution effects (Figures~\ref{f5} and \ref{f18}; Table~\ref{t2}).}

{The adopted resolution needs to be improved to capture the substructures observed in some radio relics, such as the few-kpc filamentary features reported in high-resolution radio images \citep[e.g.,][]{vanweeren2019}. In our merger simulations, the postshock synchrotron-emitting region has a width of $\sim100$ kpc (see Section \ref{s2.8}), which is sampled by only $\sim5$ grid zones at the $512^3$ resolution. However, as noted, the primary objective of this study is to investigate the global characteristics of merger-driven shocks, rather than to resolve the fine structures of specific radio relics. These global features are robustly reproduced in our simulations, as presented in the next section. Importantly, convergence tests show that the overall shock morphology, Mach number distribution, and large-scale radio emission profiles remain qualitatively consistent between the $512^3$ and $256^3$ runs (see Figures~\ref{f5} and \ref{f18}). This indicates that the current resolution is sufficient for characterizing the merger shocks and their associated synchrotron emission on scales of tens of kiloparsecs to megaparsecs.}

{If structures such as kpc-scale filaments or shock-surface rippling are to be investigated, higher spatial resolution should be implemented. However, simply increasing the number of grid zones to reach kpc-scale resolution is computationally prohibitive. In our framework, the MHD solver is coupled to an Eulerian Fokker–Planck solver that tracks the full CRe energy-density distribution, typically over $N_p = 64$ logarithmic momentum bins at every spatial grid zone (see Section \ref{s2.5}). This approach substantially increases both memory usage and computational cost, compared to tracer-based or reduced-moment approaches, as noted in the introduction. Due to this, and also owing to the high-order nature of the HOW-MHD code, simulations with $512^3$ grid zones, or at most a factor of two larger in each dimension, represent the practical limit with current computational facilities. Higher effective resolution near merger shocks and in postshock regions should be achieved through the implementation of AMR and comparable techniques, which is outside of the scope of this work.}

The simulations of binary mergers are evolved up to $t_{\rm end}/t_0 \approx 1.5$ (roughly $3.8~\mathrm{Gyr}$); our analysis focuses primarily on the period from pericenter passage at $t_p/t_0 \approx 0.6$ to approximately 1~Gyr afterward, corresponding to $t_{\rm obs}/t_0 \approx 1.0$, which we take as the representative observation epoch.

\subsection{Shock Identification}\label{s2.4}

Merger shocks in our simulations are identified using the algorithm of \citet{ryu2003}. Along each coordinate direction, a grid zone is tagged as a shock zone if it satisfies all three conditions: (1) $\mbox{\boldmath$\nabla$}\cdot \mbox{\boldmath$u$} < 0$ (converging flow), (2) $\Delta T \times \Delta \rho > 0$ (aligned temperature and density gradients), and (3) $\left|\Delta \log T\right| > 0.11$ (temperature jump larger than that of $M_s=1.3$). The Mach number is then calculated by inverting the Rankine–Hugoniot relation,
\begin{equation}
\frac{T_2}{T_1} = \frac{(5M_s^2-1)(M_s^2+3)}{16M_s^2}.
\end{equation}
Hereafter, subscripts 1 and 2 refer to preshock and postshock states, respectively, where relevant. For each shock zone, we compute directional Mach numbers along the $x$, $y$, and $z$-axes and assign $M_s = \max(M_{s,x}, M_{s,y}, M_{s,z})$. 
Only shock zones with $M_s\ge1.5$ are included in our analysis, since weaker shocks are expected to be inefficient in DSA. 

{Owing to the high-order accuracy of the WENO scheme, the numerical shock transition is typically resolved over approximately two grid zones in our simulations. Although multiple zones may therefore be tagged across a numerical shock transition, we retain only one zone, corresponding to the location of maximum compression, as the ``shock zone''. Thus, each detected shock along a given 1D computational path is represented by a single shock zone. The 3D distribution of these shock zones is illustrated by the volume-rendered image in Figure~\ref{f5}(a).}

Conventionally, radio relics have been characterized by a single Mach number inferred from X-ray or radio observations. In reality, however, their curved, spherical-cap geometries and the turbulent nature of the ICM imply that a merger-shock surface hosts a range of Mach numbers. We therefore treat each merger shock as an ensemble of shock zones with varying strengths and, when necessary, summarize its global properties using the mean Mach number, $\langle M_s\rangle$, { averaged over the shock surface area} \citep[e.g.][]{lee2025}.

\subsection{Fokker-Planck Equation for CR Electron Population}\label{s2.5}

Along with the gas-flow dynamics, we follow the evolution of CRe population by solving the Fokker-Planck equation for the isotropic component of the distribution function $f(p,\mbox{\boldmath$x$})$ \citep{schlickeiser2002}:
\begin{equation}
\begin{aligned}
\frac{\partial f}{\partial t} + \mbox{\boldmath$u$} \cdot \frac{\partial f}{\partial \mbox{\boldmath$x$}} &= \left( \frac{1}{3} \mbox{\boldmath$\nabla$} \cdot \mbox{\boldmath$u$} \right) p \frac{\partial f}{\partial p} + \frac{1}{p^2} \frac{\partial}{\partial p} \left[ p^2 b_l \frac{\partial f}{\partial p} \right] \\
&\quad + \frac{1}{p^2} \frac{\partial}{\partial p} \left[ p^2 D_{\rm pp} \frac{\partial f}{\partial p} \right] + j(\mbox{\boldmath$x$}, p).
\end{aligned}
\label{diffcon2}
\end{equation}
Here, the cooling term $b_l(p)$ describes energy losses due to Coulomb interactions, bremsstrahlung emission, inverse-Compton (IC) scattering, and synchrotron radiation \citep{sarazin1999}, while the momentum diffusion coefficient $D_{\rm pp}$ accounts for Fermi-II acceleration. {Considering that the diffusion lengths of CRe are much smaller than our grid resolution, the spatial diffusion term is ignored.} The source term $j(\mbox{\boldmath$x$}, p)$ incorporates the \textit{in situ} injection of CRe and the reacceleration of preexisting fossil CRe via DSA, which are applied locally at identified shock zones (see Section \ref{s2.6}). 

In the literature, the term ``turbulent reacceleration’’ is often used to describe Fermi-II acceleration. In this work, however, we reserve ``reacceleration’’ exclusively for the reacceleration of preexisting fossil CRe via DSA, while ``turbulent acceleration’’ refers specifically to Fermi-II acceleration governed by the momentum-diffusion term $D_{\rm pp}$.

With $p\gg m_e c$, we define the energy density of CRe as $e_c(p,\mbox{\boldmath$x$})\equiv 4\pi p^3 f(p,\mbox{\boldmath$x$})$. Introducing the logarithmic momentum variable $h \equiv \ln(p/m_e c)$, equation (\ref{diffcon2}) can be rewritten as follows:
\begin{align}
\frac{\partial e_c}{\partial t}
+ \frac{\partial}{\partial \mbox{\boldmath$x$}}\cdot (e_c \mbox{\boldmath$u$})
= \frac{\partial}{\partial h}\left(
  \frac{D_{\rm pp}}{p^2} ~ \frac{\partial e_c}{\partial h}
\right) \notag\\
- \frac{\partial}{\partial h}\left[
  \left(3 \frac{D_{\rm pp}}{p^2}
    -\frac{1}{3} \mbox{\boldmath$\nabla$}\cdot \mbox{\boldmath$u$}
    - \frac{b_l}{p}   
  \right) e_c \right] + Q(\mbox{\boldmath$x$},h),
\label{diffcon}
\end{align}
where $Q(\mbox{\boldmath$x$}, h)\equiv 4\pi p^3 j(\mbox{\boldmath$x$}, h)$. Both the spatial advection on the left-hand side and the momentum advection in the $h$-coordinate on the right-hand side are written in conservative form and are calculated using a WENO scheme. Our simulations cover a momentum range spanning from $p_{\rm min}/m_ec = 10$ to $p_{\rm max}/m_ec = 10^5$, discretized into 64 logarithmically spaced bins with a spacing of $\Delta h = 0.144$, at every spatial grid zone. The numerical scheme to solve equation (\ref{diffcon}) is described in Appendix \ref{sec:sa}.

The energy losses of CRe due to IC and synchrotron cooling can be expressed in terms of an ``effective'' magnetic field strength, $B_{\rm e}^2 = B^2 + B_{\rm rad}^2$. Here, $B$ is the physical magnetic field strength, and $B_{\rm rad} = 3.24~\mu{\rm G}\,(1+z_r)^2$ represents the equivalent magnetic field strength associated with IC cooling by the cosmic microwave background (CMB) at redshift $z_r$. Thus, the redshift dependence enters to our simulations through IC cooling. {For the redshift evolution, $z_r(t)$, we adopt the standard $\Lambda$CDM cosmology model with the following parameters: matter density $\Omega_m$= 0.315, cosmological constant $\Omega_\Lambda$ = 0.685, and Hubble constant $ H_0=67.4 {\rm km s^{-1}Mpc^{-1}}$ \citep{planck2020}.} In our models, we track mergers from the initial epoch at $z_{r,{\rm init}} \approx 0.48$ ($t=0$), through pericenter passage at $z_{r,{\rm pp}} \approx 0.3$, to the observed stage at $z_{r,{\rm obs}} \approx 0.2$ ($t=2.54$~Gyr). Over this period, $B_{\rm rad}$ decreases from $7.1~\mu{\rm G}$ to $4.7~\mu{\rm G}$. With typical ICM magnetic fields of order $\mu\mathrm{G}$ in our simulations, IC losses dominate over synchrotron cooling.

As noted in the introduction, Fermi-II acceleration of CRe can arise from scattering off magnetic turbulence at MHD and plasma kinetic scales. These fluctuations may be produced by the cascade of large-scale MHD turbulence \citep[e.g.,][]{brunetti2014} or generated through shock–turbulence interactions and microinstabilities in the shock transition zone \citep[e.g.,][]{treumann2009,kim2021,trotta2023}. However, as mentioned, the microphysics governing the excitation and evolution of MHD and plasma waves and the subsequent wave-particle interactions remains poorly constrained in the high-$\beta_p$ intracluster plasma. Furthermore, these processes cannot be treated self-consistently within our MHD framework {due to limited spatial resolution.}

To account for Fermi-II acceleration, we hence parameterize the momentum diffusion coefficient as:
\begin{equation}
\frac{D_{\rm pp}}{p^2} \approx \frac{4}{\tau_{\rm pp}(p)}.
\label{Dpf}
\end{equation}
where $\tau_{\rm pp}$ denotes the characteristic turbulent acceleration timescale. In principle, $\tau_{\rm pp}$ depends on the amplitude and spectral properties of magnetic fluctuations in the flow. For practical purposes, we treat $\tau_{\rm pp}$ as a constant. Values in the range $\tau_{\rm pp}\sim0.1$–$1~\mathrm{Gyr}$ are broadly consistent with the eddy-turnover (and decay) timescales of turbulence found in cosmological simulations of structure formation \citep[e.g.,][]{porter2015,miniati2015,vazza2017}. We adopt $\tau_{\rm pp}=1~\mathrm{Gyr}$ as our fiducial value, and explore the consequences of smaller $\tau_{\rm pp}$ and the limiting case of $D_{\rm pp}=0$.

With the CRe distribution followed across the entire computational domain, we are, in principle, able to track the turbulent acceleration of CRe throughout the full cluster volume. However, in this study, we restrict turbulent acceleration to the postshock region. {Accordingly, we apply Fermi-II acceleration only within a downstream region of $L_{\rm Dpp}=5\Delta x\approx97.7$~kpc in the $512^3$ simulations and $L_{\rm Dpp}=3\Delta x\approx117$~kpc in the $256^3$ simulations. These widths are comparable to the characteristic width of the synchrotron-emitting region discussed in Section~\ref{s2.3.2} (see also Section~\ref{s2.8}). This prescription reflects our primary focus on merger shocks and their associated radio relics, rather than on diffuse synchrotron emission throughout the cluster volume (e.g., radio halos and mega radio halos).} 

{The turbulent-acceleration region is identified as follows. Starting from each detected shock zone, we tag the downstream region extending over $5\Delta x$ in the $512^3$ simulations or $3\Delta x$ in the $256^3$ simulations, along the same 1D computational path used to determine the Mach number of the shock zone. Within these tagged postshock zones, $D_{\rm pp}$ is assigned a nonzero value according to equation~(\ref{Dpf}), while it is set to zero elsewhere. To examine the dependence of our results on the adopted width, we additionally consider a $512^3$ model with $L_{\rm Dpp}=10\Delta x\approx195$~kpc, twice the fiducial width, and present its consequences in Figure~\ref{f9}.}

With this grid-based approach, our Fokker-Planck solver approximately accounts for turbulent acceleration as the CRe population advects through the postshock volume, without explicitly reconstructing 3D shock-normal vectors.

\subsection{CR Electrons Deposited at Shock Zones}\label{s2.6}

For the source term in equation~(\ref{diffcon2}), we prescribe the distribution functions of freshly injected CRe, $f_{\rm inj}(p)$, and reaccelerated fossil CRe, $f_{\rm RA}(p)$, at the shock zones using a phenomenological approach. In the DSA theory, the term ``injection’’ typically refers to the preacceleration of suprathermal particles above the injection momentum, enabling protons and electrons to begin diffusing across the shock and participate in the DSA cycle. In this work, however, we use ``fresh injection'' to denotes the full process that encompasses both the initial leakage from the thermal pool, preacceleration, and the subsequent establishment of the DSA power-law spectrum, distinguishing it from the reacceleration of preexisting CRe. Given that merger shocks generally have modest strengths, $M_s \lesssim 5$, as noted in the introduction, we adopt the test-particle DSA to model both injection and reacceleration \citep{bell1978,drury1983,kang2020}.

Employing a thermal-leakage prescription, {the DSA power-law spectrum for CRe is assumed to emerge from the Maxwellian distribution at the electron injection momentum, $p_{\rm inj,e}= Q_e(M_s)\, p_{\rm th,e}$ with $Q_e(M_s)$ \citep[e.g.,][]{arbutina2021,kang2024}.} Then, the freshly injected component in each shock zone is given by
\begin{equation}
f_{\rm inj}(p) \approx \left[\frac{n_2}{\pi^{1.5} p_{\rm th}^3} \exp\left(-Q_e^2\right)\right] \cdot \left(\frac{p}{p_{\rm inj,e}}\right)^{-q_{\rm sh}},
\label{finj}
\end{equation}
where $q_{\rm sh} = 4M_s^2/(M_s^2 - 1)$ is the test-particle DSA slope. Here, $n_2$ and $T_2=R_T T_{2,0}$ are the postshock gas number density and temperature, respectively, and $T_{2,0}$ is the Rankine-Hugoniot jump temperature. The temperature reduction factor $R_T(M_s)$ accounts for the transfer of thermal energy to CRp energy \citep{ryu2019}. 

For $Q_e(M_s)$ and $R_T(M_s)$, we adopt the analytic fitting functions proposed by \citet{kang2024}:
\begin{equation}
Q_e(M_s) \approx 3.5 + \frac{0.4}{1 + \exp[-1.7(M_s-3.6)]},
\label{Qe}
\end{equation}
\begin{equation}
R_T(M_s) \approx 1.0 - \frac{0.08}{1 + \exp[-3(M_s-2.7)]}.
\label{RT}
\end{equation}
These forms yield a gradual decline in $R_T$ and a mild increase in $Q_e$ with increasing shock Mach number. For $M_s \lesssim 5$, $R_T \approx 0.92$–1.0 and $Q_e \approx 3.5$–3.8. As a comparison model, we also consider the case with constant values, $R_T=1$ and $Q_e=3.5$. We note that because $f_{\rm inj}(p)$ depends on $Q_e$ through $\propto \exp(-Q_e^2)$, even modest variations in $Q_e$ lead to substantial changes in the CRe injection efficiency and therefore in the postshock CRe population.

To model the reacceleration of preexisting CRe, we assume a fossil population for $p \ge p_{\rm min}$ with a spectrum
\begin{equation}
f_{\rm pre}(p) = f_{\rm o} \cdot \left(\frac{p}{p_{\rm min}}\right)^{-s}\exp\left[-\left(\frac{p}{p_{\rm cut}}\right)^2\right],
\label{fpre}
\end{equation}
where we adopt a slope $s\simeq 4.5$ and a cutoff momentum $p_{\rm cut}/m_ec \sim 300$, consistent with IC and synchrotron cooling in typical ICM conditions \citep{brunetti2014}. The normalization factor scales with the local gas density:
\begin{equation}
f_{\rm o}(\mbox{\boldmath$x$})\approx 10^{-6}n_1(\mbox{\boldmath$x$}),
\label{f_o}
\end{equation}
with $n_1(\boldsymbol{x})$ denoting the preshock gas number density. This choice corresponds to a fossil-thermal electron energy fraction of $\varepsilon_{\rm CRe}/\varepsilon_{\rm th} \approx (1.0$–$4.2)\times10^{-3}$ in the preshock gas; it leads to a substantial enhancement relative to the injection-only model, as shown in the result section.

\begin{figure}[t]
\vskip 0.2cm
\hskip -0.5 cm
\centering
\includegraphics[width=0.95\linewidth]{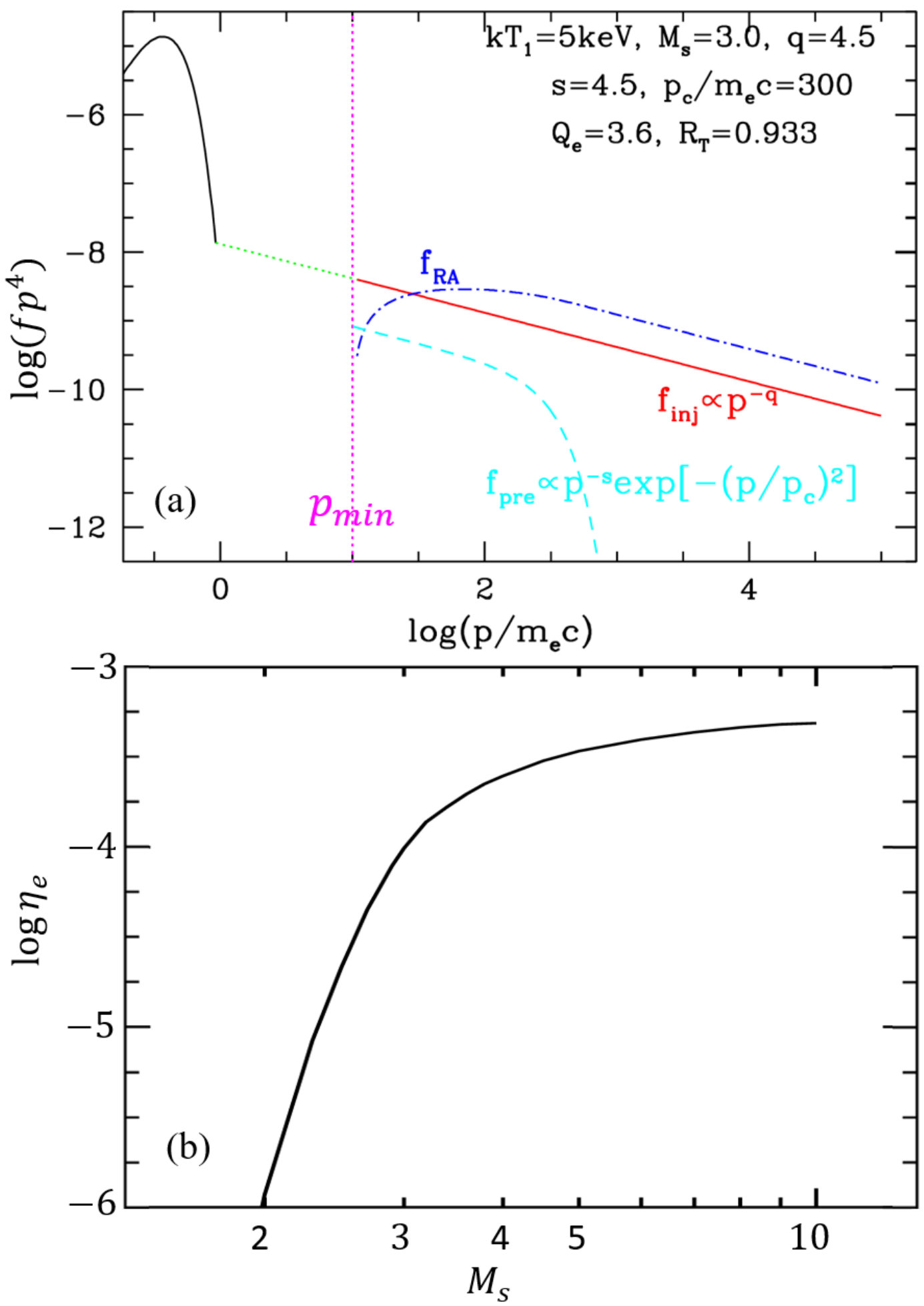}
\vskip -0.1 cm
\caption{(a) Model spectra of CRe for a shock with $M_s=3.0$. Shown are the postshock Maxwellian distribution (black), the freshly injected CRe spectrum $f_{\rm inj}$ (red), the preexisting fossil CRe spectrum $f_{\rm pre}$ (cyan), and the reaccelerated fossil CRe spectrum $f_{\rm RA}$ (blue). The logarithmic momentum axis covers $p_{\rm min}/m_ec=10$ to $p_{\rm max}/m_ec=10^5$. (b) Acceleration efficiency of CRe computed using the injection spectrum $f_{\rm inj}$ given in equation (\ref{finj}).}\label{f4}
\end{figure}

The reaccelerated component in each shock zone is computed numerically as 
\begin{equation}
f_{\rm RA}(p)= q\, p^{-q_{\rm sh}} \int_{p_{\rm min}}^{p} p^{\prime\,q-1} f_{\rm pre}(p^\prime)\, dp^\prime,
\label{fRA}
\end{equation}
where $q_{\rm sh}$ is the same as in equation (\ref{finj}) \citep{drury1983, kang2011}. At sufficiently high momenta, $f_{\rm RA}(p)$ approaches $\propto p^{-q_{\rm sh}}$, and the slope is largely insensitive to the input slope $s$, because the fossil population is concentrated at relatively low energies ($10 \lesssim p \lesssim p_{\rm cut}$). As a result, both freshly injected and reaccelerated electrons, $f_{\rm inj}(p)$ and $f_{\rm RA}(p)$, share the same asymptotic slope $q$ for $p/m_ec \gtrsim 100$ in our model; hence, the presence of fossil CRe primarily enhances the normalization of the postshock CRe spectrum without altering its slope.

In the reference models, we consider only the fresh injection of CRe at shocks with no fossil CRe. In the result section, additionally, we also present a case that has only the reacceleration of fossil CRe, as well as a case that includes both fresh injection and reacceleration, in order to demonstrate the effects of preexisting fossil CRe.

Figure~\ref{f4}(a) shows $f_{\rm inj}(p)$ and $f_{\rm RA}(p)$ for a shock with $M_s=3.0$ in a typical ICM environment, illustrating that both the spectra follow identical power-law slopes for $p/m_ec \gtrsim100$. Figure~\ref{f4}(b) plots that the acceleration efficiency of CRe, $\eta_e(M_s) \equiv \varepsilon_{\rm CRe,2} u_2 / (0.5\rho_1 V_s^3)$, calculated with the injection spectrum $f_{\rm inj}$. The efficiency lies in the range $10^{-4}$–$10^{-3}$ for $M_s \gtrsim 3.0$, but decreases to $\lesssim 10^{-4}$ for weaker shocks. Below $M_{s,\rm crit}\approx2.3$, the critical Mach number for the preacceleration of suprathermal electrons at high-beta ICM shocks \citep{ha2018b,kang2019,ha2021,ha2022}, the efficiency is prescribed to be $\lesssim 10^{-5}$, intending effectively no fresh injection.

\subsection{Synchrotron Radio Emission}\label{s2.7}

Synchrotron emission is computed in post-processing using the simulated CRe population and magnetic fields. Assuming an isotropic pitch-angle distribution of CRe, the synchrotron volume emissivity can be expressed as \citep[e.g.,][]{rybicki1979}
\begin{equation}
j_{\nu}= \frac{\sqrt{3}\, q_e^2 \nu_L}{2c} 
\int_{\tilde{p}_{\rm min}}^{\tilde{p}_{\rm max}} 
\left[4\pi \tilde{p}^2 f(\tilde{p})\right]
\, \xi \int_{\xi}^{\infty} K_{5/3}(\chi)\, d\chi\, d\tilde{p},
\label{jnu}
\end{equation}
where $\tilde{p}=p/(m_e c)$ is the dimensionless electron momentum, $\nu_L=q_e B_{\perp}/(2\pi m_e c)$ is the Larmor frequency, $\nu_c=(3/2)\tilde{p}^2 \nu_L$ is the characteristic (critical) synchrotron frequency, and $K_{5/3}$ denotes the modified Bessel function of the second kind of order $5/3$. Here, {$\nu$ is the frequency in the emitter rest frame} and $\xi \equiv \nu / \nu_c$; $B_{\perp}$ is the component of the local magnetic field, $\mbox{\boldmath$B$}(\mbox{\boldmath$x$})$, perpendicular to the LoS.

The synchrotron surface brightness on the sky plane in a synthetic observation is obtained by integrating the emissivity along the LoS:
\begin{equation}
\begin{aligned}
S_{\nu}(\mbox{\boldmath$x$}_{\perp}) &= I_{\nu}(\mbox{\boldmath$x$}_{\perp})\, \theta_{\rm beam}^2 \\
&= \left[\int j_{\nu}(\mbox{\boldmath$x$}_{\perp})\,d\mbox{\boldmath$x$}_{\parallel}\right] \theta_{\rm beam}^2,
\end{aligned}
\label{snu}
\end{equation}
where $I_{\nu}(\mbox{\boldmath$x$}_{\perp})$ is the intensity, and $\mbox{\boldmath$x$}_{\perp}$ and $\mbox{\boldmath$x$}_{\parallel}$ denote the spatial coordinates perpendicular and parallel to the LoS, respectively. $\mbox{\boldmath$x$}_{\perp}$ is included in $S_{\nu}$, $I_{\nu}$, and $j_{\nu}$ to explicitly indicates that $B_{\perp}$ is used. Here, $\theta_{\rm beam}$ is the angular resolution of the observation, and the fiducial beam size is taken to be $\theta_{\rm beam}^2 = 1'' \times 1''$, as noted in Section \ref{s2.1}.

The quantity $B_{\perp}$ depends on the viewing geometry. Therefore, for synchrotron surface brightness maps produced with LoS directions not aligned with the computational coordinate axes, $B_{\perp}$ is recalculated for each chosen LoS by applying an appropriate coordinate transformation to $\boldsymbol{B}(\boldsymbol{x})$ (Figures~\ref{f12} and \ref{f18}).

We evaluate the emissivity in 100 logarithmically spaced frequency bins over the range $\nu = 10^6-10^{11}~\mathrm{Hz}$. For illustration, we present results at four representative frequencies: $\nu_1 = 53$~MHz, $\nu_2 = 150$~MHz, $\nu_3 = 600$~MHz, and $\nu_4 = 1.33$~GHz, spanning the LOFAR LBA, LOFAR HBA, and typical GHz-band observations \citep[e.g.,][]{deGasperin2021}.

The radio spectral index between two frequencies $\nu_k$ and $\nu_{k'}$ is then computed as
\begin{equation}
\alpha_{k,k'}(\mbox{\boldmath$x$}_{\perp})
= -\,\frac{\ln S_{\nu_{k'}}(\mbox{\boldmath$x$}_{\perp}) - \ln S_{\nu_k}(\mbox{\boldmath$x$}_{\perp})}{\ln \nu_{k'} - \ln \nu_k},
\label{alpanu}
\end{equation}
and we focus on $\alpha_{1,2}$ (between 53 and 150 MHz) and $\alpha_{2,3}$ (between 150 and 600 MHz).

The resulting synchrotron surface brightness $S_{\nu}$ and the spectral index $\alpha_{k,k'}$ directly probe the properties of postshock CRe and turbulent magnetic fields. Regions with freshly injected electrons or enhanced $B_{\perp}$ appear as bright radio structures, while spectral steepening reflects radiative aging and turbulent acceleration in the downstream flow. These synthetic observations provide a quantitative basis for comparison with observed radio relic data, enabling diagnoses on the underlying shock strength, magnetic-field amplification, and CRe acceleration processes.

In the result section below, we present the frequency and synchrotron emissivity in the local rest frame of the emitting plasma. For a snapshot at $z_r(t)$, the corresponding observed frequency is given by $\nu_{\rm obs} = \nu/(1+z_r)$, while the intensity is cosmologically dimmed as $I_{\nu_{\rm obs}}=I_{\nu}/(1+z_r)^4$. Intending to examine the intrinsic properties of merger shocks and radio relics, rather than to make detailed comparisons with specific observed relics, we do not include these additional cosmological effects.

\subsection{Injected versus Integrated Radio Spectra}\label{s2.8}

In the framework of DSA, the Mach number of radio relic shocks can be estimated from the spectral index of the synchrotron emission at the shock location, $\alpha_{\rm sh}$, using
\begin{equation}
M_{\rm s,sh}=\left( \frac{2\alpha_{\rm sh}+3}{2\alpha_{\rm sh}-1} \right)^{1/2}.
\label{Msh}
\end{equation}

Alternatively, the integrated radio spectrum from the postshock region provides a useful diagnostic for estimating the shock Mach number, particularly for distant or poorly resolved relics. In the idealized case of a plane-parallel shock with constant speed and a uniform postshock magnetic field, the volume-integrated CRe spectrum, $F(p)=E_c(p)/p^3$ (where $E_c(p)=\int e_{\rm c}(p,\mbox{\boldmath$x$})\,dV$), steepens from the injection slope at the shock, $q_{\rm sh}$, at low momenta, to an integrated slope, $q_{\rm int} = q_{\rm sh}+1 = (5M_s^2 -1)/(M_s^2 - 1)$, above the break momentum $p_{\rm br}$, provided that CRe cool only via synchrotron and IC losses (i.e., neglecting Coulomb losses and turbulent acceleration). The break momentum depends on the shock age, $t_{\rm age}$, and the effective magnetic field strength, $B_e$ \citep[e.g.,][]{kang2015}, as
\begin{equation}
\frac{p_{\rm br}}{m_ec}\approx 10^3
\left(\frac{t_{\rm age}}{1~{\rm Gyr}}\right)^{-1}
\left(\frac{B_e}{5~\mu{\rm G}}\right)^{-2},
\label{pbr}
\end{equation}

\begin{figure*}[t]
\vskip -0.4 cm
\hskip 0.4 cm
\includegraphics[width=0.9\linewidth]{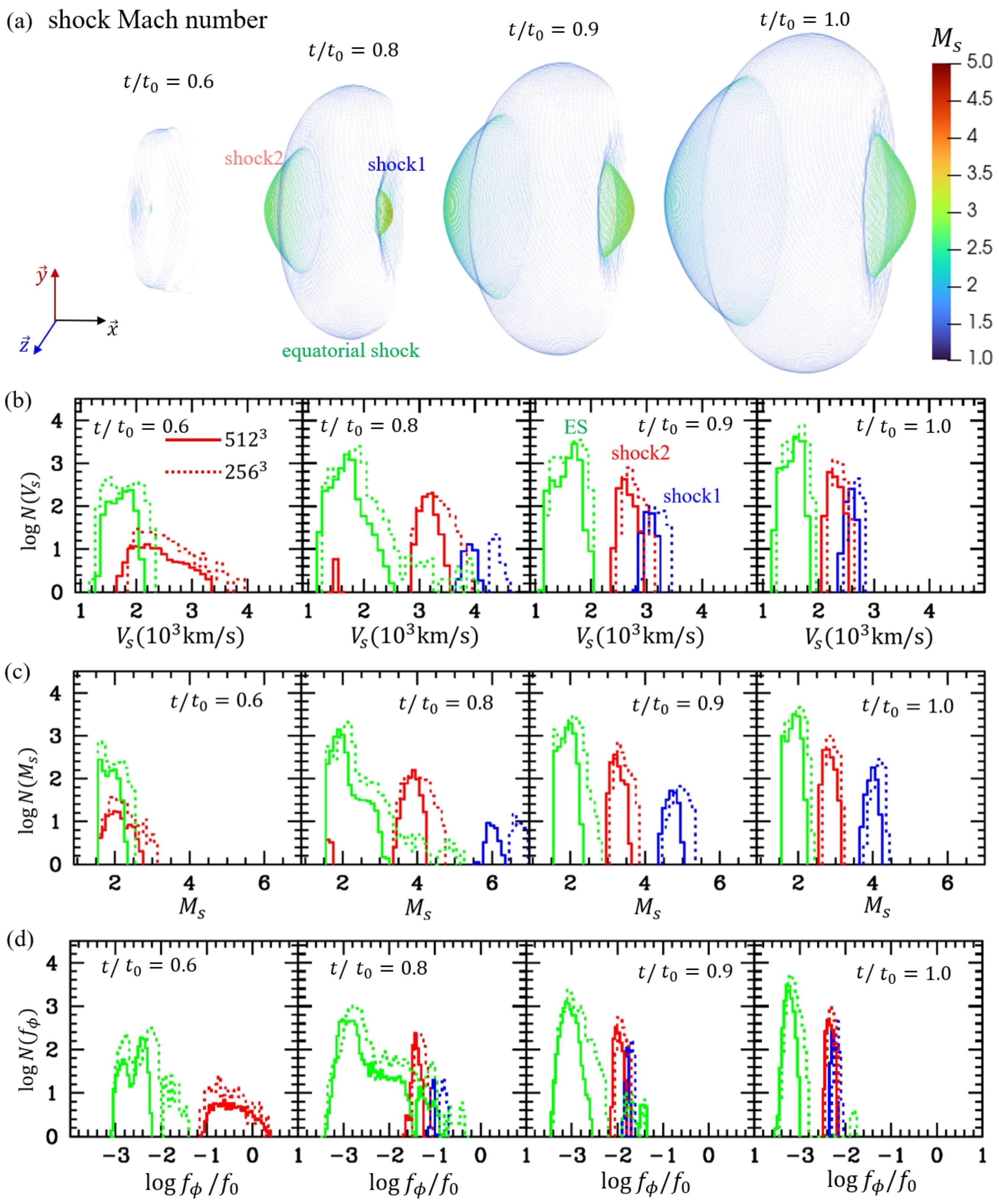}
\caption{(a) Volume-rendered images of the 3D spatial distribution of the shock Mach numbers at $t/t_0 = 0.6-1.0$ for the \texttt{m4m2$\theta$0} simulation. Equatorial shocks are present at $t/t_0 = 0.6$, near pericenter passage, after which two axial shocks propagate in opposite directions. The equatorial shocks and the two axial shocks, Shock~1 (ahead of the heavier subcluster) and Shock~2 (ahead of the lighter subcluster), are indicated. (b)--(d) Corresponding distribution functions of the shock speed $V_s$, Mach number $M_s$, and kinetic energy flux $f_{\phi} = (1/2)\,\rho_1 V_s^3$ ({in units of $f_0=\rho_0u_0^3$}). Solid and dotted lines show results from the $512^3$ and $256^3$ \texttt{m4m2$\theta$0} simulations, respectively. Blue (red) histograms represent Shock~1 (Shock~2), while green histograms denote equatorial shocks.}\label{f5}
\end{figure*}

Correspondingly, the volume-integrated synchrotron spectrum, $J_{\nu}=\int j_{\nu}\,dV \propto\nu^{-\alpha}$, steepens from the injection value at the shock, $\alpha_{\rm sh}=(q_{\rm sh}-3)/2$, at low frequencies, to the integrated spectral index, $\alpha_{\rm int}=\alpha_{\rm sh}+0.5$, above the break frequency $\nu_{\rm br}$. 
For typical parameters of radio relics, 
\begin{equation}
\nu_{\rm br}\approx 3.8~{\rm MHz}
\left(\frac{t_{\rm age}}{1~{\rm Gyr}}\right)^{-2}
\left(\frac{B_e}{5~\mu{\rm G}}\right)^{-4}
\left(\frac{B}{3~\mu{\rm G}}\right),
\label{nubr}
\end{equation}
which is generally below the frequency range probed by radio observations. Here, $\nu_{\rm br}$ is the source-frame break frequency; the observed value is $\nu_{\rm br,obs}=\nu_{\rm br}/(1+z_r)$. In this regime, the shock Mach number can be estimated from $\alpha_{\rm int}$ using
\begin{equation}
M_{\rm s,int}=\left( \frac{\alpha_{\rm int}+1}{\alpha_{\rm int}-1} \right)^{1/2}.
\label{Mint}
\end{equation}

The cooling length of CRe with $\gamma_e = 10^4$, emitting synchrotron radiation at $\nu\sim$ several hundred MHz, behind a shock with $M_s = 3$, $V_s = 3 \times 10^3~{\rm km~s^{-1}}$, and an effective magnetic field $B_e \approx 5~\mu{\rm G}$ is estimated as $l_{\rm cool} = \tau_{\rm cool} u_2 \approx 100~{\rm kpc}$. Here, $u_2 \approx 10^3~{\rm km~s^{-1}}$ for the postshock flow speed and $\tau_{\rm cool}=\tau_{\rm syn+IC} \approx 0.1~{\rm Gyr}\,(\gamma_e/10^4)^{-1}(B_e/5\muG)^{-2}$ for the radiative cooling timescale due to synchrotron and IC losses are used.

For a spherically expanding shock with age $t_{\rm age} \gg \tau_{\rm cool}$ and curvature radius $R_s \gg l_{\rm cool}$, the plane-parallel approximation may be justified. Under these conditions, equation~(\ref{Mint}) provides a reasonable estimate of the shock Mach number, provided that the mean magnetic field strength is approximately uniform in the postshock region. On the other hand, additional processes such as turbulent acceleration may lead to deviations from these expectations.

\section{Results}\label{s3}

\subsection{Merger Shocks}\label{s3.1}

We first examine the properties of typical merger shocks using the \texttt{m4m2$\theta$0} model, which represents a head-on collision between two subclusters of unequal mass. As the merger progresses, the subclusters accelerate to supersonic speeds, and their normalized relative velocity, $V_{1,2}/u_0$, approaches unity. Figure~\ref{f5}(a) shows volume-rendered snapshots of the 3D Mach-number distribution at four representative stages of the \texttt{m4m2$\theta$0} simulation.

Near pericenter passage at $t/t_0 \approx 0.6$, equatorial shocks, induced by the compression of the subclusters, propagate primarily in directions perpendicular to the merger axis. After pericenter passage, the supersonic motions of the receding DM clumps\footnote{Hereafter, the DM clumps are referred to as the gravitational halos in simulations.} drive two axial bow shocks, Shock~1 and Shock~2, which travel in opposite directions along the merger axis. As the merger further advances, these shocks expand in spatial extent; at $t/t_0=1.0$, roughly 1~Gyr after pericenter passage, the tips of the two axial shocks are separated by $\sim 4$~Mpc. These axial shocks are systematically stronger than the equatorial shocks, because they are driven by supersonic motion of DM clumps.

The heavier subcluster, $M_1$, which begins on the left-hand side, appears on the right-hand side after pericenter passage, while the lighter subcluster, $M_2$, reverses in the opposite direction. In unequal-mass mergers, the upstream flow into Shock~1 is both faster due to the stronger gravitational pull of the heavier DM clump and cooler because it originates predominantly from the lighter subcluster (see Figure~\ref{f7}(a); see also Figure~\ref{f2}(c) which correspond to the \texttt{m4m2$\theta$10} model but illustrate the same trend). As a result, Shock~1 attains a higher average Mach number than Shock~2, although its overall extent is smaller due to more rapid, directed inflow that limits its lateral expansion.

Each shock surface is composed of many ``shock zones'' that exhibit a distribution of Mach numbers rather than a single value (Section~\ref{s2.4}). Figures~\ref{f5}(b)--(d) show the distributions of the shock speeds, $V_s$, Mach numbers, $M_s$, and kinetic energy fluxes, $f_{\phi}=(1/2)\,\rho_{g,1}V_s^3$, for the shock zones comprising the equatorial shocks and the two axial shocks in the \texttt{m4m2$\theta$0} model. Shock zones are classified geometrically into three categories: (1) equatorial shocks (green) - zones within $30^\circ$ of the $y$--$z$ plane; (2) Shock~1 (blue) - zones inside the positive polar cone of $0^\circ \le \theta_p \le 30^\circ$ about the $x$-axis; and (3) Shock~2 (red) - zones inside the negative polar cone of $150^\circ \le \theta_p \le 180^\circ$ about the $x$-axis. Here, the $x$-axis is aligned with the merger axis, the $x-y$ plane is the merger plane (Figure~\ref{f2}), and $\theta_p$ is the polar angle measured from the merger axis. This geometric classification provides an approximate but useful distinction among the three shock components. Solid and dotted lines indicate the $512^3$ and $256^3$ simulations, respectively.

\begin{deluxetable}{ccccccccccc}\label{t2}
\tablecolumns{8}
\tablewidth{0pt}
\tablecaption{Mean Properties of Merger Shocks in \texttt{m4m2$\theta$0}$^a$}
\tablehead{
\colhead{resolutions} &
\colhead{$t/t_0$} &
\colhead{$\langle V_{s,1} \rangle$$^b$} &
\colhead{$\langle V_{s,2} \rangle$$^b$} &
\colhead{$\langle V_{s,\rm ES} \rangle$$^b$} &
\colhead{$\langle M_{s,1} \rangle$$^c$} &
\colhead{$\langle M_{s,2} \rangle$$^c$} &
\colhead{$\langle M_{s,\rm ES} \rangle$$^c$}&
\colhead{$(\sum f_{\phi,1}\Delta x^2)$$^d$} &
\colhead{$(\sum f_{\phi,2}\Delta x^2)$} &
\colhead{$(\sum f_{\phi,\rm ES}\Delta x^2)$}
}
\startdata
$512^3$ & 0.83 & 3.85 & 3.08 & 1.62 & 5.94 & 3.78 & 1.87 & 9.83 & 64.8 & 46.2\\
$512^3$ & 0.93 & 2.99 & 2.57 & 1.58 & 4.66 & 3.18 & 1.87 & 9.26 & 31.7 & 20.5\\
$512^3$ & 1.03 & 2.53 & 2.23 & 1.53 & 3.93 & 2.76 & 1.84 & 7.17 & 21.1 & 17.9\\
\hline
$256^3$ & 0.83 & 4.34 & 3.22 & 1.74 & 6.64 & 3.93 & 2.04 & 11.1& 72.6 & 85.0\\
$256^3$ & 0.93 & 3.15 & 2.68 & 1.61 & 4.90 & 3.32 & 1.97 & 10.1 & 36.0 & 26.8\\
$256^3$ & 1.03 & 2.64 & 2.31 & 1.58 & 4.12 & 2.85 & 1.92 & 7.62 & 22.3 & 21.9 \\
\enddata
\tablenotetext{a}{The subscripts 1, 2, and ES stand for Shock~1, Shock~2, and equatorial shock, respectively.}
\tablenotetext{b}{Mean shock speed in units of $10^3~\kms$, weighted by shock surface area, $\langle V_{s} \rangle=\sum_{i=1}^{N_s} V_s/N_s$, where $N_s$ is the total number of shock zones associated with each shock surface.}
\tablenotetext{c}{Mean shock Mach number, weighted by shock surface area, $\langle M_{s} \rangle=\sum_{i=1}^{N_s} M_s/N_s$.}
\tablenotetext{d}{{Kinetic energy flux integrated over the shock surface for Shock~1, Shock~2, and equatorial shock are given  in units of $10^{-5}\times$ $\rho_0 u_0^3r_0^2=1.27\times10^{45}{\rm erg~s^{-1}}$.}}
\vskip -0.8cm
\end{deluxetable}
 
At pericenter passage, most identified shock zones correspond to equatorial shocks (green), although early signs of the axial shocks, particularly Shock~2 (red), begin to emerge. As the merger evolves beyond pericenter passage, the DM clumps decelerate, and Shock~1 and Shock~2, which are bow shocks, fully develop along the merger axis. Shock~1 consistently exhibits higher values of $V_s$, $M_s$, and $f_\phi$ than Shock~2. During $t/t_0 = 0.8$--1.0, as the DM clumps slow down, the speeds of these axial shocks decrease (Figure~\ref{f5}(b)), leading to a gradual decline in their Mach numbers (Figure~\ref{f5}(c)). 

Table~\ref{t2} summarizes the temporal evolution of the surface-area-weighted shock speed, $\langle V_s \rangle$, and Mach number, $\langle M_s \rangle$, as well as the shock-surface-integrated kinetic energy flux, $\sum f_{\phi}\Delta x^2$, for the three types of shocks in both the $512^3$ and $256^3$ simulations of the \texttt{m4m2$\theta$0} model. {As noted in Section~\ref{s2.4}, only one grid zone is tagged as a shock zone across each numerically broadened shock transition. The identified shock surface is therefore one zone thick, and each tagged shock zone represents the same area, $\Delta x^2$, on the uniform Cartesian grid. Consequently, the total shock surface area is $A_{\rm sh}\approx N_{\rm sh}\Delta x^2$, where $N_{\rm sh}$ is the number of identified shock zones for the shock surface, and an arithmetic average over the tagged shock zones is equivalent to a shock-surface-area-weighted average.}

In the $512^3$ simulation, $\langle M_s \rangle$ for Shock~1 decreases from $5.94$ to $3.93$, while that for Shock~2 decreases from $3.78$ to $2.76$ over a time interval of $\sim 0.5$~Gyr. The $256^3$ run yields slightly higher shock speeds and correspondingly larger values of $\langle M_s \rangle$. For Shock~1, the difference in $\langle M_s \rangle$ between the $256^3$ and $512^3$ runs is $\sim 10\%$ at $t/t_0 = 0.83$, decreasing to $\sim 5\%$ at $t/t_0 = 1.03$; the discrepancy is smaller for Shock~2.

\begin{figure*}[t]
\vskip 0.1 cm
\includegraphics[width=0.99\linewidth]{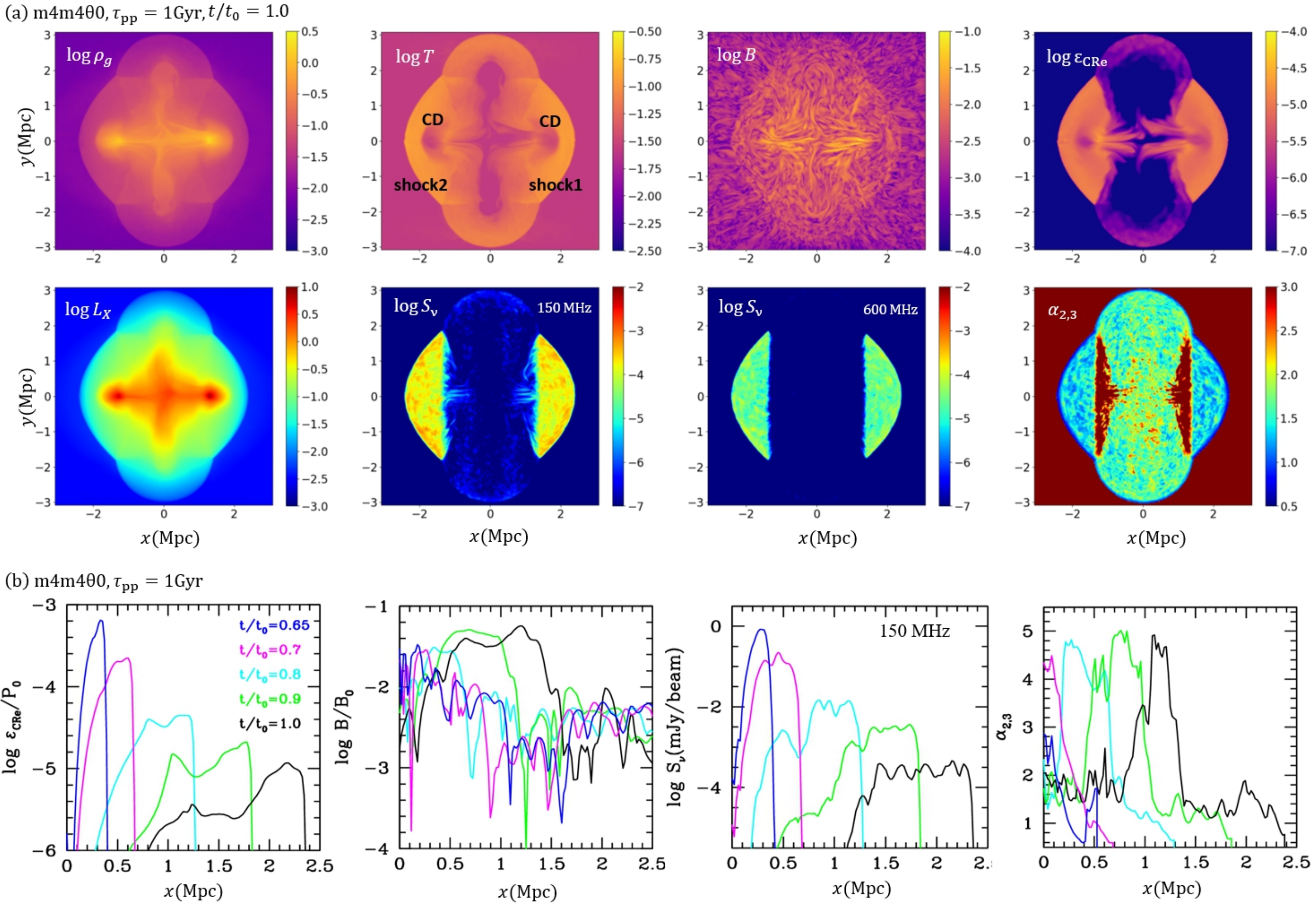}
\vskip -0.1 cm
\caption{{(a) Key diagnostic quantities for the equal-mass head-on merger model, \texttt{m4m4$\theta$0}, at $t/t_0=1.0$. The upper rows show slice maps in the merger plane ($z=0$) of $\rho_g/\rho_0$, $T/T_0$, $B/B_0$, and $\varepsilon_{\rm CRe}/P_0$. The lower rows show quantities projected along the $z$-axis, namely $L_X/L_0$, $S_{\nu}/S_0$ at 150~MHz and 600~MHz, and $\alpha_{2,3}$ in the $x-y$ plane (sky plane). The normalization factors $L_0$ and $S_0$ are defined in Section~\ref{s2.1}. Here and in the figures below, a fiducial beam size of $\theta_{\rm beam}^2 = 1'' \times 1''$ is adopted. (b) Temporal evolution of the 1D profiles of $\varepsilon_{\rm CRe}/P_0$, $B/B_0$, $S_{\nu}/S_0$ at 150~MHz, and $\alpha_{2,3}$ shown in panel (a) along the $x$-axis behind Shock~1. Profiles are shown for $t/t_0=0.65$ {(blue)}, 0.7 (magenta), 0.8 (cyan), 0.9 (green), and 1.0 (black).}}\label{f6}
\end{figure*}

\begin{figure*}[t]
\vskip 0.1 cm
\includegraphics[width=0.99\linewidth]{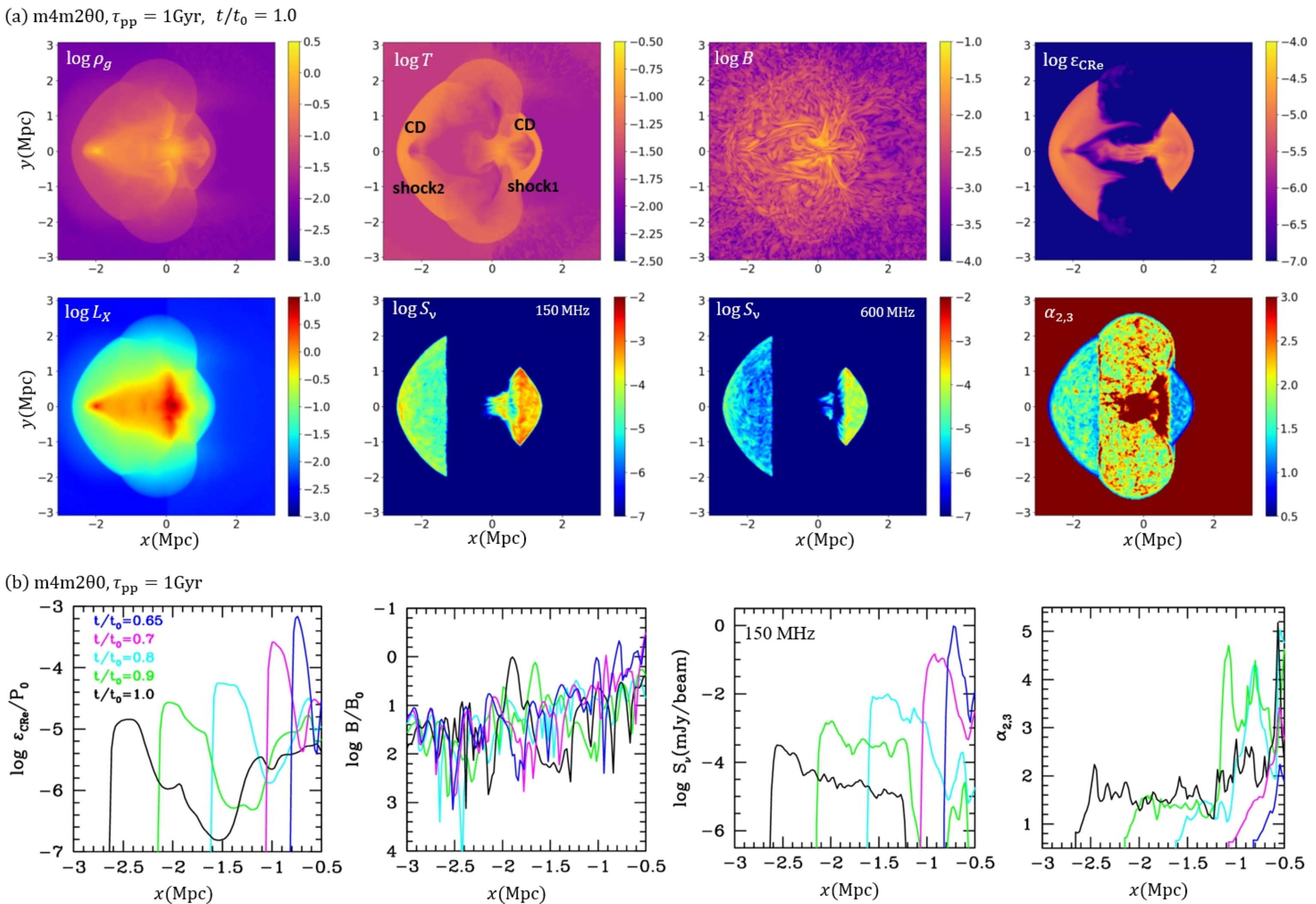}
\vskip -0.1 cm
\caption{(a) Key diagnostic quantities for an unequal-mass head-on merger, \texttt{m4m2$\theta$0}, at $t/t_0=1.0$. (b) Temporal evolution of 1D profiles behind Shock~2. The panels show the same physical quantities as in Figure \ref{f6}, and the time epochs and color schemes are identical.}\label{f7}
\end{figure*}

\begin{figure*}[t]
\vskip 0.1 cm
\includegraphics[width=0.99\linewidth]{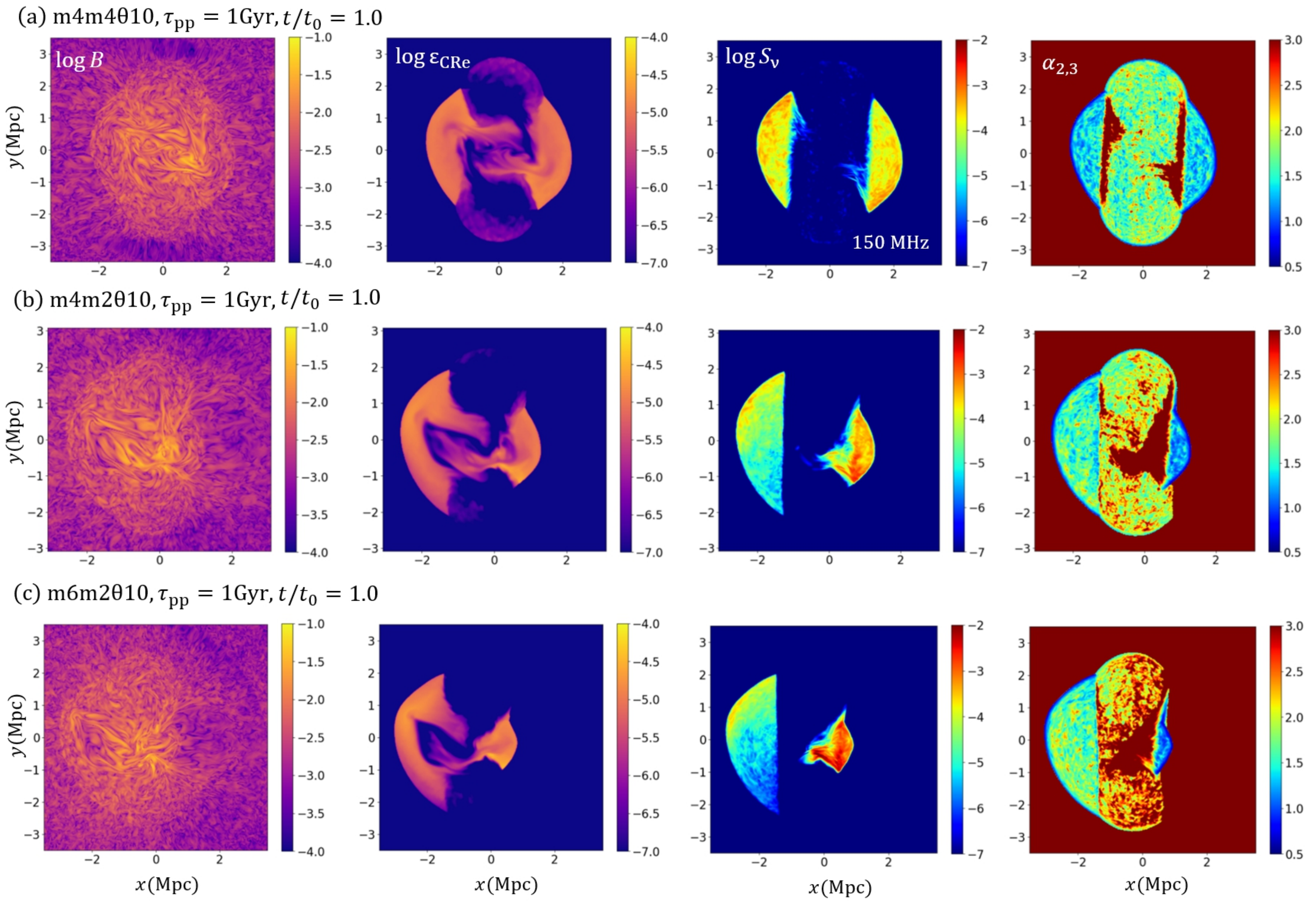}
\vskip -0.1 cm
\caption{{Comparison of off-axis mergers with three different mass ratios at $t/t_0=1.0$: (a) \texttt{m4m4$\theta$10}, (b) \texttt{m4m2$\theta$10}, and (c) \texttt{m6m2$\theta$10}. For each model, the columns display (from left to right): $B/B_0$ and $\varepsilon_{\rm CRe}/P_0$ in the merger plane ($z=0$), and the projected synchrotron surface brightness $S_{\nu}/S_0$ at 150~MHz and the resulting spectral index $\alpha_{2,3}$ in the $x-y$ plane.}}\label{f8}
\end{figure*}

The properties of merger-shocks in Figure~\ref{f5}, derived from idealized simulations, may be compared with results from cosmological structure formation simulations \citep[e.g.,][]{ha2018,lee2025}. A notable difference is that the Mach-number distributions, $N(M_s)$, of our shock zones exhibit a narrow width, $\Delta M_s \sim 0.5$, whereas cosmological simulations typically produce a log-normal Mach-number distribution. The broader distribution in cosmological simulations arises from the secular evolution of the ICM driven by successive minor mergers and the continuous infall of the warm--hot intergalactic medium (WHIM) along filaments. These processes modify the dynamical and thermal state of the ICM, generating turbulence and significant density fluctuations. Although our initial subclusters contain forced turbulence, idealized simulations do not fully capture the complexity in cosmological simulations. In cosmological settings, minor mergers and WHIM infall broaden the $M_s$ distribution, while in our idealized merger, the width of $N(M_s)$ is set primarily by shock-surface curvature and by subsonic turbulent motions with $u_{\rm turb}/c_s \lesssim 0.5$.

Nevertheless, our finding that Shock~1 is systematically stronger, $\langle M_{s,1}\rangle > \langle M_{s,2}\rangle$, while Shock~2 is more extended, $ A_{ss,2} >  A_{ss,1}$ (where $A_{ss}$ denotes the integrated shock surface area), is broadly consistent with cosmological simulations, though the ratios $\langle M_{s,1}\rangle / \langle M_{s,2}\rangle$ and $ A_{ss,2} / A_{ss,1}$ exhibit substantial event-to-event variability in fully cosmological environments (see Figure 5 in \citet{lee2025}).

Figure~\ref{f5}(d) shows that the kinetic energy flux, $f_{\phi}$, has larger values for Shock~1 than for Shock~2, although the two become comparable by $t/t_0 = 1.0$. {In contrast, the shock-surface-integrated energy flux, $\sum f_{\phi}\Delta x^2$, is larger for Shock~2, as can be seen in Table \ref{t2}; this is because Shock~2 has a larger surface area. The differences in $\sum f_{\phi}\Delta x^2$ between the $256^3$ and $512^3$ runs are at the level of $\sim 5\%$ for both Shock~1 and Shock 2 at $t/t_0 = 1.03$. Resolution effects are discussed further in Appendix \ref{sc3}.}

\subsection{Simulated Radio Relics}\label{s3.2}

{We next examine synthetic radio observations of meger shocks in our simulations, which reveal radio relics. Figures~\ref{f6}(a) present key diagnostic quantities for the \texttt{m4m4$\theta$0} model, an equal-mass head-on merger, at $t/t_0 = 1.0$, $\sim1$~Gyr after pericenter passage. The upper panels display slice maps in the merger plane ($x-y$, $z=0$), showing the normalized gas density $\rho_g/\rho_0$, temperature $T/T_0$, turbulent magnetic field strength $B/B_0$, and total CRe energy density, $\varepsilon_{\rm CRe}(\boldsymbol{x})/P_0 = \int e_c(p,\boldsymbol{x})\,dp/P_0$. The lower panels display the projected quantities, i.e., LoS-integrals, along the $z$-direction, showing the bolometric X-ray surface brightness $L_X(x,y)/L_0=\int \rho_g^2 T^{1/2}dz/L_0$, the synchrotron radio surface brightness $S_{\nu}(150~\mathrm{MHz})/S_0$ and $S_{\nu}(600~\mathrm{MHz})/S_0$, and the two-frequency spectral index $\alpha_{2,3}$ derived from these maps.}

{The density and temperature maps clearly show two merger shocks: Shock~1 on the right-hand side and Shock~2 on the left. Behind each shock front, a contact discontinuity (cold front) is also visible. The magnetic field map exhibits intermittent filamentary structures produced by turbulent dynamo amplification, even though turbulence forcing is not applied in the main merger simulations. The radio surface-brightness maps display two crescent-shaped bow shocks. Along the merger shock axis, the apparent relic width reaches $\sim1$~Mpc due to projection effects, even though the postshock cooling length of radio-emitting electrons is only of order $\sim100$~kpc. As illustrated in Figure \ref{f5}(a), the 3D geometry of the axial shocks resembles a smooth cap in these idealized setup. LoS integration through these curved shock surfaces therefore produces a significantly broadened projected radio-emission region. The equatorial shocks are not visible in the $S_\nu$ maps with the color range used, because their Mach numbers and $f_\phi$ are relatively small and hence CRe production and magnetic-field amplification are not large enough.}

{Figure \ref{f6}(b) plots the temporal evolution of 1D profiles for $\varepsilon_{\rm CRe}$, $B$, $S_{\nu}(150~\mathrm{MHz})$, and $\alpha_{2,3}$ along the $x$-axis behind Shock~1 in the \texttt{m4m4$\theta$0} model, which propagates up to $x\approx 2.35$~Mpc at $t/t_0 = 1.0$. As discussed with Figure~\ref{f5}, the axial shocks begin to develop after pericenter passage at $t/t_0\approx0.6$. At $t/t_0=0.65$ (yellow curves), the CRe population is confined to a narrow region immediately behind the shock. As the shock propagates outward, downstream advection transports the CRe population away from the shock front, gradually increasing the spatial extent of the radio-emitting region, while the amplitude of $\varepsilon_{\rm CRe}$ decline. By $t/t_0=1.0$ (black curves), the peak of $\varepsilon_{\rm CRe}$ extends to $\Delta x \sim0.4$~Mpc  behind the shock front}

{The characteristic scale of the downstream radio-emitting region is expected to be governed by the advection and radiative cooling of CRe. The postshock advection length is given by $l_{\rm adv}\approx u_2 t_{\rm age}$, while the cooling length can be estimated as $l_{\rm cool}\approx u_2 \tau_{\rm cool}$. Adopting approximate values relevant for the shock at $t/t_0 = 1.0$ in Figures~\ref{f6}(b), $u_2 \approx 10^3~{\rm km~s^{-1}}$ and $t_{\rm age} = t-t_p \approx 1$~Gyr yield $l_{\rm adv}\approx 1$~Mpc. The cooling timescale and length are $\tau_{\rm cool}\approx 0.1~{\rm Gyr}$ and $l_{\rm cool} \approx 100~{\rm kpc}$, respectively, for CRe with $\gamma_e = 10^4$ and an effective magnetic field of $B_e \approx 5~\mu{\rm G}$, as noted in Section \ref{s2.8}.}

{We point out that the shock properties change substantially over the interval $t/t_0 \approx 0.6$ to 1.0, and thus the values of $l_{\rm adv}$ and $l_{\rm cool}$ change. And the spatial and momentum distributions of downstream CRe would be further affected by flow dynamics such as adiabatic expansion and the curved geometry of the shock, in addition to advection and cooling. Nevertheless, $l_{\rm cool} \ll l_{\rm adv}$ implies that the spatial extent of the radio-emitting region is primarily limited by radiative cooling. On the other hand, the width of the synchrotron emission regions in the projected surface brightness, $S_{\nu}$, is of order $\sim1$~Mpc, and it is due to projection broadening.}

{By $t/t_0=1.0$, Shock~1 has decelerated to $\langle V_{s,1}\rangle\approx 2.64\times 10^3 \kms$ and $\langle M_{s,1}\rangle\approx 3.26$, corresponding to a DSA power-law slope of $q\approx 4.42$ and an injection spectral index at the shock of $\alpha_{\rm sh} = 0.708$. The spectral indices derived from the simulated $S_{\nu}$ maps at the midpoint ($x=0$, $y=0$) of the shock surface are $\alpha_{1,2}\approx 0.724$ and $\alpha_{2,3}\approx 0.754$, which are slightly steeper than, but still close to, the injection index estimated with the mean Mach number. This indicates that the spectral indices measured near the shock location at these frequencies provide a reasonable estimate of the shock Mach number. Downstream, the spectral index steepens due to radiative cooling of the CRe population, as seen in the $\alpha_{2,3}$ profile in Figure \ref{f6}(b), producing a characteristic steepening scale of $\sim0.3$~Mpc. The large fluctuations in $\alpha_{2,3}$ farther downstream arise from the low radio emissivity in those regions.}

\begin{figure*}[t]
\vskip 0.1cm
\includegraphics[width=0.99\linewidth]{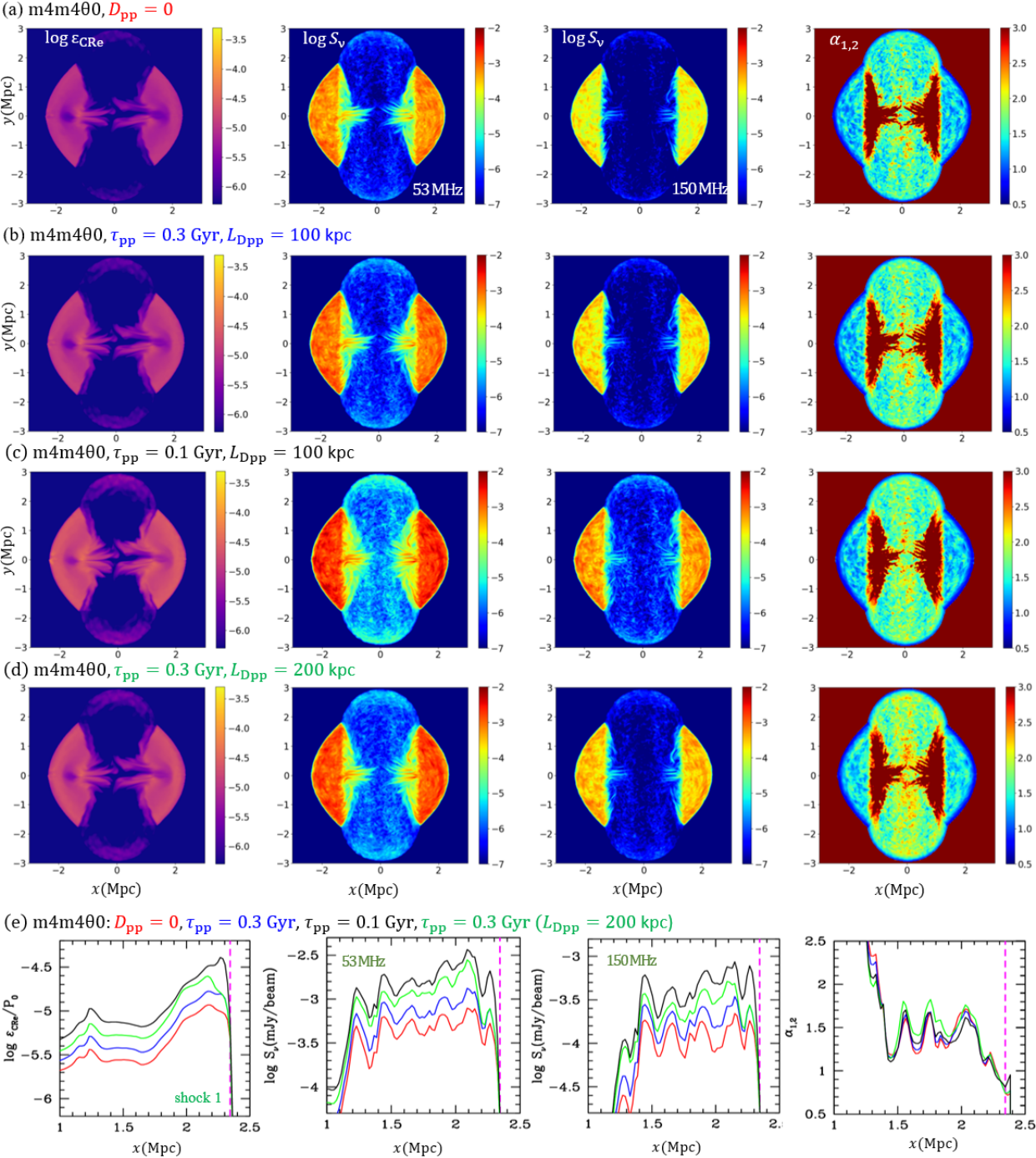}
\vskip -0.1cm
\caption{Equal-mass head-on merger model (\texttt{m4m4$\theta$0}) at $t/t_0=1.0$ for different momentum-diffusion timescales, $\tau_{\rm pp}$, and different widths of the postshock turbulent-acceleration region, $L_{\rm Dpp}$. The upper four rows show 2D maps of $\varepsilon_{\rm CRe}/P_0$ in the merger plane, the projected synchrotron surface brightness, $S_{\nu}/S_0$, at 53 and 150 MHz, and the resulting spectral index, $\alpha_{1,2}$, for (a) $D_{\rm pp}=0$, (b) $\tau_{\rm pp}=0.3~{\rm Gyr}$ and $L_{\rm Dpp}=100$ kpc, (c) $\tau_{\rm pp}=0.1~{\rm Gyr}$ and $L_{\rm Dpp}=100$ kpc, and (d) $\tau_{\rm pp}=0.3~{\rm Gyr}$ and $L_{\rm Dpp}=200$ kpc. The bottom row (e) shows the corresponding 1D profiles behind Shock~1 along the $x$-axis for the four models: $D_{\rm pp}=0$ (red), $\tau_{\rm pp}=0.3~{\rm Gyr}$ and $L_{\rm Dpp}=100$ kpc (blue), $\tau_{\rm pp}=0.1~{\rm Gyr}$ and $L_{\rm Dpp}=100$ kpc (black), and $\tau_{\rm pp}=0.3~{\rm Gyr}$ and $L_{\rm Dpp}=200$ kpc (green). The magenta vertical line marks the position of Shock~1.}\label{f9}
\end{figure*}

{The 2D maps of $S_{\nu}$ and $\alpha_{2,3}$ exhibit small-scale fluctuations. These features arise primarily from intermittent structures of the turbulent magnetic fields, whereas the spatial distribution of $\varepsilon_{\rm CRe}$ is relatively smooth, as indicated by the slice maps of $B$ and $\varepsilon_{\rm CRe}$. This suggests that the small-scale variations in radio relics are likely a consequence of magnetic-field structures rather than inhomogeneities in the CRe population or variations in the Mach number distribution on the shock surfaces.}

{Figure \ref{f7}(a) shows the same quantities as in Figure \ref{f6}(a), but for the unequal-mass head-on merger model \texttt{m4m2$\theta$0}. Shock~1 (ahead of the heavier subcluster) has larger $M_s$ and $f_{\phi}$, as noted with Figure \ref{f5}, resulting in higher $\varepsilon_{\rm CRe}$ and brighter $S_{\nu}$. In contrast, Shock~2 (ahead of the lighter subcluster) is weaker but spatially more extended.}

{Figure \ref{f7}(b) presents the temporal evolution of the 1D profiles behind Shock~2 in the \texttt{m4m2$\theta$0} model. At $t/t_0=1.0$ when the shock is located at $x\approx -2.6$~Mpc (black curves), while the peak of $\varepsilon_{\rm CRe}$ extends to $\Delta x \sim0.3$~Mpc, $S_{\nu}$ exhibits the bright region of a width of $\sim0.2$~Mpc immediately behind the shock, followed by fainter emission extending to $\Delta x \sim1$~Mpc. Thus, projection broadening manifests differently from the equal-mass merger case.}

{By $t/t_0=1.0$, the shock velocity and Mach number decrease to $\langle V_{s,2}\rangle\approx2.23\times10^3~{\rm km~s^{-1}}$ and $\langle M_{s,2}\rangle\approx2.76$, corresponding to $q\approx4.60$ and $\alpha_{\rm sh}\approx0.802$. The spectral indices derived from the $S_{\nu}$ maps at the mid point ($x=0$, $y=0$) of the shock surface are $\alpha_{1,2}\approx0.776$ and $\alpha_{2,3}\approx0.814$, marginally consistent with that of the mean Mach number. This further demonstrates that the spectral indices estimated at these frequencies near the shock location provide a reasonable estimate of the Mach number of merger shocks.}

Figure~\ref{f8} compares three off-axis merger models with the same velocity inclination, $\theta=10^{\circ}$, but with different mass ratios: \texttt{m4m4$\theta$10}, \texttt{m4m2$\theta$10}, and \texttt{m6m2$\theta$10}. 
We present slice maps of $B/B_0$ and $\varepsilon_{\rm CRe}/P_0$ in the merger plane, together with projected synchrotron surface brightness, $S_{\nu}/S_0$, at 150~MHz, and the spectral index, $\alpha_{2,3}$, at $t/t_0 = 1.0$. The shock geometry and velocity field of the \texttt{m4m2$\theta$10} model are illustrated in Figures~\ref{f2}(b)-(c). As in the head-on merger cases, two axial shocks dominate the $S_\nu$ maps, whereas equatorial shocks are not visible within the adopted color range.

In both unequal-mass merger models, \texttt{m4m2$\theta$10} and \texttt{m6m2$\theta$10}, the downstream CRe energy density behind Shock~1 is higher than that behind Shock~2. This difference reflects the systematically larger Mach numbers and higher $f_\phi$ values of Shock~1, resulting from stronger infall toward the more massive DM core. 
Consequently, as in the head-on merger cases, the radio relic associated with Shock~1 appears brighter than that produced by the more extended Shock~2. This brightness contrast arises not only from the higher $\varepsilon_{\rm CRe}$ but also from the stronger turbulent magnetic fields in the downstream region of Shock~1. In addition, at earlier times, the radio relics associated with merger shocks are brighter, although more compact, as can be inferred from Figures~\ref{f6}(b) and \ref{f7}(b).
 
We note that the trends from our simulations, such as $\langle M_{s,1}\rangle > \langle M_{s,2}\rangle$ and $A_{\rm ss,2} > A_{\rm ss,1}$, may not represent the characteristics of all observed double radio relic systems. As mentioned above, our idealized binary-merger models do not include additional environmental influences such as more frequent minor mergers and WHIM accretion, which stir the ICM flow dynamics and hence can modify the shock geometry and properties. In realistic environments, radio relics are therefore expected to display a broader diversity of morphologies and brightness distributions. {In cosmological simulations, for example, \citet{lee2025} identified merger shocks in 12 clusters and found that 11 clusters exhibit $A_{\rm ss,2}/A_{\rm ss,1}>1$ (see their Figure 5). However, the Mach number does not always show $\langle M_{s,1}\rangle > \langle M_{s,2}\rangle$ (see their Figure 9(c)). On the observational side, while $A_{\rm ss,2}/A_{\rm ss,1}>1$ for CIZAJ2242, $A_{\rm ss,2}/A_{\rm ss,1}<1$ for RCXJ1314, although the mass ratio of two clumps, estimated using weak lensing, is less than 2 for those clusters \citep{degasperin2014,finner2025}.}

At the surfaces of Shock~1 and Shock~2, the radio spectral indices are $\alpha_{2,3}\approx 0.58$–0.75, corresponding to shock Mach numbers $M_s\approx 3$–5. Downstream of the shocks, the synchrotron spectrum steepens as a result of radiative cooling of the CRe population, producing the larger $\alpha_{2,3}$ values visible behind the shock fronts, as in the head-on mergers.

\subsection{Impacts of Turbulent Acceleration}\label{s3.3}

\begin{figure*}[t]
\vskip 0.1cm
\hskip -0.8cm
\includegraphics[width=1.05\linewidth]{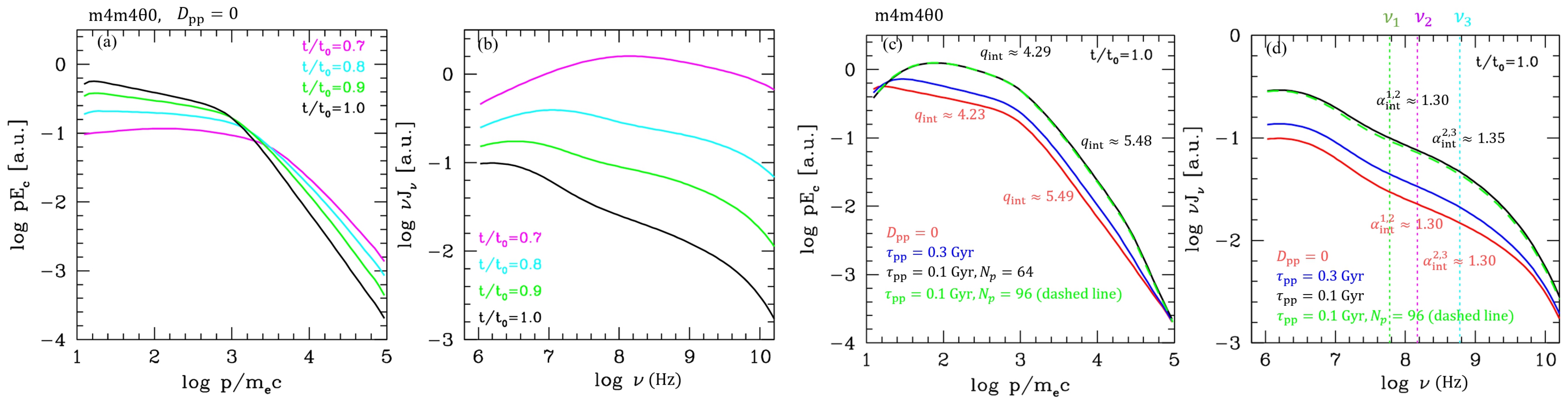}
\vskip -0.1 cm
\caption{{ Volume-integrated CRe energy spectra, $E_c(p)=\int e_c(p)\,dV$, and synchrotron radio spectra, $J_{\nu}=\int j_{\nu}\,dV$, (both in arbitrary units) for the \texttt{m4m4$\theta$0} models with various momentum-diffusion strengths. (a)-(b) Temporal evolution ($t/t_0=0.7-1.0$) for the model with no turbulent acceleration ($D_{\rm pp}=0$). (c)-(d) Comparison of three models with different $D_{\rm pp}$ at $t/t_0=1.0$ (shown in Figure~\ref{f9}), illustrating the effects of different momentum-diffusion strengths. In panel (c), the power-law slopes for the integrated distribution function, $F(p)=E_c(p)/p^3 \propto p^{-q_{\rm int}}$, are given for the cases of $D_{\rm pp}=0$ and $\tau_{\rm pp}=0.1$ Gyr. In panel (d), the integrated spectral indices, $\alpha_{\rm int}^{1,2}$ and $\alpha_{\rm int}^{2,3}$, are given for the same two cases; the vertical dotted lines mark $\nu_1=53$~MHz (green), $\nu_2=150$~MHz (purple), and $\nu_3=600$~MHz (cyan). In panels (c) and (d), for the model with $\tau_{\rm pp}=0.1$~Gyr, the results from an additional run with $N_p=96$ are shown using green dashed lines to demonstrate convergence in terms of the momentum bin size; the green dashed lines almost overlap with the black lines for the same run with $N_p=64$ (see Section \ref{sec:sb2}).}}\label{f10}
\end{figure*}

{To assess the effects of Fermi-II (turbulent) acceleration, we examine variants of the turbulent-acceleration model (Section \ref{s2.5}) using the \texttt{m4m4$\theta$0} merger at $t/t_0=1.0$. The fiducial model, shown in Figure~\ref{f6}, adopts $\tau_{\rm pp}=1.0~\mathrm{Gyr}$ and $L_{\rm Dpp}=100$ kpc. Figure~\ref{f9} presents the following four variants: (1) $D_{\rm pp}=0$ (equivalent to $\tau_{\rm pp}=\infty$); (2) $\tau_{\rm pp}=0.3~\mathrm{Gyr}$ with $L_{\rm Dpp}=100$ kpc; (3) $\tau_{\rm pp}=0.1~\mathrm{Gyr}$ with $L_{\rm Dpp}=100$ kpc; and (4) $\tau_{\rm pp}=0.3~\mathrm{Gyr}$ with $L_{\rm Dpp}=200$ kpc.}

{Figures~\ref{f9}(a)-(d) show the {spatial maps} of $\varepsilon_{\rm CRe}$, $S_{\nu}$ at 53 and 150 MHz, and $\alpha_{1,2}$ derived from the two radio maps. These can be compared with the fiducial model case shown in Figure~\ref{f6}(a). Again, $\varepsilon_{\rm CRe}$ is presented as a slice in the $x$--$y$ merger plane, whereas $S_{\nu}$ is projected along the $z$-direction. The figures show that introducing turbulent acceleration with $\tau_{\rm pp}\lesssim1~\mathrm{Gyr}$ enhances the postshock CRe population relative to the $D_{\rm pp}=0$ case by partially
compensating for radiative cooling losses, thereby producing brighter radio relics. 

A comparison of Figures 9(b) and (d) demonstrates that extending the turbulent-acceleration region from {$L_{\rm Dpp}\approx 97.7$ to 195 kpc} further increases the downstream CRe energy density and radio surface brightness. Nevertheless, varying $\tau_{\rm pp}$ and extending $L_{\rm Dpp}$ result in only modest changes in the spatial extent of the radio emission.} 
{Furthermore, the 2D maps of $\alpha_{1,2}$ likewise exhibit only modest differences among the models, indicating that the radio spectral index is relatively insensitive to the adopted $D_{\rm pp}$ and $L_{\rm Dpp}$ within the range explored here.}

{Figure~\ref{f9}(e) presents 1D profiles of the quantities shown in Figures~\ref{f9}(a)–(d) along the $x$-axis behind Shock~1, located at $x \approx 2.35$~Mpc. The $\varepsilon_{\rm CRe}$ and $S_{\nu}$ panels confirm that turbulent acceleration enhances $\varepsilon_{\rm CRe}$, and consequently $S_{\nu}$, by larger amounts for smaller $\tau_{\rm pp}$ and also for larger $L_{\rm Dpp}$. On the other hand, the thicknesses of the $S_{\nu}$ profiles along the $x$-axis are similar for different values of $\tau_{\rm pp}$ and $L_{\rm Dpp}$. We note that turbulent acceleration is applied only within a downstream region of {$L_{\rm Dpp} \approx 97.7 - 195$~kpc} in our models. However, the spatial extent of the brightest $S_{\nu}$ region is governed primarily by projection effects, as discussed above. In all models with different $D_{\rm pp}$ and $L_{\rm Dpp}$, the widths of the relics in both the 53~MHz and 150~MHz maps are of order $\sim 1$~Mpc. This indicates that while turbulent acceleration mainly enhances the synchrotron surface brightness of radio relics, it does not significantly broaden their spatial extent. The $\alpha_{1,2}$ panel shows that while the spectral index steepens in the postshock region due to radiative cooling of the CRe population, {the differences in the 1D profiles of $\alpha_{1,2}$ among models with different $D_{\rm pp}$ and $L_{\rm Dpp}$ remain modest.} This is partly because the turbulent-acceleration model adopted here assumes a fixed acceleration timescale, $\tau_{\rm pp}$, that is independent of particle momentum; consequently, turbulent acceleration does not significantly alter the spectral slope.}

\begin{figure*}[t]
\vskip 0.1cm
\includegraphics[width=0.99\linewidth]{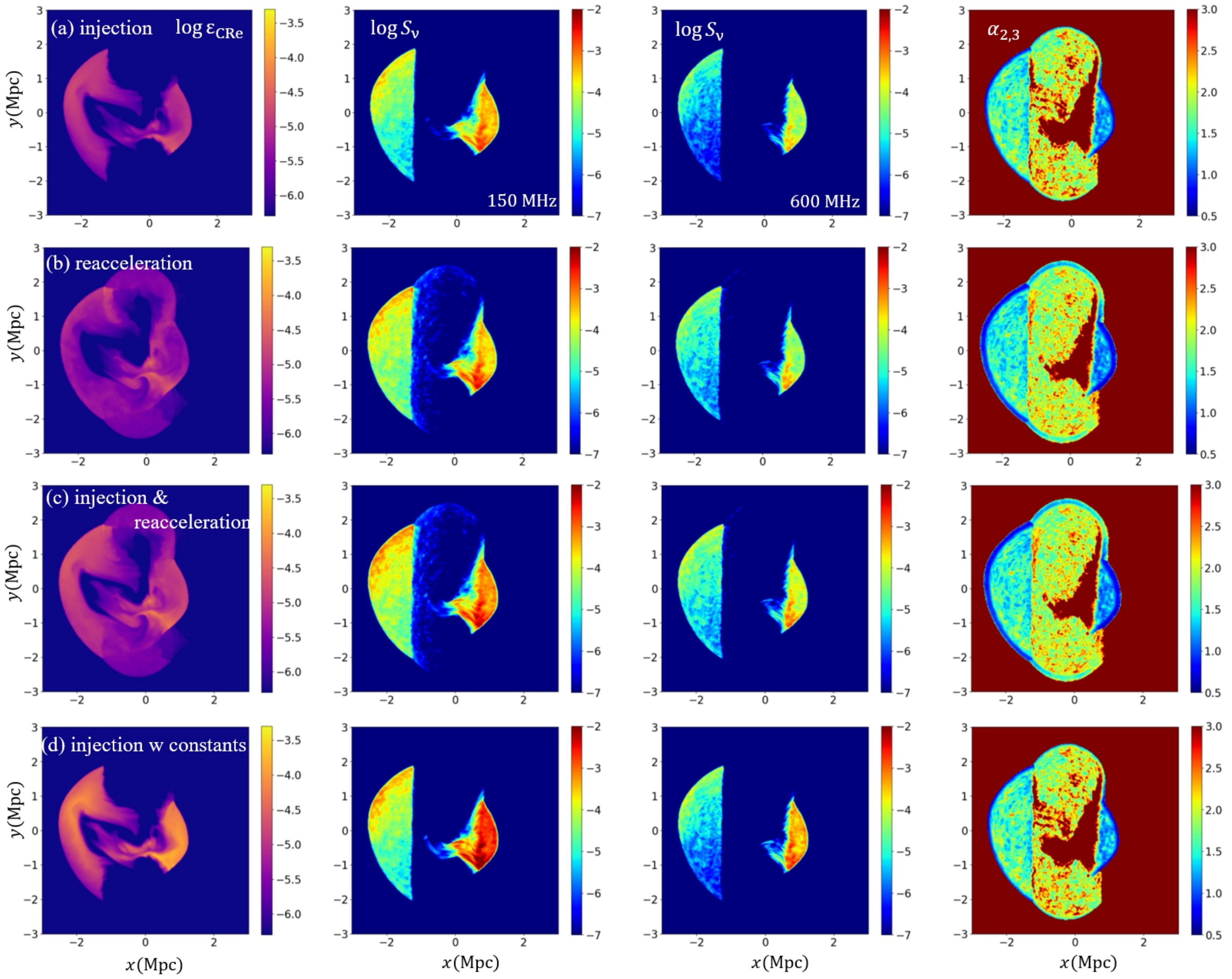}
\vskip -0.1cm
\caption{2D maps of $\varepsilon_{\rm CRe}/P_0$ in the merger plane, the projected synchrotron surface brightness, $S_{\nu}/S_0$, at 150 and 600 MHz, and the resulting spectral index, $\alpha_{2,3}$, at $t/t_0=1.0$, for \texttt{m4m2$\theta$10} model variants with four different CRe production scenarios: (a) fresh shock injection only (the same model as in Figure \ref{f8}(b)); (b) shock reacceleration of fossil CRe only; (c) both injection and reacceleration; and (d) fresh injection using constant parameters, $R_T=1.0$ and $Q_e=3.5$, instead of the parameterized formulas in equations (\ref{Qe})-(\ref{RT}).}\label{f11}
\end{figure*}

{Figures~\ref{f10}(a)–(b) show the temporal evolution of the volume-integrated CRe momentum spectrum, $E_c(p)$, and the volume-integrated radio spectrum, $J_{\nu}$, for the \texttt{m4m4$\theta$0} model without turbulent acceleration ($D_{\rm pp}=0$). As the merger shocks evolve, they weaken with the mean Mach number decreasing from $\langle M_s\rangle\approx 5.08$ at $t/t_0=0.7$ (magenta) to $3.26$ at $t/t_0=1.0$ (black), while the shock surface area increases. Accordingly, the integrated CRe spectrum steepens at low energies. At the same time, the integrated radio spectrum becomes steeper and its amplitude decreases, reflecting both the evolution of $E_c(p)$ and the decline of the magnetic-field strength in the cluster outskirts.}

{We highlight the following features in the figures: 
(1) $E_c(p)$ steepens above $p_{\rm br}\sim 10^3$; 
(2) the break momentum of $E_c(p)$, $p_{\rm br}$, decreases with the shock age, $t_{\rm age}=t-t_p$; 
(3) the slope of $E_c(p)$ below $p_{\rm br}$ increases as the shock weakens over time; 
(4) the break frequency of $J_{\nu}$, $\nu_{\rm br}$, decreases with $t_{\rm age}$; and 
(5) the overall amplitude of $J_{\nu}$ declines. 
In summary, $J_{\nu}$ deviates from a simple power-law form due to several factors, including the spherical-cap geometry of the shock surface, the temporal decline of the shock speed, and the radial decrease of the magnetic field strength in the cluster outskirts.}

Figures~\ref{f10}(c)–(d) compare $E_c(p)$ and $J_{\nu}$ for the \texttt{m4m4$\theta$0} models with different momentum-diffusion strengths at $t/t_0=1.0$. For the mean Mach number $\langle M_s\rangle\approx 3.26$, the slopes of $F(p)$ and $J_{\nu}$ at the shock are $q_{\rm sh}\approx 4.42$ and $\alpha_{\rm sh}\approx 0.71$, respectively; thus, in the idealized planar-shock case, the integrated slopes above the break are expected to be $q_{\rm int}\approx 5.42$ and $\alpha_{\rm int}\approx 1.21$.

{In Figure~\ref{f10}(c), for the model with $D_{\rm pp}=0$ (red), the volume-integrated CRe spectrum shows slopes of $q_{\rm int}\approx 4.23$ at $\gamma_e\sim10^{2.5}$ (below the break) and $q_{\rm int}\approx 5.49$ at $\gamma_e\sim10^{3.5}$ (above the break). For the model with $\tau_{\rm pp}=0.1~\mathrm{Gyr}$ (black), the corresponding slopes are $q_{\rm int}\approx 4.29$ and $q_{\rm int}\approx 5.48$. These values are close to the idealized expectations, differing by only a few to several percent, and exhibit only a modest dependence on the momentum-diffusion strength.}

{In contrast, deviations from the idealized case are more pronounced in the integrated spectral indices, $\alpha_{\rm int}$, and hence in the Mach numbers inferred from them. In Figure~\ref{f10}(d), for $D_{\rm pp}=0$, the indices between 53 and 150 MHz and between 150 and 600 MHz are $\alpha_{\rm int}^{1,2}\approx 1.30$ and $\alpha_{\rm int}^{2,3}\approx 1.30$, respectively, while for $\tau_{\rm pp}=0.1$~Gyr they are $\alpha_{\rm int}^{1,2}\approx 1.30$ and $\alpha_{\rm int}^{2,3}\approx 1.35$. These values exceed the idealized prediction, $\alpha_{\rm int}\approx 1.21$, by $\sim 5$–$10\%$, a somewhat larger discrepancy than for $q_{\rm int}$, partly because $\alpha_{\rm int}$ is measured over frequency intervals.}

{The values $\alpha_{\rm int}\approx 1.30$ and $1.35$ correspond to Mach numbers of $M_{\rm s,int} \approx 2.76$ and $2.59$, respectively, in equation (\ref{Mint}), underestimating the actual mean value of $\langle M_s\rangle\approx 3.26$ by $\sim 15$–$20\%$. These results indicate that combined effects, such as time-dependent shock properties, spherical shock geometry, spatial variations in the magnetic field, and turbulent acceleration, can bias Mach number estimates derived from radio relic observations at the level of tens of percent. However, turbulent acceleration alone does not significantly affect the inferred Mach number if $D_{\rm pp}$ is independent of particle momentum.}

\subsection{Impacts of Preexisting Fossil Population}\label{s3.4}

Next, to examine the influence of a preexisting fossil CRe population and the adopted injection model, we compare four variants of the \texttt{m4m2$\theta$10} model at $t/t_0 = 1.0$: (1) fresh shock injection only (the default setting); (2) shock reacceleration of fossil CRe only (with injection turned off); (3) both fresh injection and reacceleration; and (4) fresh injection only but with constant injection parameters, $Q_e=3.5$ and $R_T=1.0$. Figure~\ref{f11} presents the corresponding maps of $\varepsilon_{\rm CRe}$, $S_{\nu}(150~\mathrm{MHz})$, $S_{\nu}(600~\mathrm{MHz})$, and $\alpha_{2,3}$.

Comparing panel (a) (injection only) with panel (d) (injection with constant $Q_e$ and $R_T$) reveals notable differences; CRe acceleration becomes more efficient at high-Mach-number shocks when fixed injection parameters are adopted. This result highlights that the modeling of $Q_e$ and $R_T$ can significantly influence theoretical predictions of radio relic properties. While our prescriptions in equations~(\ref{Qe}) and (\ref{RT}) are physically motivated, observations of radio relics may potentially help refine these parameterizations and thereby advance our understanding of particle acceleration in cluster merger shocks.

Additionally, the comparison of the figures in panels (a), (b), and (c) demonstrates that the presence of a fossil CRe population, as described by equations~(\ref{fpre}) and (\ref{f_o}), enhances the downstream CRe energy density relative to that produced by fresh injection alone. {However, the overall morphology of the radio relics remains largely unchanged, indicating that fossil electrons primarily increase the brightness of the radio relics, rather than altering their observed structures.} 

\begin{figure*}[t]
\vskip 0.1cm
\hskip 0.2 cm
\includegraphics[width=0.95\linewidth]{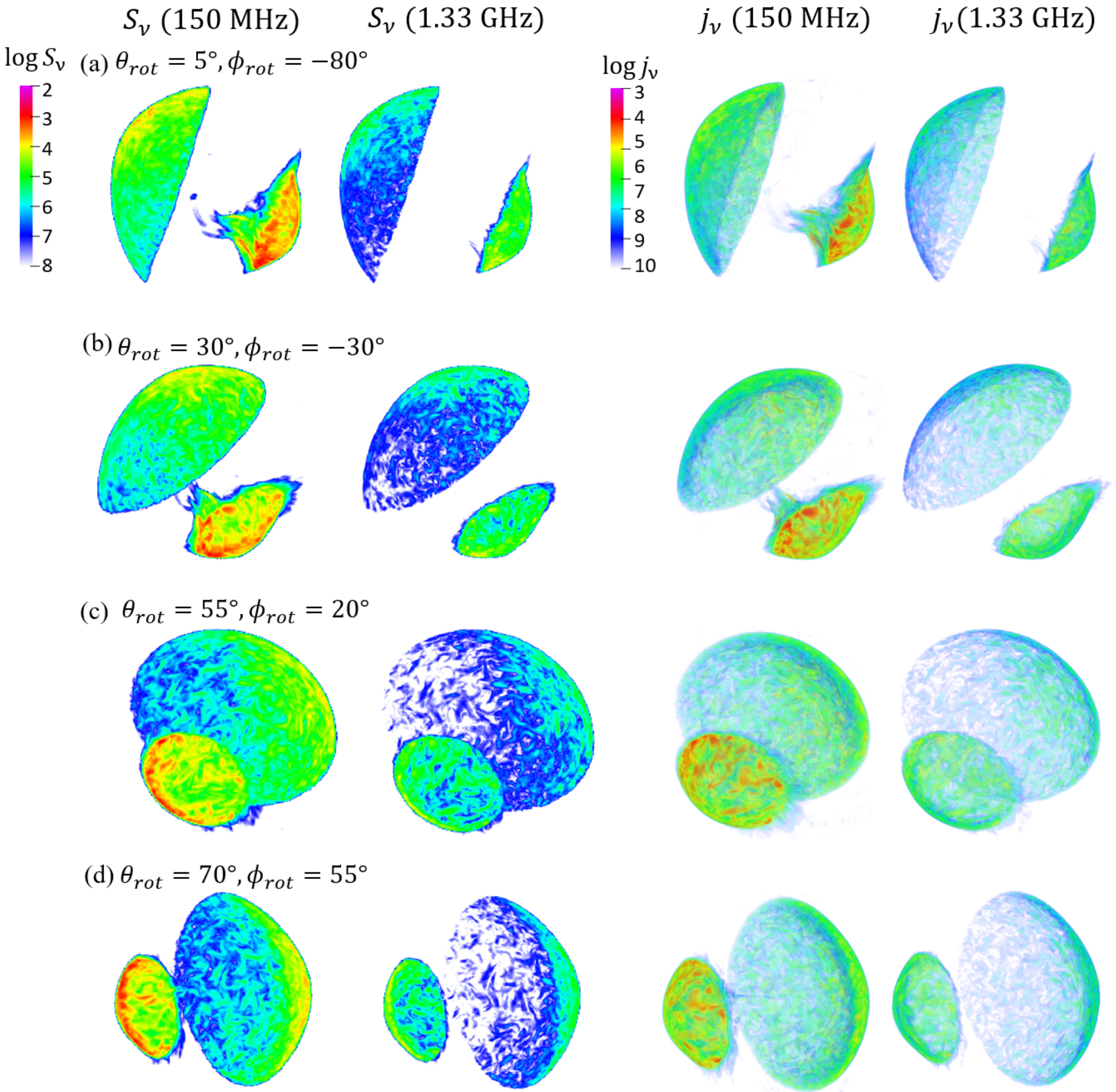}
\vskip -0.1cm
\caption{{Synthetic radio relic images along the LoS of different viewing angles for the \texttt{m4m2$\theta$10} model at $t/t_0=1.0$. The first two columns present the synchrotron surface brightness, $S_{\nu}(x',y')$, projected along the $z'$-axis at 150~MHz and 1.33~GHz. The last two columns show volume-rendered images of the 3D emissivity distribution, $j_{\nu}(\mbox{\boldmath$x'$})$, at the same frequencies. Each row displays images in the $x'-y'$ plane of the rotated frame specified by the polar and azimuthal angles $(\theta_{\rm rot}, \phi_{\rm rot})$: (a) $(5^{\circ},-80^{\circ})$; (b) $(30^{\circ},-30^{\circ})$; (c) $(55^{\circ},20^{\circ})$; and (d) $(70^{\circ},55^{\circ})$. Here, $S_{\nu}$ is given in units of mJy~beam$^{-1}$ with a beam size of $\theta_{\rm beam}^2 = 1'' \times 1''$, while $j_{\nu}$ is shown in units of $10^{-23}~{\rm erg~s^{-1}~cm^{-3}~Hz^{-1}~sr^{-1}}$.}}\label{f12}
\end{figure*}

\subsection{Projection Effects}\label{s3.4}

To illustrate how the observed morphology of radio relics depends on the viewing direction, we examine synthetic radio images constructed for different LoS, using the simulation data for the \texttt{m4m2$\theta$10} model at $t/t_0=1.0$. For this, we first take the 3D data cubes of $\mbox{\boldmath$B$}(\mbox{\boldmath$x$})$ and $e_c(p, \mbox{\boldmath$x$})$ in the simulation coordinate system $\mbox{\boldmath$x$}=(x,y,z)$, and produce the remapped cubes in a rotated frame $\mbox{\boldmath$x'$}=(x',y',z')$ specified by the angles $\theta_{\rm rot}$ and $\phi_{\rm rot}$. Here, $\theta_{\rm rot}$ is the polar angle measured from the $+z$-axis, and $\phi_{\rm rot}$ is the azimuthal angle measured in the $x-y$ plane from the $+x$-axis. Accordingly, the three components of $\mbox{\boldmath$B$}(\mbox{\boldmath$x'$})$ are redefined in this rotated coordinate system. Figure~\ref{f12} presents the surface brightness, $S_{\nu}(x',y')$, integrated along the $z'$-axis, alongside volume-rendered images of the 3D distribution of $j_{\nu}(\mbox{\boldmath$x'$})$ at 150~MHz and 1.33~GHz for four sets of rotation angles ($\theta_{\rm rot}$,$\phi_{\rm rot}$). Here, the synchrotron emissivity $j_{\nu}(\mbox{\boldmath$x'$})$ is computed from $e_c(p, \mbox{\boldmath$x'$})$ and $B_{\perp}(\mbox{\boldmath$x'$})=(B_{x'}(\mbox{\boldmath$x'$})^2+B_{y'}(\mbox{\boldmath$x'$})^2)^{1/2}$, where $B_{x'}(\mbox{\boldmath$x'$})$ and $B_{y'}(\mbox{\boldmath$x'$})$ denote the components in the rotated frame. In this setup, while the LoS is along the $z'$-direction, the $x'-y'$ plane corresponds to the sky plane.

Figure~\ref{f12} illustrates how the apparent morphology of radio relics, including their shape, extent, and brightness contrast, depends on the viewing direction, despite the underlying shock structure is identical in all panels. For example, when the LoS is perpendicular to the merger plane, the relics appear as spherical caps, with arc-like boundaries tracing the shock surfaces in the plane (see Figures \ref{f8}(b) and \ref{f11}(a)). For other viewing angles, especially in the third-row case, on the other hand, emissions from different parts of the shock surfaces overlap along the LoS, producing broader and more irregular features. Therefore, such projection effects can potentially complicate the reconstruction of the true shape of the underlying merger shocks. However, these projection effects remain qualitatively similar across the observed frequency range. In addition, the comparison between the 2D $S_{\nu}$ maps and the 3D $j_{\nu}$ structures highlights how complex volumetric synchrotron features are projected into the observed relic morphologies.

Figure~\ref{f12} further demonstrates the presence of patchy, fine features in both the projected and volume-rendered maps, generated largely by highly intermittent filamentary magnetic fields. We point out that the numerical resolution of the $512^3$ simulation, $\Delta x = 19.5$ kpc, is close to or slightly larger than the Coulomb mean free path in the ICM \citep[e.g.,][]{brunetti2014}. Consequently, whereas the physics operating near and below the Coulomb scale remains uncertain, physical processes operating on plasma scales are not resolved in the present simulations. On the other hand, current and future radio observations, such as those from the Low-Frequency Array (LOFAR), the Square Kilometre Array (SKA; \citealt{Dewdney2009}), or the next-generation Very Large Array (ngVLA; \citealt{Murphy2018}), are expected to achieve increasingly high sensitivity and spatial resolution. Such observations may provide valuable constraints on turbulent magnetic fields and the underlying plasma processes, thereby improving simulation studies such as the one presented in this paper.

\section{Summary}\label{s4}

We have carried out 3D simulations of idealized binary mergers of galaxy clusters, in which the initial subclusters have total (DM plus gas) virial masses of $M_{200}=(2$--$6)\times10^{14}\,M_{\odot}$. The gas component, described by a $\beta$-profile, is initially in hydrostatic equilibrium (HSE) and is permeated by turbulent magnetic fields. The merger dynamics and the formation and evolution of merger-driven shocks are followed using the HOW-MHD code, which employs a high-order WENO scheme.

In parallel with the MHD evolution, we track the acceleration and transport of cosmic-ray electrons (CRe) by solving the Fokker-Planck equation {on a uniform Eulerian grid}. CRe are supplied through diffusive shock acceleration (DSA) at shock zones via fresh injection and reacceleration of preexisting fossil CRe. Injection is modeled using a thermal-leakage prescription with Mach-number-dependent parameters $R_T(M_s)$ and $Q_e(M_s)$, which account for shock modification of the postshock temperature and the Mach-dependent injection momentum. In the downstream regions, the evolution of CRe, subject to Coulomb, bremsstrahlung, IC, and synchrotron cooling, as well as Fermi-II (turbulent) acceleration parameterized by a constant acceleration time $\tau_{\rm pp}=0.1$--1~Gyr, is followed.

This paper details the numerical framework, and then presents the first set of simulation results {Below, we summarize the results quoting values for the unequal-mass merger models, \texttt{m4m2$\theta$0} and \texttt{m4m2$\theta$10}, and highlight several general findings that apply broadly across the full set of models.}

1) As the merger evolves, the subclusters accelerate toward each other, launching equatorial shocks near pericenter passage. As they subsequently recede, each subcluster drives an axial bow shock (Shock~1 and Shock~2). The surfaces of these shocks consist of multiple shock zones with varying Mach numbers, leading to spatially nonuniform CRe injection. The equatorial shocks are weak and contribute negligibly to CRe production. In the unequal-mass merger models, Shock~1 (ahead of the heavier subcluster) is stronger but less spatially extended than Shock~2 (ahead of the lighter subcluster). At our representative observational epoch, approximately 1~Gyr after pericenter passage, the axial shocks in the \texttt{m4m2$\theta$0} model have expanded to $R_s\sim1$--3~Mpc, with mean Mach numbers of $\langle M_{s,1}\rangle \sim 3.9$ for Shock~1 and $\langle M_{s,2}\rangle \sim 2.8$ for Shock~2. Both shocks exhibit relatively narrow Mach-number distributions with widths of $\Delta M_s\sim0.5$.

2) In the unequal-mass merger models, Shock~1 produces a brighter, although more compact, radio relic than Shock~2 due to its higher $\langle M_s\rangle$ and stronger downstream magnetic fields. While the radio spectral index measured at the shock surface estimates the shock Mach number broadly consistent with the actual Mach number, the radio spectrum steepens downstream because of radiative cooling.

3) Merger-driven turbulence amplifies magnetic fields, making them highly intermittent and spatially nonuniform. These fields imprint patchy, fine-scale structures on the synchrotron emissivity, $j_{\nu}$. As a result, both the 2D synchrotron surface-brightness map, $S_{\nu}$, and the 3D rendering of $j_{\nu}$ exhibit small-scale variations associated with turbulent magnetic fields.

4) Turbulent acceleration with $\tau_{\rm pp}\lesssim1$~Gyr affects the CRe energy spectrum and, consequently, the synchrotron radio spectrum. In our models, which assume constant $\tau_{\rm pp}$, turbulent acceleration primarily boosts the amplitudes of the spectra, thereby increasing the surface brightness of radio relics. The presence of a preexisting fossil CRe population further enhances the downstream CRe energy density and the radio surface brightness.

{5) Both the volume-integrated CRe spectrum, $E_c(p)$, and the volume-integrated radio spectrum, $J_{\nu}$, show departures from the canonical steepening relations due to CRe cooling, $q_{\rm int}=q_{\rm sh}+1$ and $\alpha_{\rm int}=\alpha_{\rm sh}+0.5$, which are expected for an idealized plane-parallel shock with uniform magnetic fields and no turbulent acceleration. The is due to the combined effects of time-dependent shock evolution, spherical shock geometry, spatially varying magnetic fields, and turbulent acceleration. In our simulations, while the deviation in the integrated CRe slope is modest, the deviation in the integrated radio spectral index is somewhat larger. As a result, estimating the merger-shock Mach number in radio observations using the commonly employed relation, $M_{\rm s,int}=[(\alpha_{\rm int}+1)/(\alpha_{\rm int}-1)]^{1/2}$, can mislead the actual shock Mach number by up to a few tens of percent.}

6) Although the thermal-leakage injection model with Mach-number-dependent $R_T(M_s)$ and $Q_e(M_s)$ is physically motivated, CRe acceleration remains sensitive to the adopted parameters. A test case with constant values of these injection parameters, still within plausible ranges, produces noticeably higher $\varepsilon_{\rm CRe}$, highlighting the importance of postshock thermodynamics and injection modeling.

7) The apparent morphology and brightness distribution in both the 2D maps of  $S_{\nu}$ and the 3D renderings of $j_{\nu}$ depend strongly on the viewing angle, even for the same underlying shock. This implies that projection effects can complicate the reconstruction of the true geometry of merger shocks, as well as the retrieval of their correct properties.

While our idealized binary merger simulations are designed to capture key physical processes for radio relics, they do not fully reproduce the complexity seen in observations and cosmological simulations. {For example, in our models, Shock~1 is stronger, with a higher Mach number, and is spatially less extended than Shock~2, whereas some observed and simulated merging systems show the opposite behavior.} Moreover, in cosmological simulations, the Mach number distributions of shock zones across merger shock surfaces are typically broader and log-normal, in contrast to the narrower distributions obtained here. Such differences arise because realistic environments involve frequent minor mergers and continuous WHIM infall, which are absent from our idealized setup. Nevertheless, our results highlight the interconnected roles of merger dynamics, MHD turbulence, and CRe physics in shaping radio relics in cluster outskirts.

{Furthermore, the default simulation configuration adopted in this work employs a uniform grid of $N_{\rm grid}=512^3$ zones in a cubic box of a side length of $r_0=10$ Mpc and $N_p=64$ logarithmically spaced momentum bins spanning $p/m_ec=10-10^5$. For the merger shocks considered here, the typical postshock cooling length is $\sim100$ kpc, so the $512^3$ simulations resolve the synchrotron-emitting region with approximately 5 zones. Although modest, this resolution appears adequate for capturing the global features of radio luminosity and spectral aging. Therefore, the present exploratory study provides a validated framework for modeling the macrophysics of cluster mergers. Nevertheless, we point out that these resolutions will need to be further enhanced to fully reproduce the fine-scale substructures observed in radio relics like the Sausage. Studies aimed at modeling such radio relics will likely require kpc-scale resolution to resolve small-scale turbulent magnetic fluctuations. The kpc spatial resolution can be achieved only by employing AMR or comparable techniques, and we leave such implementation for future work.}

Finally, this work is the first in a series aimed at modeling binary cluster mergers. Future efforts will relax the fixed ``gravity halo'' approximation by implementing a particle--mesh gravity scheme, enabling the self-consistent DM--gas gravitational interaction. We also plan to further explore the observational manifestations of simulated merger shocks in both radio and X-ray bands, including the Mach numbers quoted in observations and the polarized radio signatures, as well as the impacts of minor mergers and WHIM accretion.

\begin{acknowledgments}
The authors thank the anonymous referee for constructive comments and suggestions that have improved the clarity of the paper. HK's work was supported by the National Research Foundation (NRF) of Korea through grant RS-2023-NR076397. DR's work was supported by NRF of Korea through grant RS-2025-00556637, and the Korea Astronomy and Space Science Institute (KASI) under the R\&D program (Project No. 2025-9-844-00) supervised by the Korea AeroSpace Administration (KASA). JS’s work was supported by the Los Alamos National Laboratory (LANL) through the Center for Space and Earth Science (CSES), funded by LANL’s Laboratory Directed Research and Development (LDRD)program under project No. 20240477CR. {The authors used ChatGPT (OpenAI) during the preparation of this manuscript to improve language clarity and grammatical correctness. Data visualization was done using \texttt{SM} \citep{lupton1991}, \texttt{Matplotlib} \citep{hunter2007}, and \texttt{ParaView} \citep{Ahrens2005, Woodring2011}.} 
\end{acknowledgments}

{\software{Matplotlib \citep{hunter2007}, NumPy \citep{harris2020}, SM \citep{lupton1991}, ParaView \citep{Ahrens2005, Woodring2011}, HOW-MHD \citep{seo2023}}}

\bibliography{reference}{}
\bibliographystyle{aasjournal}

\appendix

\twocolumngrid

\setcounter{equation}{0}
\renewcommand{\theequation}{\thesection\arabic{equation}}


\section{Numerical Scheme to Solve the Fokker--Planck Equation}\label{sec:sa}

{The Fokker--Planck (FP) equation for the CRe energy density, $e_c(p,\mbox{\boldmath$x$}) = 4\pi p^3 f(p,\mbox{\boldmath$x$})$, in equation (\ref{diffcon}) can be rewritten as
\begin{equation}
\frac{\partial e_c}{\partial t} 
+ \frac{\partial}{\partial \mbox{\boldmath$x$}} \cdot (e_c \mbox{\boldmath$u$}) 
=  \frac{\partial}{\partial h}\!\left( D'_{\rm pp}\, \frac{\partial e_c}{\partial h} \right) 
-\frac{\partial}{\partial h}\!\left( A_p\, e_c \right)
+ Q(\mbox{\boldmath$x$},h),
\label{FP}
\end{equation}
where $h=\ln (p/m_ec)$, and
\begin{align}
A_p &\equiv  3 \frac{D_{\rm pp}}{p^2} -\frac{1}{3}\,\mbox{\boldmath$\nabla$} \cdot \mbox{\boldmath$u$} 
       - \frac{b_l}{p}, \\[4pt]
D'_{\rm pp} &\equiv \frac{D_{\rm pp}}{p^2}.
\end{align}
In our FP solver for equation (\ref{FP}), $e_c(p,\mbox{\boldmath$x$})$ is assigned on all spatial grid zones, and its temporal evolution is followed using an ``Eulerian'' scheme. Here, $\mbox{\boldmath$u$}$ is the velocity of the background MHD flow, and $D_{\rm pp}$ is the momentum diffusion coefficient. The total energy loss rate is given by $b_l=b_{\rm IC} + b_{\rm syn}+b_{\rm Coul} + b_{\rm brem}$; for the IC, synchrotron, Coulomb, and bremsstrahlung losses, we adopt the formulations of \citet{sarazin1999}. These loss terms are evaluated at the instantaneous redshift $z_r(t)$, using the local MHD quantities at each spatial grid zone, namely $B(\mbox{\boldmath$x$})$ and the gas number density, $n_H(\mbox{\boldmath$x$})$. The source term $Q(\mbox{\boldmath$x$},h)$ accounts for particle injection and reacceleration at shock zones.}

\subsection{Discretization}\label{a1}

{Equation~(\ref{FP}) is written in a discretized form as
\begin{equation}
\frac{e_{c,i,l}^{n+1} - e_{c,i,l}^{n}}{\Delta t} = \mathcal{A}^s_{i,l} + \mathcal{A}^p_{i,l} + \mathcal{D}_{i,l} +Q_{i,l},
\label{disc}
\end{equation}
where $\mathcal{A}^s_{i,l}$ and $\mathcal{A}^p_{i,l}$ denotes operators for the advection terms in physical space and momentum space, respectively, while $\mathcal{D}_{i,l}$ denotes the diffusion term in momentum space. Here, the index $n$ represents the timestep, and $i$ and $l$ label the spatial grid zones and the logarithmically spaced momentum bins, respectively. For brevity, only the spatial index along the $x$-direction is retained, and the indices for the $y$ and $z$-directions are dropped.

The spatial advection term is given by
\begin{equation}
\mathcal{A}^s_{i,l} = - \frac{1}{\Delta x}\left(F^s_{i+\frac{1}{2},l}-F^s_{i-\frac{1}{2},l}\right),
\label{advs}
\end{equation}
where $F^s_{i\pm\frac{1}{2},l}$ are the numerical fluxes computed for $F^s = e_c u_x$ at the interfaces between grid zones $i$ and $i\pm1$. Only the formulation along the $x$-direction is presented; the corresponding expressions along the $y$ and $z$-direction are given analogously. The momentum advection term is given by
\begin{equation}
\mathcal{A}^p_{i,l} = - \frac{1}{\Delta h}\left(F^p_{i,l+\frac{1}{2}}-F^p_{i,l-\frac{1}{2}}\right),
\label{adv}
\end{equation}
where $F^p_{i,l\pm\frac{1}{2}}$ are the numerical fluxes computed for $F^p = A_p e_c$ at the interfaces between momentum bins $l$ and $l\pm1$. The numerical fluxes, $F^s_{i\pm\frac{1}{2},l}$ and $F^p_{i,l\pm\frac{1}{2}}$, are reconstructed using a WENO (Weighted Essentially Non-Oscillatory) scheme as described in the next subsection.

The momentum diffusion term is given by
\begin{equation}
\mathcal{D}_{i,l} = \frac{1}{\Delta h}\left(G_{i,l+\frac{1}{2}}-G_{i,l-\frac{1}{2}}\right),
\label{momdiff1}
\end{equation}
where
\begin{equation}
G_{i,l+\frac{1}{2}} = (D'_{\rm pp})_{i,l+\frac{1}{2}}
\left[\frac{(e_c)_{i,l+1}-(e_c)_{i,l}}{\Delta h} \right].
\label{momdiff2}
\end{equation}

The advection and diffusion contributions to the evolution of $e_{c,i,l}$ are numerically integrated as described in Section \ref{a4}. On the other hand, with our scheme for fresh injection and reacceleration of CRe at the shock (see Section \ref{s2.6}), the source term $Q$ is not time-integrated. Instead, at the beginning of each FP step, the CRe energy density is reset in ``shock zones'' according to equations (\ref{finj}) and (\ref{fRA}), $e_{c,\rm sh}(p)= 4\pi p^3 [f_{\rm inj}(p)+f_{\rm RA}(p)]$, so the source term acts independently of the explicit integration timestep, $\Delta t$.}

\subsection{WENO Scheme}\label{a2}

{We adopt a third-order finite-difference scheme described in \citet{jiang1996}. Here, we briefly outline the reconstruction of the numerical flux, leaving details to the original reference. Below, the index $m$ denotes either the spatial grid zone for the spatial flux $F^s$ or the momentum bin for the momentum flux $F^p$. The numerical flux at the zone or bin interface is given by
\begin{equation}
F_{m+\frac{1}{2}} = F^{+}_{m+\frac{1}{2}} + F^{-}_{m+\frac{1}{2}},
\end{equation}
where
\begin{align}
F^{+}_{m+\frac{1}{2}} & = \omega_0 \left(F^{+}_m + \frac{1}{2}\Delta F^{+}_{m-\frac{1}{2}} \right) + \omega_1 \left(F^{+}_m + \frac{1}{2}\Delta F^{+}_{m+\frac{1}{2}} \right), \\
F^{-}_{m+\frac{1}{2}} & = \omega_0 \left(F^{-}_{m+1} - \frac{1}{2}\Delta F^{-}_{m+\frac{3}{2}} \right) + \omega_1 \left(F^{-}_{m+1} - \frac{1}{2}\Delta F^{-}_{m+\frac{1}{2}} \right).
\end{align}

The local Lax--Friedrichs splitting is used to obtain
\begin{equation}
\Delta F^{\pm}_{m+\frac{1}{2}} = \frac{1}{2}\left(\Delta F_{m+\frac{1}{2}} \pm a_{\max}\Delta e_{c,m+\frac{1}{2}} \right),
\end{equation}
where $\Delta F_{m+\frac{1}{2}} = F_{m+1}-F_m$, and $\Delta e_{c,m+\frac{1}{2}} = e_{c,m+1}-e_{c,m}$. The quantities $\Delta F^+_{m-\frac{1}{2}}$ and $\Delta F^-_{m+\frac{3}{2}}$ are calculated analogously. Here, $a_{\max}$ is the maximum of $a$ in the stencil $m-1,m,m+1,m+2$, where $a=u_x$, $u_y$, or $u_z$ for the spatial flux and $a=A_p$ for the momentum flux. Accordingly, 
\begin{align}
F^{\pm}_m & = \frac{1}{2}\left(F_m \pm a_{\max} e_{c,m} \right), \\
F^{\pm}_{m+1} & = \frac{1}{2}\left(F_{m+1} \pm a_{\max} e_{c,m+1} \right).
\end{align}

The weight functions are defined as
\begin{equation}
\omega_0 = \frac{\alpha_0}{\alpha_0+\alpha_1}, \qquad
\omega_1 = \frac{\alpha_1}{\alpha_0+\alpha_1},
\end{equation}
with
\begin{equation}
\alpha_0 = (\beta_0+\epsilon)^2, \qquad
\alpha_1 = 2(\beta_1+\epsilon)^2.
\end{equation}
The smoothness indicators are given by $\beta_0=(\Delta F^{+}_{m+\frac{1}{2}})^2$ and $\beta_1=(\Delta F^{+}_{m-\frac{1}{2}})^2$ for the reconstruction of $F^{+}_{m+\frac{1}{2}}$, and $\beta_0=(\Delta F^{-}_{m+\frac{1}{2}})^2$ and $\beta_1=(\Delta F^{-}_{m+\frac{3}{2}})^2$ for $F^{-}_{m+\frac{1}{2}}$. The parameter $\epsilon$ is a small positive constant introduced to avoid division by zero.}

\subsection{Timestep}\label{a3}

{The temporal evolution of MHD quantities in the MHD solver (Section \ref{s2.3}) is advanced with the usual Courant-Friedrichs-Levy (CFL) timestep, determined globally over all spatial grid zones, as
\begin{equation}
\Delta t_{\rm MHD} = \min\!\left[\frac{f_{\rm CFL}\,\Delta x}{\max\left(\lambda_x,\lambda_y,\lambda_z\right)}\right],
\label{tMHD}
\end{equation}
with $f_{\rm CFL}=0.8$. Here, $\lambda_x$, $\lambda_y$, $\lambda_z$ are the maximum speeds of  characteristic modes along the $x$, $y$, and $z$-directions, respectively.

On the other hand, the FP solver for equation~(\ref{FP}) involves timesteps associated with $\mathcal{A}^p_{i,l}$ and  $\mathcal{D}_{i,l}$, in addition to that for $\mathcal{A}^s_{i,l}$ which is essentially the MHD timestep; hence we set the FP timestep as
\begin{equation}
\Delta t_{\rm FP} = \min\left[\min\!\left(\frac{0.5\,\Delta h}{|A_p|}, \, \frac{0.5\,(\Delta h)^2}{D'_{\rm pp}}\right), \Delta t_{\rm MHD}\right],
\label{tFP}
\end{equation}
where the minimum inside the bracket is evaluated over all momentum bins across all spatial zones. For the representative parameters adopted here (e.g., $z_r\sim 0.2$ and $B\sim 1 \muG$), the IC cooling timescale, $\tau_{\rm IC}\approx 0.1~{\rm Gyr} (\gamma_e/10^4)^{-1}$, controls the FP timestep. For instance, with $\gamma_{\rm e,max}=10^5$ and $N_p=64$ momentum bins, $\Delta t_{\rm FP} \sim 0.5-1\times 10^{-3}$~Gyr, whereas $\Delta t_{\rm MHD} \sim 1.3-1.5\times 10^{-3}$~Gyr for $512^3$ MHD runs. Hence, for our numerical configuration, the two timesteps are typically compared as $\Delta t_{\rm MHD} \sim 2-5 \Delta t_{\rm FP}$, requiring a few to several FP sub-steps per each MHD step. To maintain high fidelity during FP sub-cycling, the background MHD variables ($B$, $n_H$, and $\mbox{\boldmath$u$}$) are linearly interpolated in time between the current and subsequent MHD timesteps.}

\begin{figure*}
\vskip 0.1 cm
\hskip 0.5 cm
\includegraphics[width=0.9\linewidth]{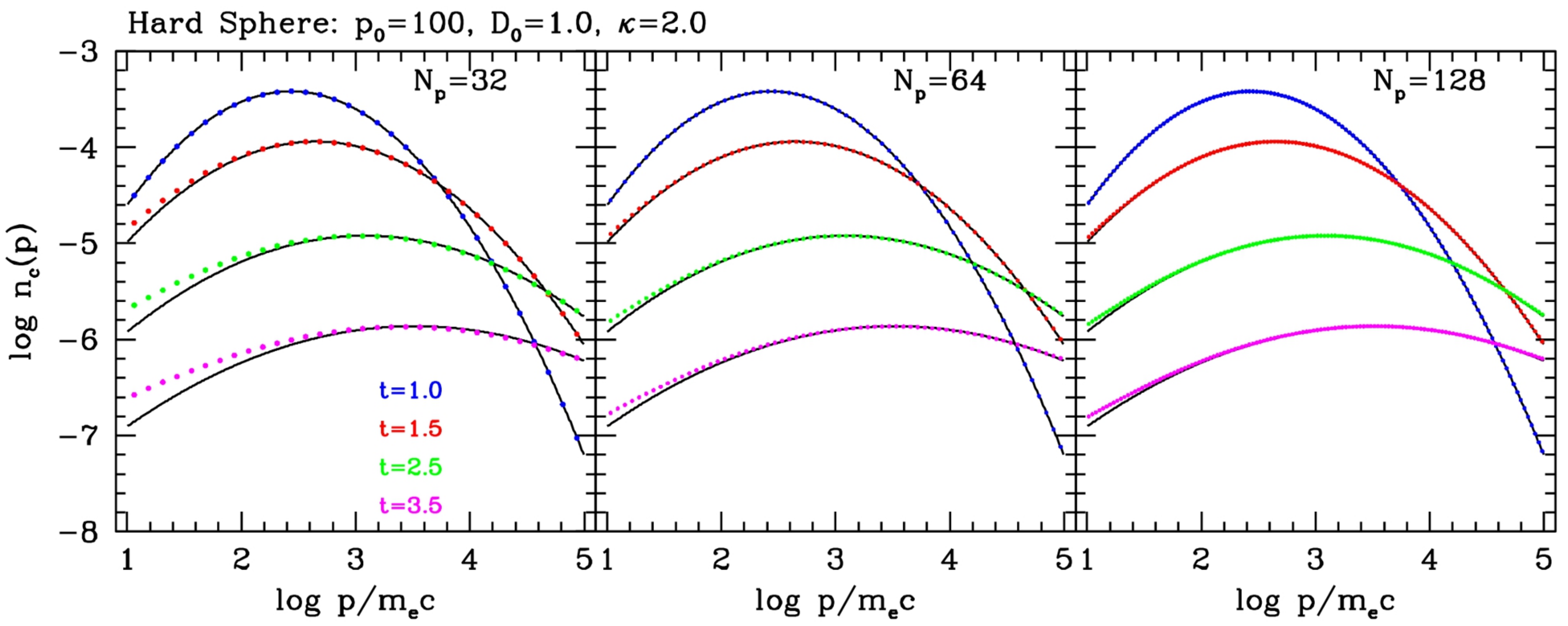}
\vskip -0.1 cm
\caption{{Comparison between the numerical results of our FP solver (colored dots) and the analytical solution of equation (\ref{FP3}) (black solid lines) for $p_0=10^2 m_e c$, $D_0=1.0$, and $\kappa=2.0$ (hard-sphere approximation) at different times $t$. Results for $N_p = 32, 64,$ and $128$ demonstrate convergence of the numerical solution toward the analytic solution with increasing resolution.}}\label{f13}
\end{figure*}

\subsection{Time Integration}\label{a4}

Equation~(\ref{disc}) comprises operators that act separately on the spatial and momentum coordinates. Based on the operator-splitting approach, we first integrate the spatial advection term with the second-order Runge--Kutta (RK) scheme and then integrate the momentum advection and diffusion terms using the Strong Stability Preserving Runge--Kutta Implicit--Explicit (SSP-IMEX) scheme \citep[e.g.,][]{pareschi2005,kundu2021}. The combination of operator splitting, WENO reconstruction, and RK and SSP-IMEX time integrations ensures stable and accurate reproduction of $e_c$ in smooth regions while getting $e_c$ around shocks without introducing spurious oscillations in our FP solver.

In the first part of the FP step from $n$ to $n+1$, $e_c$ is updated to account for spatial advection as
\begin{equation}
\begin{aligned}
e_c^{1} &= e_c^{n} + \Delta t_{\rm FP}\,\mathcal{A}^{s,n}, \\
e_c^{n'} &= e_c^{n} + \frac{\Delta t_{\rm FP}}{2}\left(\mathcal{A}^{s,n}+\mathcal{A}^{s,1}\right),
\end{aligned}
\end{equation}
using the RK scheme. Here and below, indices denoting spatial zones and momentum bins are omitted for clarity. The superscript $1$ denotes an intermediate value, and $n'$ indicates the value in which only spatial advection has been updated. $\mathcal{A}^s$ is computed with the numerical fluxes reconstructed using the WENO scheme. In this part, the spatial boundary conditions are the same as those for the MHD solver, which are periodic in all spatial directions for the simulations presented in this paper.

In the second part, the momentum advection and diffusion terms are integrated as
\begin{equation}
\begin{aligned}
e_c^{1'} &= e_c^{n'} + \Delta t_{\rm FP}\,\alpha \mathcal{D}^{1'}, \\
e_c^{2'} &= e_c^{n'} + \Delta t_{\rm FP}\left( \mathcal{A}^{p,1'} + (1 - 2\alpha)\mathcal{D}^{1'} + \alpha \mathcal{D}^{2'} \right), \\
e_c^{n+1} &= e_c^{n'} + \frac{\Delta t_{\rm FP}}{2} \left( \mathcal{A}^{p,1'} + \mathcal{A}^{p,2'} + \mathcal{D}^{1'} + \mathcal{D}^{2'} \right),
\end{aligned}
\end{equation}
employing the SSP-IMEX scheme. The superscripts $1'$ and $2'$ denote intermediate values, and $\alpha = 1 - 1/\sqrt{2}$. In the first stage to obtain $e_c^{1'}$, the diffusion is treated implicitly via the Crank-Nicolson scheme over a  time interval $\alpha \Delta t_{\rm FP}$. In the second stage to obtain $e_c^{2'}$, the advection term $\mathcal{A}^{p,1'}$ and the diffusion term $\mathcal{D}^{1'}$ are evaluated explicitly with the WENO scheme and equations (\ref{momdiff1}) and (\ref{momdiff2}) using $e_c^{1'}$; then, the diffusion involving $\mathcal{D}^{2'}$ is computed implicitly via the Crank--Nicolson scheme over the time interval $\alpha \Delta t_{\rm FP}$. In the final stage to obtain $e_c^{n+1}$, the advection and diffusion terms are evaluated explicitly using $e_c^{1'}$ and $e_c^{2'}$.

For the implicit treatment of momentum diffusion, zero-gradient conditions ($\partial e_c/\partial h=0$), equivalently continuous conditions, are imposed at both the lowest and highest momentum bins. For the explicit treatment of momentum advection and diffusion, a continuous condition is applied at the lowest momentum bin, while a power-law condition is enforced at the highest momentum bin, i.e., $e_c(N_p+1)/e_c(N_p)= (p_{N_p+1}/p_{N_p})^b$, where $b=\ln [e_c(N_p)/e_c(N_p-1)]/\Delta h$, considering the typical behavior of $e_c(p)$ near the momentum boundaries (see Figure \ref{f10}).

\begin{figure*}
\includegraphics[width=0.99\linewidth]{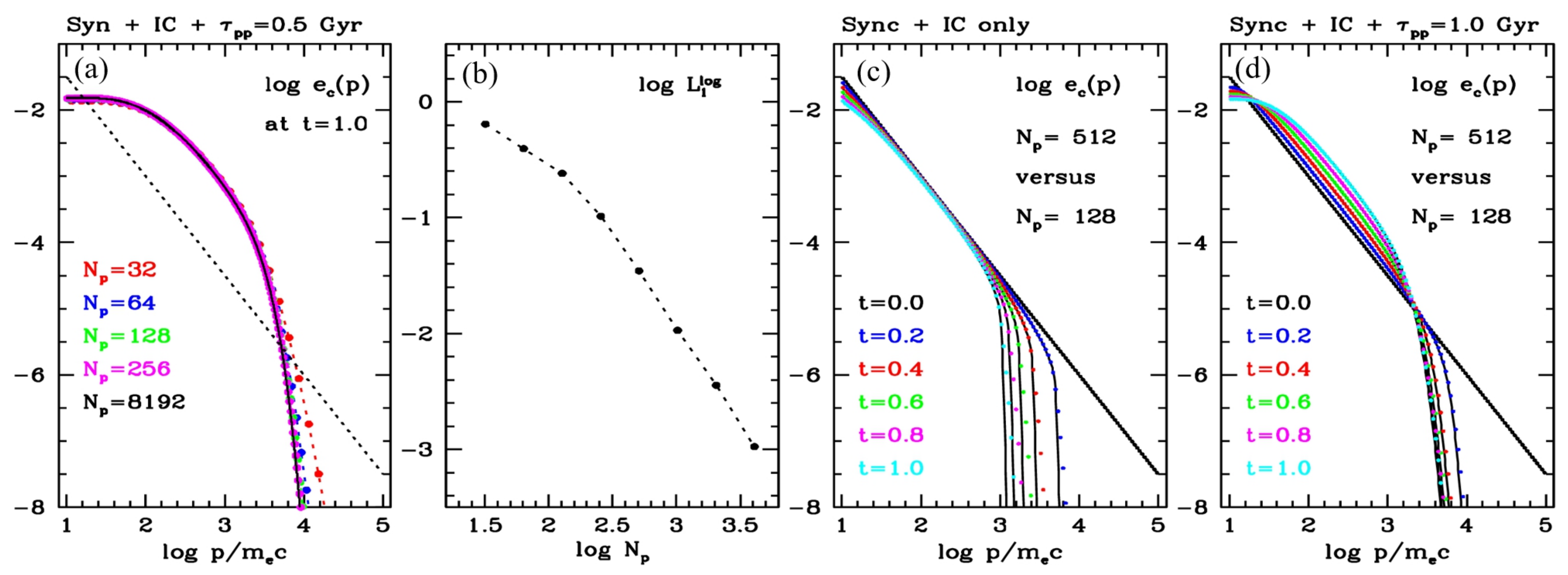}
\vskip -0.1 cm
\caption{
{Resolution convergence tests for one spatial zone simulations of postshock aging with different numbers of uniform logarithmic momentum bins, $N_p=32 - 8192$ ($n_H=10^{-4} \rm cm^{-3}$, $B=1 \muG$, $z_r=0.2$). (a) Spectrum of $e_c(p)$ at $t=1.0$~Gyr for the case with $\tau_{\rm pp}=0.5$~Gyr. The black dotted line shows the initial spectrum, $e_c(p)=p^{-1.5}$. (b) Log-relative $L_1$ error, $L_1^{\rm log}$, estimated for the numerical results shown in panel (a). The results with $N_p=8192$ are taken as the reference solution. (c) Spectrum of $e_c(p)$ at $t=0.0,~0.2, ..,1.0$~Gyr simulated with $N_p=128$ (colored dots) and $N_p=512$ (black solid lines) for the case with $D_{\rm pp}=0$. (d) Same as Panel (c) except $\tau_{\rm pp}=1.0$~Gyr.}}\label{f14}
\end{figure*}

\begin{figure}  
\vskip 0.1 cm
\hskip -0.2 cm
\centering
\includegraphics[width=1.01\linewidth]{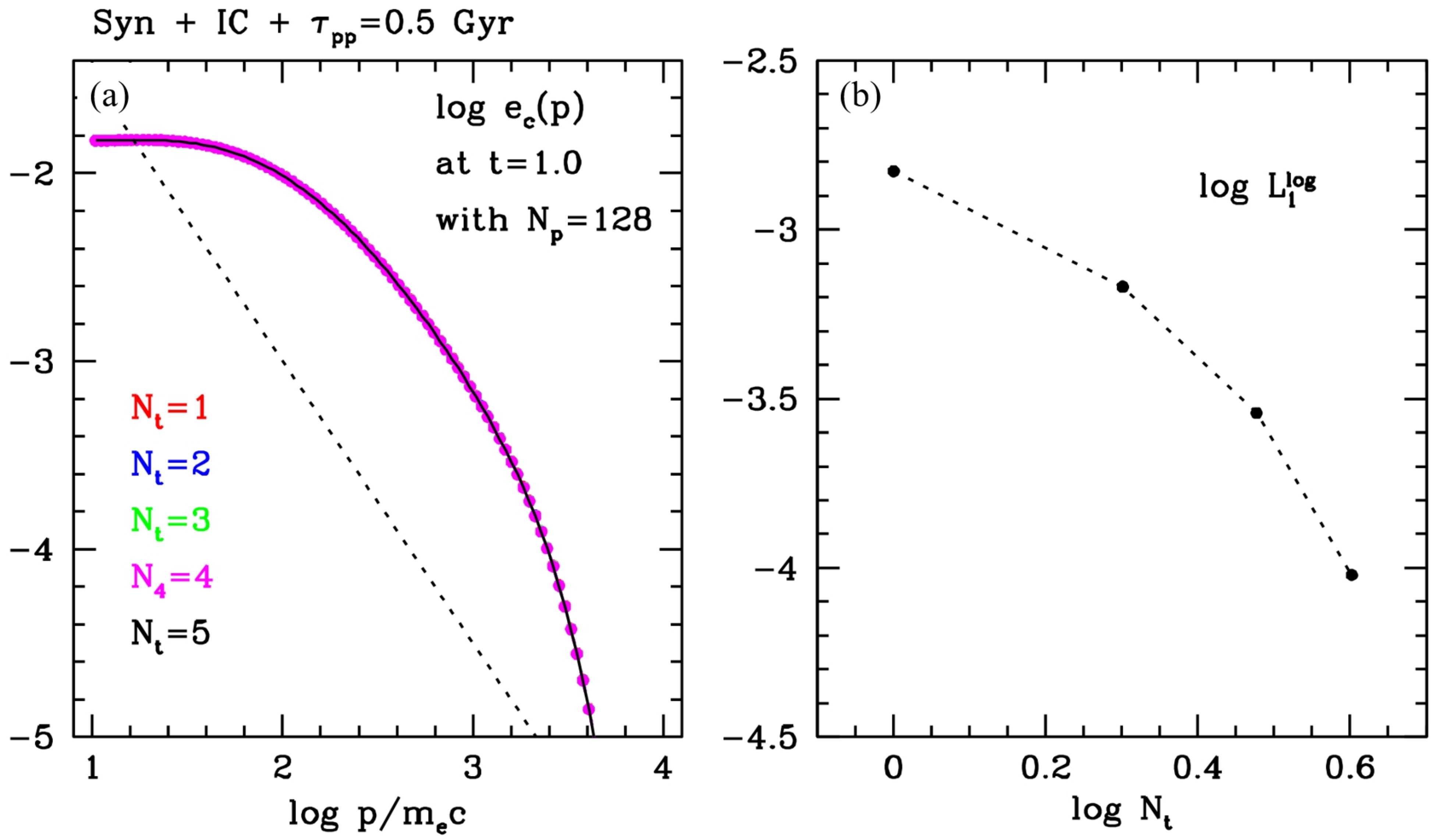}
\vskip 0.0 cm
\caption{{Time-stepping convergence tests for one spatial zone simulations of postshock aging with $N_p=128$ ($n_H=10^{-4} \rm cm^{-3}$, $B=1 \muG$, $z_r=0.2$ and $\tau_{\rm pp}=0.5$~Gyr). Different FP timesteps are used as follows: $\Delta t_{\rm FP} = (1/2)^{N_t} (\Delta h/\max(|A_p|))$, where $N_t=1-5$. (a) Spectrum of $e_c(p)$ at $t=1.0$~Gyr. The black dotted line shows the initial spectrum, $e_c(p)=p^{-1.5}$. (b) Log-relative $L_1$ error, $L_1^{\rm log}$, as a function of $N_t$, estimated for the numerical results shown in panel (a). The results with $N_t=5$ are taken as the reference solution.}}\label{f15}
\end{figure}

\section{Tests for the Fokker-Planck Solver}\label{sec:sb}

The FP solver is validated through tests conducted in one spatial zone, where $e_c(h)$ evolves only in the momentum coordinate.

\subsection{Hard-Sphere Approximation}\label{sec:sb1}

{We first assess the performance of our FP solver by comparing its solution with a known analytic solution, specifically that presented by \citet[][PP95, hereafter]{park1995}. They considered a simplified FP equation for the CRe number density $n_c(p)=4\pi p^2f(p)$,
\begin{equation}
\frac{\partial n_c}{\partial t}
= \frac{\partial}{\partial p}\!\left(D\, \frac{\partial n_c}{\partial p} \right)-\frac{\partial}{\partial p}\!\left(A\, n_c \right),
\label{FP2}
\end{equation}
where only diffusion and advection of $n_c(p)$ in momentum space are included. These terms are specified with $D=D_0 p^2$ and $A=A_0 p$ using constant $D_0$ and $A_0$. Then, the analytic solution is given by
\begin{equation}
n_c(p,p_0,t) =  \frac{1}{p\sqrt{4 \pi D_0 t}} \exp\left\{-\frac {[\ln(p/p_0)-(\kappa+1) D_0t]^2}{4D_0t}\right\},
\label{FP3}
\end{equation}
where $\kappa=A_0/D_0$ (see equation (43) of PP95). 

In terms of $e_c=p\, n_c$ and $h=\ln(p/m_ec)$, equation (\ref{FP2}) can be rewritten as
\begin{equation}
\frac{\partial e_c}{\partial t}
= \frac{\partial}{\partial h}\!\left(D_0\, \frac{\partial e_c}{\partial h} \right)-\frac{\partial}{\partial h}\!\left[\left(D_0+A_0\right)\, e_c \right].
\end{equation}
We solve this equation using our FP solver for the ``hard-sphere approximation'' with $D_0=1.0$ and $\kappa=2.0$. Figure \ref{f13} presents a comparison between the numerical results and the analytic solution in equation (\ref{FP3}). In this test, where momentum diffusion dominates, the primary source of numerical errors is the finite-domain implementation of the zero-gradient boundary conditions. The comparison indicates that the resolution of $N_p \gtrsim 64$ uniform logarithmic momentum bins over $p/m_ec = 10-10^5$ yields reasonably converged results.}

\begin{figure*}[t]
\vskip 0.1 cm
\includegraphics[width=0.99\linewidth]{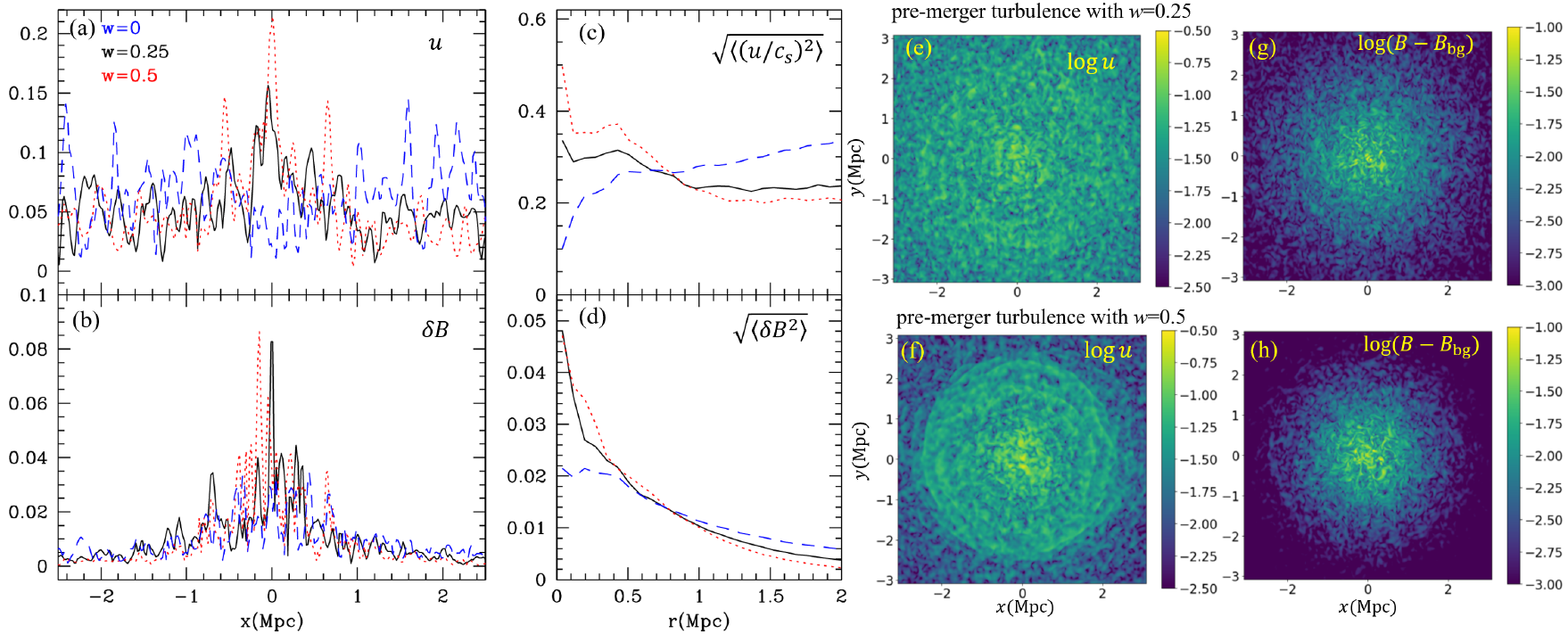}
\vskip -0.1 cm
\caption{{Pre-merger turbulence generated with density scaling parameter values $w=0$ (blue), 0.25 (black), and 0.5 (red) in the \texttt{m4} subcluster with $512^3$ grid zones. (a)-(b) 1D profiles of $u/u_0$ and $\delta B/B_0$  (where $\mathbf{\delta B} = \mathbf{B}-\mathbf{B_{bg}}$) along the $x$-axis. (c)-(d) Radial profiles of $\sqrt{\langle M_{s,\rm turb}^2 \rangle}$ and $\sqrt{\langle(\delta B/B_0)^2 \rangle}$ as a function of distance from the cluster center. (e)-(h) 2D spatial maps of $u/u_0$ and $\delta B/B_0$ in the $x-y$ plane ($z=0$) for models with $w=0.25$ (upper) and $w=0.5$ (lower). All quantities are normalized by their respective reference units in Section \ref{s2.1}, as in the figures in the main part.}}\label{f16}
\end{figure*}

\subsection{Convergence Tests}\label{sec:sb2}

Next, we conduct a convergence test by solving equation (\ref{FP}) without the terms involving spatial variation, $\mbox{\boldmath$\nabla$} \cdot (e_c\mbox{\boldmath$u$})$ and $\mbox{\boldmath$\nabla$} \cdot \mbox{\boldmath$u$}$. We consider a realistic setup that includes energy losses due to synchrotron and IC coolings, as well as turbulent acceleration with $D_{\rm pp}=D_0p^2$, with the following parameters: $n_H=10^{-4}~\rm cm^{-3}$, $B=1~\muG$, $z_r=0.2$, and $\tau_{\rm pp}=0.5-1$~Gyr. The initial CRe energy density is taken to be $e_c(p)=(p/m_ec)^{-1.5}$ at $t=0$, which corresponds to the shock injection spectrum $f_{\rm inj}(p)\propto p^{-4.5}$ for a $M_s=3$ shock.  

Figure \ref{f14}(a) presents the spectrum of $e_c(p)$ at $t=1$~Gyr, obtained from test simulations with $\tau_{\rm pp}=0.5$~Gyr, computed using different resolutions of uniform logarithmic momentum bins, $N_p=32-8192$. Figure \ref{f14}(b) shows a log-relative $L_1$ error of the spectrum,
\begin{equation} 
L_1^{\rm log}(N_p)\equiv \frac{1}{N_p} \sum_{i=1}^{N_p} \left| \log_{10}\left[\frac{e_c(i)}{e_{c,\rm ref}(p_i)}\right]\right|, \label{ll1}
\end{equation}
where $e_{c,\rm ref}(p_{i})$ is the reference solution obtained from the highest resolution simulation using $N_p=8192$. Given that $e_c(p)$ typically varies several orders of magnitude over the momentum range $p/m_ec = 10 - 10^5$, the standard linear $L_1$ norm would be dominated by errors in the low-momentum bins where $e_c$ is largest. On the other hand, $L_1^{\rm log}$, as defined in equation (\ref{ll1}), is expected to provide a more uniform weighting of the relative error across the entire momentum range, thereby effectively measuring the fidelity of the simulated spectral index and the cooling tail. Panel (a) shows that with $\tau_{\rm IC}\propto p^{-1}$, the highest momentum bin imposes the most stringent constraint. The numerical results are reasonably converged for $N_p\gtrsim 64$. Panel (b) indicates that the $L_1^{\rm log}$ scaling is close to $L_1^{\rm log}(N_p)\propto N_p^{-2}$ for $N_p\gtrsim 256$, while it is shallower at lower resolution, possibly reflecting numerical stiffness introduced by strong radiative cooling near the exponential cutoff.

Figures \ref{f14}(c) - (d) show a comparison between results obtained with $N_p=128$ and $N_p=512$ for test simulations including (c) only synchrotron and IC coolings and (d) both synchrotron and IC coolings and turbulent acceleration with $\tau_{\rm pp}=1$~Gyr. Panel (c) indicates that our FP solver exhibits somewhat slow convergence in the case where the energy spectrum cuts off exponentially due to strong radiative cooling without momentum diffusion. This sharp exponential cutoff generates a large second derivative of $h$, degrading formal convergence. In contrast, even moderate turbulent acceleration with $\tau_{\rm pp}\sim 1.0$~Gyr can delay rapid radiative cooling, so the convergence at $N_p=128$ looks significantly improved, as shown in panel (d). This test demonstrates that reasonably converged solutions are achieved with an affordable resolution of $N_p=64-128$ for the merger shock problems considered in this work.  

Furthermore, we examine the convergence with different FP timesteps for test simulations with $\tau_{\rm pp}=0.5$~Gyr; $\Delta t_{\rm FP} = (1/2)^{N_t} (\Delta h/\max(|A_p|))$ with $N_t=1-5$ is employed, and $N_p=128$ is used. As shown in Figure \ref{f15}(a), the results with different $\Delta t_{\rm FP}$ are almost indistinguishable, implying that the FP timestep in equation (\ref{tFP}) therefore provides a conservative stability criterion. Figure \ref{f15}(b) shows the $L_1^{\rm log}$ error, estimated with different $\Delta t_{\rm FP}$; the error decreases with timestep refinement roughly with third-order temporal accuracy \citep{pareschi2005}.

Simulations presented in the main part of this paper employ $N_p=64$ logarithmically spaced momentum bins covering $p/m_ec=10-10^5$. For a convergence test, we additionally perform a higher-resolution simulation for the \texttt{m4m4$\theta$0} model with $\tau_{pp}=0.1$ Gyr (the strong momentum diffusion case) using $N_p=96$. The results of this higher momentum resolution simulation are compared to those with $N_p=64$ in Figure \ref{f10}(c) and (d). The close agreement between the two runs indicates that the solutions are nearly converged, confirming that $N_p=64$ provides sufficient momentum resolution for the present study.

\begin{figure*}[t]
\vskip 0.2 cm
\includegraphics[width=0.99\linewidth]{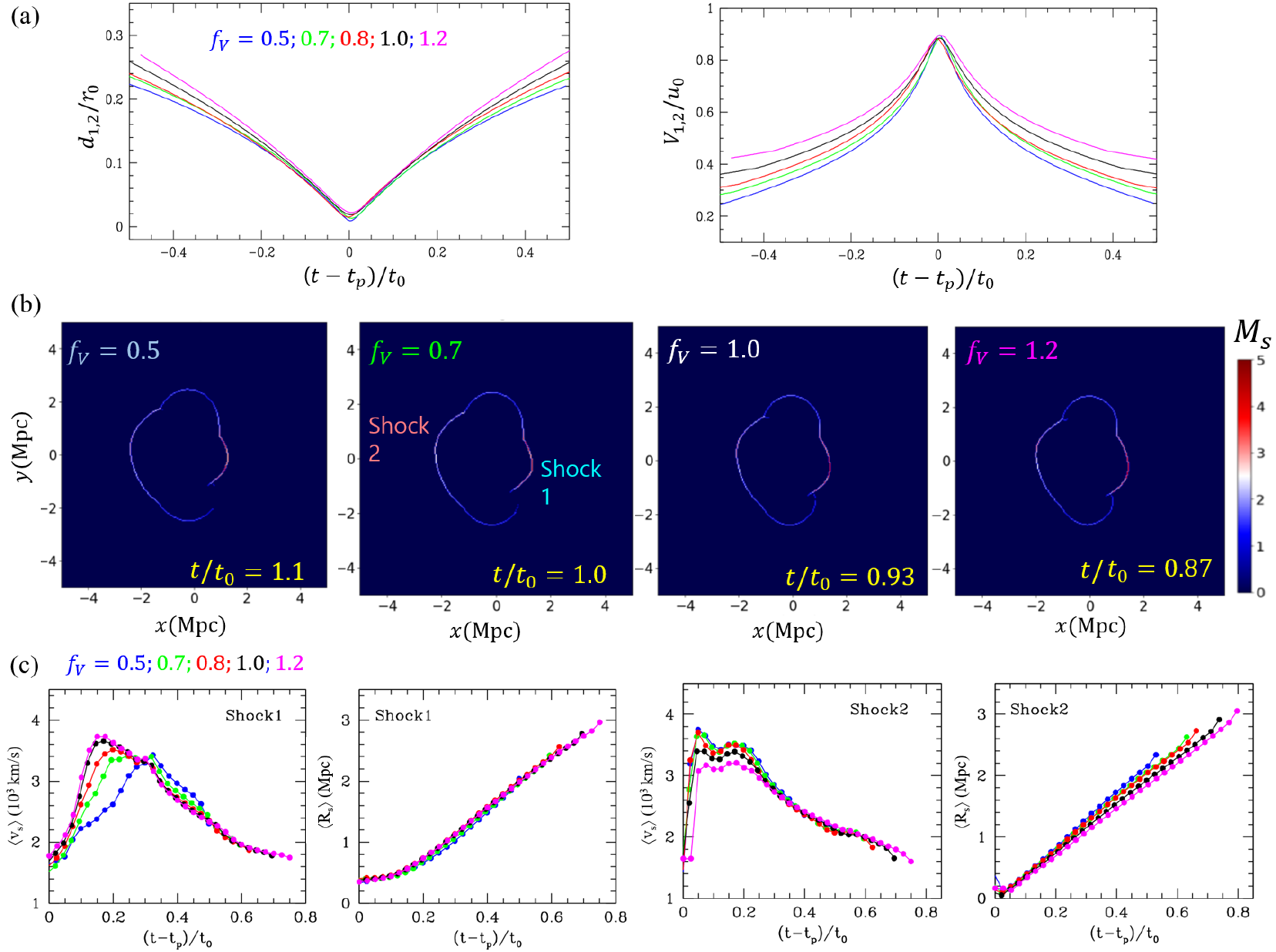}
\vskip -0.1 cm
\caption{(a) Time evolution of the relative separation, $d_{1,2}$, and relative velocity, $V_{1,2}$, between the two DM clumps as a function of time since pericenter passage, $t-t_p$, for the \texttt{m4m2$\theta$10} model with different velocity factors: $f_V=0.5$ (blue), 0.7 (green), 0.8 (red), 1.0 (black), and 1.2 (magenta). (b) 2D maps of the shock Mach number in the $x$--$y$ merger plane ($z=0$) for four corresponding models, shown approximately 1~Gyr after pericenter passage ($(t-t_p)/t_0\approx0.4$). (c) Time evolution of the shock velocity, $\left<V_s\right>$, and shock radius, $\left<R_s\right>$, for Shock~1 and Shock~2 as functions of $t-t_p$ for different values of $f_V$. The same color scheme is used as in (a). All results are obtained from simulations with $256^3$ grid zones.}\label{f17}
\end{figure*}

\begin{figure*}[t]
\vskip 0.1 cm
\includegraphics[width=0.99\linewidth]{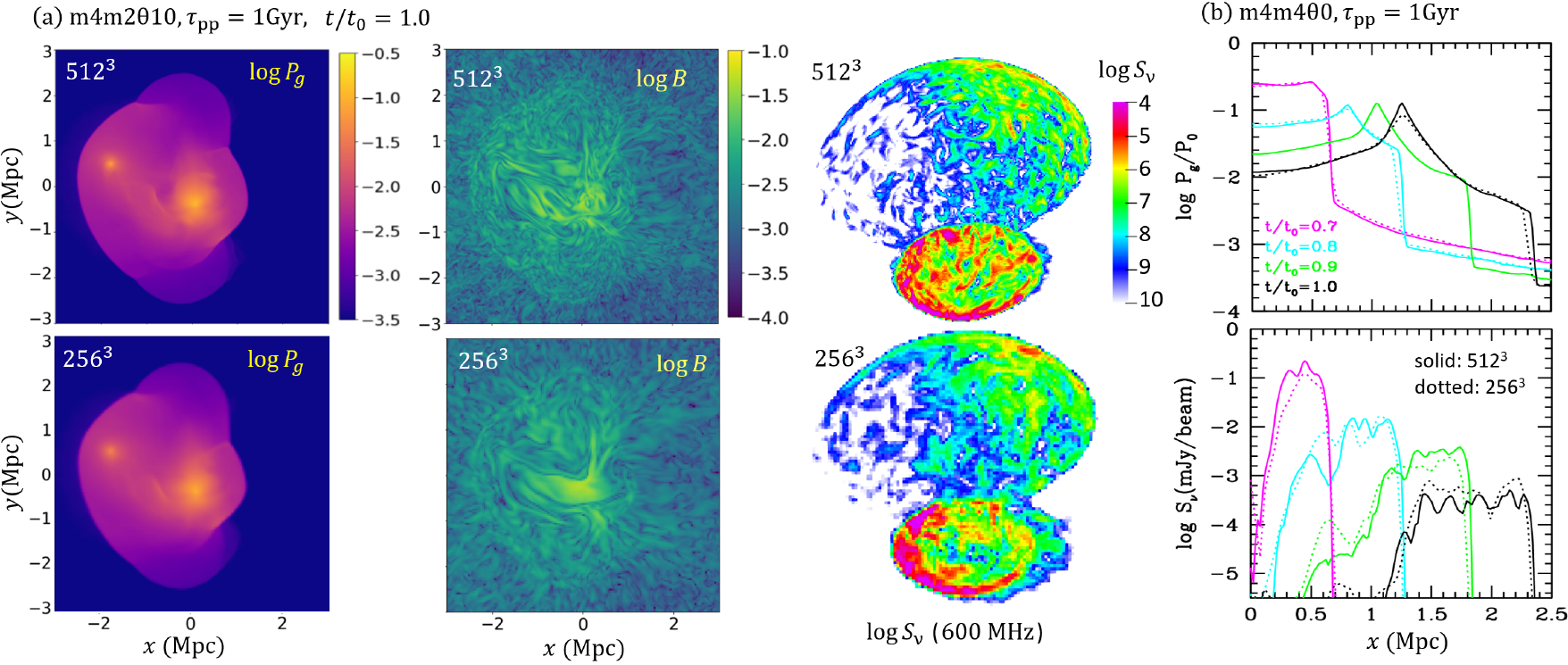}
\vskip -0.1 cm
\caption{{Resolution convergence test comparing the $512^3$ ($\Delta x \approx 19.5$ kpc) and $256^3$ ($\Delta x \approx 39$ kpc) simulations. (a) 2D maps of $P_g/P_0$ and $B/B_0$ in the $x–y$ merger plane ($z=0$), and the synchrotron surface brightness, $S_{\nu}(x',y')$ at 600 MHz, for the \texttt{m4m2$\theta$10} model (unequal-mass off-axis merger) at $t/t_0=1.0$. Here, $S_{\nu}(x',y')$(600 MHz) is calculated by projecting the synchrotron emissivity, $j_{\nu}(\mbox{\boldmath$x'$})$, along the $z'$-axis in the rotated frame $(x',y',z')$ specified by $\theta_{\rm rot}=40^{\circ}$ and $\phi_{\rm rot}=-5^{\circ}$. (b) Temporal evolution of 1D profiles of $P_g/P_0$ along the $x$-axis ($y=z=0$) and $S_{\nu}(x,y)$ at 150~MHz along the $x$-axis ($y=0$), behind Shock 1, for the \texttt{m4m4$\theta$0} model (equal-mass head-on merger). Here, $S_{\nu}(x,y)$(150~MHz) is calculated by projecting $j_{\nu}(\mbox{\boldmath$x$})$ along the $z$-axis in the computational frame $(x,y,z)$. Profiles are shown at $t/t_0=0.7$ (magenta), 0.8 (cyan), 0.9 (green), and 1.0 (black). Solid lines denote the $512^3$ results, while dotted lines denote the $256^3$ results. In both panels (a) and (b), $S_{\nu}$ is given in mJy~beam$^{-1}$ for a beam size of $\theta_{\rm beam}^2 = 1'' \times 1''$.}}\label{f18}
\end{figure*}

\section{Dependence on Model Parameters}\label{sec:sc}

\subsection{Pre-merger Turbulence Generation}\label{sc1}

{For the pre-merger turbulence simulations described in Section \ref{s2.3.1}, velocity perturbations are weighted by the gas density as $\delta\mbox{\boldmath$u$} \propto \rho_g^{w}$, where $w$ is a free parameter that controls the relative strength of turbulence injection in different parts of the clusters. While we primarily present results using $w=0.25$ in Section \ref{s3}, we also examined models with $w=0.5$ and $w=0$. Figure \ref{f16} illustrates the distributions of turbulent velocity and magnetic field strength with these three values of $w$ in the \texttt{m4} subcluster with $512^3$ grid zones. The forcing amplitude was renormalized to ensure $\langle M_{s,\rm turb} \rangle \approx 0.25$ in all cases. As expected, higher values of $w$ lead to stronger velocities and magnetic fields in the dense core region. While the difference in turbulent magnetic field strength between the $w=0.25$ and $w=0.5$ models is relatively small, the $w=0$ case exhibits somewhat weaker central magnetic fields. Nevertheless, in the outskirt region (e.g., at $r\sim1$ Mpc in panel (d)), the strength of turbulent magnetic fields is only weakly affected by the assumed radial scaling of the pre-merger turbulent forcing. Consequently, although not shown here, the large-scale morphology of the radio relics appearing in the outskirts of merged clusters remains broadly similar among the models with different values of $w$.}

\subsection{Effects of Varying Approach Speeds}\label{sc2}

To examine how variations in the velocity factor, $f_V$, affect the properties of merger shocks, Figure~\ref{f17}(a) shows the time evolution of the relative separation, $d_{1,2}$, and the relative velocity, $V_{1,2}$, between the two DM clumps {as functions of the time since pericenter passage, $t-t_p$,} for a series of \texttt{m4m2$\theta$10} simulations performed with $256^3$ grid zones and different values of $f_V$. As the clumps initially approach each other, their relative velocity increases, reaches a maximum of $V_{1,2}/u_0\approx0.9$ near pericenter passage, and subsequently declines. Increasing the initial approach speed (i.e., adopting larger $f_V$) mainly shifts the timing of pericenter passage to earlier epochs, but the peak value of $V_{1,2}$ remains nearly unchanged over the range $f_V=0.5-1.2$.

This behavior indicates that the mutual gravitational attraction of the two subclusters dominates the acceleration of the DM clumps and, consequently, the merger dynamics near pericenter. As a result, the properties of the merger shocks are relatively insensitive to the exact choice of $f_V$ within the range considered. Figure~\ref{f17}(b) confirms that the overall morphology, spatial extent, and Mach-number distribution of Shock~1, Shock~2, and the equatorial shocks remain very similar when the simulations are compared at equivalent stages of the merger evolution.

{Figure~\ref{f17}(c) further shows the evolution of the shock speed, $\left<V_s\right>$, and shock radius, $\left<R_s\right>$, for Shock~1 and Shock~2 as functions of $t-t_p$ for different values of $f_V$. Here, $\left<V_s\right>$ and $\left<R_s\right>$ are the surface-area-weighted values, calculated over the shock zones classified to Shock~1 and Shock~2 (see Section \ref{s3.1}). There are notable differences in $\left<V_s\right>$ around pericenter passage, reflecting the flow histories caused by different initial approach velocities. However, around our representative observation epoch, $(t-t_p)/t_0\approx0.4$, the values of both $\left<V_s\right>$ and $\left<R_s\right>$ nearly overlap, implying that the resulting radio relic would appear similarly. The similarity in $\left<V_s\right>$ provides a dynamical explanation for the nearly identical Mach-number distributions shown in Figure~\ref{f17}(b).}

These results demonstrate that the shock properties and the physical interpretations presented in Section~\ref{s3.1} are robust against order-unity variations in the initial approach speed (i.e., $f_V\sim1$).

Although not shown, we find that a similar conclusion applies to order-unity variations in the initial separation factor, $f_d\sim1$, provided that the comparison is made at comparable stages of the merger evolution.

\subsection{Effects of Numerical Resolution}\label{sc3}

To illustrate the impact of numerical resolution on our results, Figure~\ref{f18}(a) presents the {2D maps} of $P_g/P_0$, $B/B_0$, and $S_{\nu}/S_0$ at 600 MHz for the \texttt{m4m2$\theta$10} model at $t/t_0=1.0$; these quantities are drawn from two simulations with different grid resolutions, $256^3$ and $512^3$.  This particular model is also shown in Figures \ref{f8}(b), \ref{f11}(a), and \ref{f12}. The 2D distributions of $P_g$ and $B$ are taken in the $x$–$y$ merger plane ($z=0$), while the surface brightness $S_{\nu}$ shows the projected synchrotron emission along a tilted LoS.

The gas pressure {maps} indicate that, although the higher resolution run yields sharper pressure peaks, the large-scale hydrodynamic structures depend only weakly on resolution within this range. In particular, the spatial positions and extents of the merger shocks are nearly identical in the two simulations. The $256^3$ run, however, produces slightly higher shock speeds and correspondingly larger mean Mach numbers than the $512^3$ run, although the differences remain modest, as mentioned in Section~\ref{s3.1}.

In contrast, the magnetic-field strength maps clearly show that turbulent magnetic fluctuations extend down to the grid scale, highlighting the importance of numerical resolution in reproducing small-scale magnetic structures. As a result, the fine structures in $S_{\nu}$ inherit this resolution dependence, since both synchrotron emissivity and radiative cooling directly respond to the local magnetic field strength. With current and future radio facilities such as LOFAR, SKA, and ngVLA, it will be of considerable interest to assess whether such fine-scale features in $S_{\nu}$ are observable. If they are detectable, the capability of high-accuracy, high-resolution simulations to reproduce these features will become increasingly important.

Figure~\ref{f18}(b) compares the temporal evolution of 1D profiles of $P_g/P_0$ and $S_{\nu}/S_0$ at 150~MHz along the $x$-axis ($y=z=0$) behind Shock 1 for the \texttt{m4m4$\theta$0} model. This particular model is also shown in Figure \ref{f6}. The convergence between the $512^3$ and $256^3$ runs is reasonable for both quantities, although $S_{\nu}/S_0$ in the $256^3$ run is slightly larger at $t/t_0=1.0$, reflecting the larger shock Mach numbers mentioned above.

\end{document}